\documentclass[twoside,12pt]{article}
\usepackage{epsfig}

\newcommand{\be}{\begin{equation}}
\newcommand{\ee}{\end{equation}}
\newcommand{\bea}{\begin{eqnarray}}
\newcommand{\eea}{\end{eqnarray}}

\usepackage{amsmath}
\usepackage{amsfonts}
\usepackage{amssymb}
\usepackage{amsxtra}
\usepackage{supertabular}
\begin{document}
\title{ \vspace{1cm} 
How Clifford algebra can help understand second quantization
of fermion and boson fields}

\author{N.S.\ Manko\v c Bor\v stnik$^{1}$ 
\\
$^1$Department of Physics, University of Ljubljana\\
SI-1000 Ljubljana, Slovenia } 
\maketitle

\begin{abstract}
In the review article in Progress in Particle and Nuclear Physics~\cite{nh2021RPPNP} 
the authors present a rather detailed review of the achievements so far of the 
{\it spin-charge-family} theory that offers the explanation for the observed 
properties of elementary fermion and boson fields, if the space-time is higher than 
$d=(3+1)$, it must be $d\ge (13+1)$, while fermions only interact with gravity. Ref.~\cite{nh2021RPPNP} presents also an explanation for the second quantization postulates for the fermion fields, since the internal space of fermions is in this theory 
described with the "basis vectors" determined by the Clifford odd objects. The anticommutativity of the "basis vectors" namely transfers to their creation and 
annihilation operators. 
This paper shows that the "basis vectors" determined by the Clifford even objects, 
if used to describe the internal space of boson fields, not only manifest all the 
known properties of the observed boson fields, but offer as well the explanation 
for the second quantization postulates for boson fields. 
Properties of fermion and boson fields with the internal spaces described by the 
Clifford odd and even objects, respectively, are demonstrated on the toy model 
with $d=(5+1)$.
 \end{abstract}




%
\section{Introduction}
\label{introduction}

In a long series of works~\cite{norma92,%
norma93,pikanorma,IARD2016,n2014matterantimatter,JMP2013,normaJMP2015,%
n2012scalars} 
the author 
has found, together with the collaborators~(\cite{nh02,nh03,nd2017,%
nh2018,n2019PIPII,2020PartIPartII,nh2021RPPNP} and the references therein), and in long 
discussions with participants during the annual workshops ''What comes beyond the 
standard models'', the phenomenological success with the model named the 
{\it spin-charge-family}  theory with the properties: \\ 
{\bf a.} The internal space of fermions are described by the ''basis  vectors'' which are 
superposition of odd products of anticommuting objects $\gamma^a$'s in 
$d=(13+1)$~\cite{prd2018,n2019PIPII,2020PartIPartII,nh2021RPPNP}.  
Correspondingly the ''basis vectors'' of one Lorentz irreducible representation in internal 
space of fermions, together with their Hermitian conjugated partners, anticommute,
fulfilling (on the vacuum state) all the requirements for the second quantized fermion fields~\cite{prd2018,n2019PIPII,nh02,nh03,nh2021RPPNP}. \\
{\bf b.} The second kind of anticommuting objects, $\tilde{\gamma}^a$'s, equip each  
 irreducible representation  of  odd ''basis vectors'' with the family quantum number~%
\cite{2020PartIPartII,nh03}.\\
{\bf c.} Creation operators for single fermion states --- which are tensor products, $*_{T}$,
of a  finite number of  odd ''basis vectors'' appearing in $2^{\frac{d}{2}-1}$ families,
each family with $2^{\frac{d}{2}-1}$ members, and the (continuously) infinite 
momentum/coordinate basis applying on the vacuum state~\cite{2020PartIPartII,
nh2021RPPNP} --- inherit anticommutativity of ''basis vectors''. Creation operators 
and their Hermitian conjugated partners correspondingly anticommute.\\
{\bf d.} The Hilbert space of second quantized fermions is represented by the tensor 
products of all possible number of creation operators, from zero to 
infinity~\cite{n2019PIPII}, applying on a vacuum state.\\
{\bf e.} In a simple starting action  massless fermions carry only spins and interact with 
only gravity --- with the vielbeins and the two kinds of the spin connection  fields (the 
gauge fields of  momenta, of $S^{ab}=\frac{i}{4}(\gamma^a \gamma^b- \gamma^b 
\gamma^a)$ and of $\tilde{S}^{ab}=\frac{1}{4} (\tilde{\gamma}^a \tilde{\gamma}^b - 
\tilde{\gamma}^b  \tilde{\gamma}^a)$, respectively~%
\footnote{
If there are no fermions present the two kinds of the spin connection fields are 
uniquely expressible by the vielbeins~\cite{prd2018}.}). The starting action includes
only even products of $\gamma^a$'s and $\tilde{\gamma^a}$'s~(\cite{prd2018} and
references therein). \\
{\bf f.} Spins from higher dimensions, $d>(3+1)$, described by $\gamma^a$'s,  
manifest in $d=(3+1)$ all the charges of the {\it standard model} quarks and 
leptons and antiquarks and antileptons of particular handedness. \\
{\bf g.} Gravity --- the gauge fields of $S^{ab}$, ($(a,b)=(5,6,....,d)$), with the space 
index $m=(0,1,2,3)$ --- manifest as the {\it standard model} vector gauge 
fields~\cite{nd2017}. The scalar gauge fields of $\tilde{S}^{ab}$ and  of some of 
superposition of $S^{ab}$, with the space index $s=(7,8)$  manifest as the scalar 
higgs  and Yukawa couplings~\cite{n2012scalars,JMP2013,IARD2016,nh2021RPPNP},
determining mass matrices (of particular symmetry) and correspondingly the masses 
of quarks and leptons 
and of the weak boson fields after (some of) the scalar fields with the space index 
$(7,8)$ gain constant values. The scalar gauge fields of $\tilde{S}^{ab}$ and of 
$S^{ab}$ with the space index $s=(9,10,...,14)$ and $(a,b)=(5,6,....,d)$ offer the 
explanation for the observed matter/antimatter asymmetry~\cite{n2014matterantimatter,normaJMP2015,
nh2018,nh2021RPPNP,normaBled2020} in the universe.\\
{\bf h.} The theory predicts the fourth family to the observed three~\cite{mdn2006,
gmdn2007,gmdn2008,gn2013,gn2014} and the stable fifth family of heavy quarks 
and leptons. The stable fifth family nucleons offer the explanation for the appearance 
of the dark matter.
Due to heavy masses  of the fifth family quarks the nuclear interaction among hadrons 
of the fifth family members is verry different than the ones so far observed~\cite{gn2009,nm2015}. \\
{\bf i.} The theory offers a new understanding of the second quantized fermion 
fields (explained in Ref.~\cite{nh2021RPPNP}) as well as of the second quantized 
boson fields. The second quantization of boson fields, the gauge fields of the second
quantized fermion fields, is the main topic of this paper~\cite{n2021SQ}.\\
{\bf j.} The theory seems promising to offer a new insight into Feynman diagrams.

\vspace{2mm}

The more work  is put into the theory the more phenomena the theory is able to 
explain.  
\vspace{2mm}

In this paper we shortly overview the description  of the internal space of the second 
quantized massless fermion fields with the ''basis vectors'' which are the superposition 
of odd products of the Clifford  algebra objects (operators) $\gamma^a$'s. Tensor products
of ''basis vectors'' with the basis in ordinary space form the creation operators for fermions
which fulfil the anticommutation relations of the Dirac second quantized fermion fields,  
without postulating them~\cite{Dirac,BetheJackiw,Weinberg}.
We kindly ask the reader to read the explanations in Ref.~\cite{nh2021RPPNP}, Sect.~3.

The ''basis vectors'', which are the superposition of {\it even} products of the Clifford 
algebra objects $\gamma^a$'s and are in a tensor product , $*_{T}$, with the basis 
in ordinary space, have all the properties of the second quantized boson fields, the 
gauge fields of the corresponding second quantized fermion fields.
The main part of this paper discusses properties of the internal space of the second
quantized boson fields described by the Clifford even ''basis vectors'' in interaction 
among the boson fields themselves and with the second quantized fermion fields 
with the internal space described by the Clifford odd ''basis vectors'', 
Ref.~\cite{nh2021RPPNP}.

In both cases when describing the second quantized either fermion or boson fields 
the creation operators are considered to be a tensor, $*_T$, product of 
$2^{\frac{d}{2}-1}\times 2^{\frac{d}{2}-1}$ of either anticommuting  Clifford odd 
(in the case of fermion fields) or commuting Clifford even (in the case of boson fields) 
''basis vectors'' and of (continuously) infinite commuting basis of ordinary space. 

While in the case of the Clifford odd ''basis vectors'' the Hermitian conjugated partners 
belong to another group with $2^{\frac{d}{2}-1}\times 2^{\frac{d}{2}-1}$ members 
(which is not reachable by either $S^{ab}$ or $\tilde{S}^{ab}$) or both, in the 
case of the Clifford even ''basis vectors'' each of the two groups with $2^{\frac{d}{2}-1}
\times 2^{\frac{d}{2}-1}$ members  have their Hermitian conjugated partners among 
themselves, that is within the group reachable by  either $S^{ab}$ or $\tilde{S}^{ab}$. \\

Subects.\ref{GrassmannClifford},~\ref{basisvectors},~\ref{reduction} of 
Sect.\ref{creationfermionsbosons} are a short overview of the Clifford odd and the
Clifford  even  algebra, used to described in the {\it spin-charge-family} theory the 
internal space of fermions ---  as already presented in Ref.~(\cite{nh2021RPPNP} in 
Sect. 3) --- and in this paper in particular 
the internal space of bosons, as the author started in Ref.~\cite{n2021SQ}.
In Subsect.~\ref{GrassmannClifford} the anticommuting Grassmann algebra and the 
two Clifford subalgebras, each 
algebra with $2\times 2^{d}$ elements, are presented, and the relations 
among them discussed. 

In Subsect.~\ref{basisvectors} the ''basis vectors'' of either odd or even character are 
defined as eigenvectors of all the members of the Cartan subalgebra of the Lorentz 
algebra for the Grassmann and the two Clifford subalgebras~(\cite{nh2021RPPNP}, Sect. 3). 

The ''basis vectors'' are products of nilpotents and projectors, each nilpotent and each 
projector is chosen to be the eigenvector of one member of the Cartan subalgebra.
The anticommuting ''basis vectors'' have an odd number of nilpotents, the commuting
''basis vectors'' have an even number of nilpotents.

The anticommutation relations  (for fermions, with the odd number of nilpotents)  and commutation relations (for bosons, with the even number of nilpotents)  are presented. 

There are obviously only one kind of fermion fields and correspondingly also of 
their gauge fields observed. There is correspondingly no need for two Clifford 
subalgebras.

In Subsect.~\ref{reduction}  this problem is solved by the reduction of the two  Clifford subalgebras to only one, what enables also to give the family quantum numbers to  
the Clifford odd anticommuting ''basis vectors'', belonging to different  irreducible 
representations of the Lorentz algebra.
The reduction enables as well to define to the Clifford even commuting ''basis vectors''   
 the generators of the Lorentz transformations in the internal space
of bosons.

In Subsect.~\ref{cliffordoddevenbasis5+1} the ''basis vectors'' for fermions, \ref{odd5+1}, 
and bosons, \ref{even5+1}, are discussed in details for the ''toy model'' in $d=(5+1)$  to 
make differences in the properties of the Clifford odd and Clifford even ''basis vectors''  
transparent and correspondingly easier to understand. The algebraic application of the
Clifford even ''basis vectors'' on the Clifford odd ''basis vectors'' is demonstrated, as well as
the algebraic application of the Clifford even ''basis vectors'' on themselves. This subsection
is the main part of the article.

In Subsect.~\ref{generalbasisinternal} the generalization  of the Clifford odd and 
Clifford even ''basis vectors'' to any even $d$ is discussed.

In Sect.~\ref{fermionsbosons} the creation operators of the second quantized 
fermion and boson fields offered by the {\it spin-charge-family} theory are studied.

Sect.~\ref{conclusions} reviews shortly what one can learn in this article and what
remains to study.

In App.~\ref{13+1representation}  ''basis vectors'' of one family of quarks and leptons 
are presented. 

In App.~\ref{grassmannandcliffordfermions} some useful relations are presented.

\section{Properties of creation and annihilation operators for fermions and bosons}
\label{creationfermionsbosons}

Second quantization postulates for fermion and boson fields~\cite{Dirac,BetheJackiw,Weinberg} require that the creation and annihilation 
operators for fermions and bosons depend on finite number of spins and other quantum numbers determining internal space of fermions and bosons and  on infinite number of momenta (or coordinates).  While fermions carry half integer spins and charges in fundamental representations of the corresponding groups bosons carry integer spins
and charges in adjoint representations of the groups.
Ref.~\cite{nh2021RPPNP} reports in Subsect. ~3.3.1. second quantization postulates 
for fermions.

The  first quantized fermion states are  in the Dirac's theory vectors which do not 
anticommute. There are the  creation operators  of the second quantized fermion 
fields which are postulated to anticommute.
The second quantized fermion fields commute with $\gamma^a$ matrices, allowing
the second quantization of the Dirac equation which includes the mass term. 

Creation and annihilation of boson fields  are postulated to fulfil commutation relations.

\vspace{5mm}

 In the {\it spin-charge-family} theory the internal space of fermions  and bosons 
in even dimensional spaces $ d=2(2n+1)$ is described by the algebraic, $*_{A}$, 
products of $\frac{d}{2}$ nilpotents and projectors, which are superposition of 
odd (nilpotents) and even (projectors) numbers of anticommuting operators 
$\gamma^a$'s. Nilpotents and projectors are chosen to be eigenvectors of 
$\frac{d}{2}$ Cartan subalgebra members of the Lorentz algebra of $S^{ab}
=\frac{i}{4}\{ \gamma^{a}\,, \gamma^{b}\}_{-}$, determining the internal 
space of fermions and bosons. \\
There are two groups of $2^{\frac{d}{2}-1}$ members appearing in 
$2^{\frac{d}{2}-1}$ irreducible representations which have an odd number of 
nilpotents (at least one nilpotent and the rest projectors). The members of 
one of the groups are (chosen to be) called ''basis vectors''. The other group
contains the Hermitian conjugated partners of the ''basis vectors''. The group of
these odd ''basis vectors'' have all the properties needed to describe internal 
space of fermions.\\
There are two groups with an even number of nilpotents, each with 
$2^{\frac{d}{2}-1}$ $2^{\frac{d}{2}-1}$  members. Each of these two groups 
have their Hermitian conjugated partners within the same group. Each of the
two groups with an even number of nilpotents have all the properties needed
to describe the internal space of boson fields, as we shall see in this article.

N.S.M.B. made a choice of $d=(13+1)$ since for such $d$ the theory offers the 
explanation for all the assumptions of the {\it standard model}, that is  for the 
charges, handedness, families of quarks and leptons and antiquarks and antileptons,
for all the observed vector gauge fields, as well as for the scalar higgs and Yukawa 
couplings. 
 
 A simple starting action~(\cite{nh2021RPPNP} and the references therein)  for the 
second quantized massless fermion and antifermion fields, and the corresponding 
massless boson fields in  $d=2(2n+1)$-dimensional space is assumed to be
%
\begin{eqnarray}
{\cal A}\,  &=& \int \; d^dx \; E\;\frac{1}{2}\, (\bar{\psi} \, \gamma^a p_{0a} \psi) 
+ h.c. +
\nonumber\\  
               & & \int \; d^dx \; E\; (\alpha \,R + \tilde{\alpha} \, \tilde{R})\,,
\nonumber\\
               p_{0a } &=& f^{\alpha}{}_a p_{0\alpha} + \frac{1}{2E}\, \{ p_{\alpha},
E f^{\alpha}{}_a\}_- \,,\nonumber\\
          p_{0\alpha} &=&  p_{\alpha}  - \frac{1}{2}  S^{ab} \omega_{ab \alpha} - 
                    \frac{1}{2}  \tilde{S}^{ab}   \tilde{\omega}_{ab \alpha} \,,
                    \nonumber\\                    
R &=&  \frac{1}{2} \, \{ f^{\alpha [ a} f^{\beta b ]} \;(\omega_{a b \alpha, \beta} 
- \omega_{c a \alpha}\,\omega^{c}{}_{b \beta}) \} + h.c. \,, \nonumber \\
\tilde{R}  &=&  \frac{1}{2} \, \{ f^{\alpha [ a} f^{\beta b ]} 
\;(\tilde{\omega}_{a b \alpha,\beta} - \tilde{\omega}_{c a \alpha} \,
\tilde{\omega}^{c}{}_{b \beta})\} + h.c.\,.               
\label{wholeaction}
\end{eqnarray}
Here~\footnote{$f^{\alpha}{}_{a}$ are inverted vielbeins to 
$e^{a}{}_{\alpha}$ with the properties $e^a{}_{\alpha} f^{\alpha}{\!}_b = 
\delta^a{\!}_b,\; e^a{\!}_{\alpha} f^{\beta}{\!}_a = \delta^{\beta}_{\alpha} $, 
$ E = \det(e^a{\!}_{\alpha}) $.
Latin indices  
$a,b,..,m,n,..,s,t,..$ denote a tangent space (a flat index),
while Greek indices $\alpha, \beta,..,\mu, \nu,.. \sigma,\tau, ..$ denote an Einstein 
index (a curved index). Letters  from the beginning of both the alphabets
indicate a general index ($a,b,c,..$   and $\alpha, \beta, \gamma,.. $ ), 
from the middle of both the alphabets   
the observed dimensions $0,1,2,3$ ($m,n,..$ and $\mu,\nu,..$), indexes from 
the bottom of the alphabets
indicate the compactified dimensions ($s,t,..$ and $\sigma,\tau,..$). 
We assume the signature $\eta^{ab} =
diag\{1,-1,-1,\cdots,-1\}$.} 
$f^{\alpha [a} f^{\beta b]}= f^{\alpha a} f^{\beta b} - f^{\alpha b} f^{\beta a}$.
The $\gamma^a$  operators appear in the Lagrangean for second quantized massless
fermion fields in pairs. \\

Fermions, appearing in families, carry only spins and only interact with gravity, what 
manifests in $d=(3+1)$ as spins and all the observed charges. Vielbeins, 
$f^a_{\alpha}$ (the gauge field of momenta), and two kinds of the spin connection 
fields, $\omega_{ab \alpha}$ (the gauge fields of $S^{ab}$) and 
$\tilde{\omega}_{ab \alpha}$  (the gauge fields of $\tilde{S}^{ab}$), manifest 
in $d=(3+1)$ as the known vector gauge fields~\cite{nd2017} and the 
scalar gauge fields taking care of masses of quarks and leptons and antiquarks and 
antileptons~\cite{nh2021RPPNP} and the weak boson fields~\footnote{
Since the multiplication with either $\gamma^a$'s or $\tilde{\gamma}^a$'s  
changes the Clifford odd ''basis vectors'' into 
the Clifford even ''basis vectors'', and even ''basis vectors'' commute, the action for 
fermions can not include an odd numbers of $\gamma^a$'s or $\tilde{\gamma}^a$'s, 
what the simple starting action of Eq.~(\ref{wholeaction}) does not. In the starting 
action $\gamma^a$'s and $\tilde{\gamma}^a$'s appear as
$\gamma^0 \gamma^a \hat{p}_a$  or as $\gamma^0 \gamma^c \, S^{ab}\omega_{abc}$
 and  as $\gamma^0 \gamma^c \,\tilde{S}^{ab}\tilde{\omega}_{abc} $.}.

Internal space of fermions is described in any even dimensional space by ''basis vectors''
with an odd number of nilpotents, the rest are projectors, as mentioned above. 
Nilpotents, described by an 
odd number of $\gamma^a$'s, anticommute with $\gamma^a$, projectors, described 
by an even number of $\gamma^a$'s, commute with $\gamma^a$. 
Correspondingly the ''basis vectors'', describing the internal space of fermions, 
anticommute among themselves.

Internal space of bosons is described in any even dimensional space by ''basis vectors''
with an even number of nilpotents and of projectors. The ''basis vectors'' describing 
bosons therefore commute.

\vspace{3mm}

Creation operators for either second quantized fermions or bosons are tensor products, 
$*_{T}$, of the Clifford odd $2^{\frac{d}{2}-1} \times$ $2^{\frac{d}{2}-1}$ ''basis 
vectors'' describing the internal space of fermions or of the Clifford even 
$2^{\frac{d}{2}-1} \times$ $2^{\frac{d}{2}-1}$ ''basis vectors'' describing the internal 
space of bosons, and of an (continuously) {\it infinite number of basis in ordinary 
momentum (or coordinate) space.}

Creation operators for fermions, represented by anticommuting ''basis vectors'' 
in a tensor product, $*_{T}$, with the basis in ordinary space,  anticommute among 
themselves and with $\gamma^a$'s, creation operators for bosons, represented by 
commuting ''basis vectors'' in a tensor product, $*_{T}$, with the basis in ordinary 
space, commute among themselves and with $\gamma^a$'s.  

''Basis vectors'' describing the internal space of fermion and boson fields transfer their
anticommutativity or commutativity into creation operators, since the basis in ordinary 
space commute.


\vspace{3mm}

One irreducible representation of  ''basis vectors'', reachable with $S^{ab}$, and
determining the internal space of fermions in $d=(13+1)$-dimensional space, 
includes quarks and leptons and antiquarks and antileptons, together with the 
right handed neutrinos and the left handed antineutrinos, as can be seen in 
Table~\ref{Table so13+1.}. There are $64$ ($=2^{\frac{d}{2}-1}$) members 
of one irreducible representation, represented as the eigenvectors of the Cartan 
subalgebra of the $SO(13+1)$ group, analysed with respect to the subgroups 
$SO(3,1), SU(2), SU(2), SU(3)$  and $U(1)$, with $7$ commuting 
operators~\footnote{The 
$SO(7,1)$ part is identical for quarks  of any of the three colours and for the 
colourless leptons, and identical for antiquarks  and the colourless antileptons. 
They differ only in the $SO(6)$ part of the group $SO(13,1)$.}.

These Clifford odd  anticommuting ''basis vectors'' transfer the anti-commutativity 
to the creation operators for quarks and leptons and antiquarks and antileptons.

The  {\it standard model} subgroups ($SO(3,1), SU(2), SU(3)$, $U(1)$) have one 
$SU(2)$ group less, and correspondingly the right handed neutrinos and the left 
handed antinutrinos, having no charge, are in the {\it standard model} assumed not
 to exist.

\vspace{2mm}

$S^{ab}$'s transform each member of one irreducible representation of fermions 
into all the members of the same irreducible representation, $\tilde{S}^{ab}$'s  
transform each member of one irreducible representation to the same member 
of another irreducible representation.
The postulate, presented in Eq.~(\ref{tildegammareduced}), equips each irreducible 
representation  with the {\it family quantum number.}

The mass terms appear in the {\it spin-charge-family} theory after the scalar fields with 
the space index $s\ge 5$ ($s=(7,8)$ indeed), which are the gauge fields of the two kinds 
of the spin connection fields $\omega_{ab s}$  and $\tilde{\omega}_{ab s}$ (the gauge 
fields of $S^{ab}$ and $\tilde{S}^{ab}$, respectively), gain the constant values, what 
makes particular charges (hypercharge $Y=\tau^4 +\tau^{23}$ and  the weak charge 
$\tau^{13}$, explained in Table~\ref{Table so13+1.}) non conserved quantities. The appearance of the mass term in $d=(3+1)$ is discussed in Subsects.  6.1, 6.2.2, 7.3 and
7.4 in Ref.~\cite{nh2021RPPNP}. \\


\vspace{2mm} 

The  ''basis vectors'', represented by the superposition of even products $\gamma^a$'s 
(with an even number of nilpotents and the rest of projectors, the eigenvectors of the 
Cartan subalgebra members) have properties of the internal space of boson fields: \\
{\bf i.} $2^{\frac{d}{2}-1}\times 2^{\frac{d}{2}-1}$ members of the Clifford even 
''basis vectors'' are Hermitian conjugated to each other or  are self adjoint~\footnote{ 
The Clifford  odd ''basis vectors'' presented in  Table~\ref{Table so13+1.} are 
orthogonal, that means that the algebraic, $*_{A}$,  product of any two ''basis 
vectors'' is equal to zero. The Hermitian conjugated partners of  ''basis 
vectors'', contributing to annihilation operators, form a separated group of  
$2^{\frac{d}{2}-1}\times 2^{\frac{d}{2}-1}\times $  
$2^{\frac{d}{2}-1}\times 2^{\frac{d}{2}-1}$ members. 

The Clifford odd ''basis vectors'' form in a tensor product with the basis in ordinary 
space the creation operators which determine, applying on the vacuum state, the 
Hilbert space of fermions. 
The Clifford even ''basis vectors'', applying algebraically on the Clifford odd ''basis 
vectors'', ''offer''  the interaction among fermions, transforming one ''basis vector'' 
into the other, Subsect.~\ref{even5+1}, manifesting properties
of the boson fields.}. 
They appear  in two separated groups. \\
{\bf ii.} Any member of these two groups of the Clifford even ''basis vectors'' carries 
with respect to ${\bf {\cal S}}^{ab}(= S^{ab} + \tilde{S}^{ab})$ the quantum 
numbers in the adjoint representations --- either spins of the group 
$SO(13,1)$ or spins and charges with respect to  the subgroups 
$SO(3,1)\times SU(2)\times SU(2)\times SU(3)\times U(1)$)  of the group 
$SO(13,1)$, 
if we make this analyses.\\
{\bf iii.}  The algebraic application of even ''basis vectors'' of one  of the two groups 
on the odd ''basis vectors'' (representing the internal space of fermion fields) 
transforms members of one odd irreducible representation into all the other 
members of the same representation, keeping the family quantum number 
unchanged.\\
{\bf iv.} The algebraic application of an even ''basis vector''  to another  even 
''basis vector'' of the same group leads to an even ''basis vectors'' (or zero).\\
{\bf v.} The algebraic application of any even ''basis vector''  to an even 
''basis vector'' of the second group gives or zero.\\

 The properties of even ''basis vectors'' are demonstrated  in the 
 Subsects.~(\ref{even5+1},                                                                                                                                                                                                                                                                                                                                                                                                   \, \ref{generalbasisinternal})                                                                                                                                                                                                                                                                                                                                                                                                         of this section. \\

\subsection{Grassmann and Clifford algebras}
\label{GrassmannClifford}

The internal space of fermions and bosons can be described by using either the 
Grassmann or the Clifford algebras. A part of this section, the one which concerns 
the Clifford odd ''basis vectors'' is a short overview of the Subsect. 3.2, 
of Ref.~\cite{nh2021RPPNP}.
 
In Grassmann $d$-dimensional space there are $d$ anticommuting operators 
$\theta^{a}$,
 and $d$ anticommuting operators which are derivatives with respect to $\theta^{a}$,
$\frac{\partial}{\partial \theta_{a}}$, 
%
\begin{eqnarray}
\label{thetaderanti0}
\{\theta^{a}, \theta^{b}\}_{+}=0\,, \, && \,
\{\frac{\partial}{\partial \theta_{a}}, \frac{\partial}{\partial \theta_{b}}\}_{+} =0\,,
\nonumber\\
\{\theta_{a},\frac{\partial}{\partial \theta_{b}}\}_{+} &=&\delta_{ab}\,, (a,b)=(0,1,2,3,5,\cdots,d)\,.
\end{eqnarray}

Defining~\cite{nh2018} 
\begin{eqnarray}
(\theta^{a})^{\dagger} &=& \eta^{a a} \frac{\partial}{\partial \theta_{a}}\,,\quad
{\rm leads  \, to} \quad
(\frac{\partial}{\partial \theta_{a}})^{\dagger}= \eta^{a a} \theta^{a}\,,
\label{thetaderher0}
\end{eqnarray}
with $\eta^{a b}=diag\{1,-1,-1,\cdots,-1\}$.

$ \theta^{a}$ and $ \frac{\partial}{\partial \theta_{a}}$ are, up to the sign, Hermitian conjugated to each other. The identity is the self adjoint member of the algebra. We 
make the choice for the following complex properties of $\theta^a$, and correspondingly 
of $\frac{\partial}{\partial \theta_{a}}$,
%
$$ \{\theta^a\}^* =  (\theta^0, \theta^1, - \theta^2, \theta^3, - \theta^5,
\theta^6,...,- \theta^{d-1}, \theta^d)\,,$$ 
$$\{\frac{\partial}{\partial \theta_{a}}\}^* = (\frac{\partial}{\partial \theta_{0}},
\frac{\partial}{\partial \theta_{1}}, - \frac{\partial}{\partial \theta_{2}},
\frac{\partial}{\partial \theta_{3}}, - \frac{\partial}{\partial \theta_{5}}, 
\frac{\partial}{\partial \theta_{6}},..., - \frac{\partial}{\partial \theta_{d-1}}, 
\frac{\partial}{\partial \theta_{d}})\,. $$

The operators $\theta^{a}$ are offering $2^d$ superposition of products of  $\theta^{a}$, 
the Hermitian conjugated partners of which are the corresponding superposition of products 
of $\frac{\partial}{\partial \theta_{a}}$.

In $d$-dimensional space of anticommuting Grassmann coordinates and of their Hermitian conjugated partners derivatives, Eqs.~(\ref{thetaderanti0}, \ref{thetaderher0}), there 
exist two kinds of the Clifford algebra elements (operators) --- $\gamma^{a}$ and 
$\tilde{\gamma}^{a}$ --- both expressible in terms of $\theta^{a}$ and their conjugate momenta $p^{\theta a}= i \,\frac{\partial}{\partial \theta_{a}}$ ~\cite{norma93}.
\begin{eqnarray}
\label{clifftheta1}
\gamma^{a} &=& (\theta^{a} + \frac{\partial}{\partial \theta_{a}})\,, \quad 
\tilde{\gamma}^{a} =i \,(\theta^{a} - \frac{\partial}{\partial \theta_{a}})\,,\nonumber\\
\theta^{a} &=&\frac{1}{2} \,(\gamma^{a} - i \tilde{\gamma}^{a})\,, \quad 
\frac{\partial}{\partial \theta_{a}}= \frac{1}{2} \,(\gamma^{a} + i \tilde{\gamma}^{a})\,,
\nonumber\\
\end{eqnarray}
offering together  $2\cdot 2^d$  operators: $2^d$ of those which are products of 
$\gamma^{a}$  and  $2^d$ of those which are products of $\tilde{\gamma}^{a}$.
Taking into account Eqs.~(\ref{thetaderher0}, \ref{clifftheta1}) it is easy to prove 
that they form two anticommuting Clifford subalgebras, 
$\{\gamma^{a}, \tilde{\gamma}^{b}\}_{+} =0$, Refs.~(\cite{nh2021RPPNP} and 
references therein)
\begin{eqnarray}
\label{gammatildeantiher}
\{\gamma^{a}, \gamma^{b}\}_{+}&=&2 \eta^{a b}= \{\tilde{\gamma}^{a}, 
\tilde{\gamma}^{b}\}_{+}\,, \nonumber\\
\{\gamma^{a}, \tilde{\gamma}^{b}\}_{+}&=&0\,,\quad
 (a,b)=(0,1,2,3,5,\cdots,d)\,, \nonumber\\
(\gamma^{a})^{\dagger} &=& \eta^{aa}\, \gamma^{a}\, , \quad 
(\tilde{\gamma}^{a})^{\dagger} =  \eta^{a a}\, \tilde{\gamma}^{a}\,.
\end{eqnarray}

 \vspace{3mm}

While the Grassmann algebra offers the description of the ''anticommuting integer 
spin second quantized fields'' and of the ''commuting integer spin second quantized fields''~\cite{%
nh2021RPPNP,n2021MDPIsymmetry}, the Clifford algebras which are superposition of 
odd products of either $\gamma^a$'s or $\tilde{\gamma}^a$'s, offer the description 
of the second quantized half integer spin fermion fields, and from the point of view
subgroups of $SO(d-1,1)$ group manifest spins and charges of fermions in the 
fundamental representations of the group and subgroups.

The superposition of even products of either $\gamma^a$'s or $\tilde{\gamma}^a$'s 
offer the description of the commuting second quantized boson fields with integer spins, 
as we shall see in this contribution and also in~\cite{n2021SQ}, which from the point of 
the subgroups of the $SO(d-1,1)$ group manifest spins and charges in the adjoint 
representations of the group and subgroups.

The {\it reduction}, Eq.~(\ref{tildegammareduced}) of Subsect.~(\ref{reduction}), of 
the {\it two Clifford algebras} --- $\gamma^a$'s and $\tilde{\gamma}^a$'s --- 
{\it to only one } --- $\gamma^a$'s are chosen --- {\it reduces the possibilities to 
describe either fermions or bosons to only one possibility.}

After the decision  that only  
$\gamma^a$'s are used to describe the internal space of fermions, the remaining ones, 
$\tilde{\gamma}^a$'s,  equip the irreducible representations of the Lorentz 
group (with the infinitesimal generators $S^{ab}=\frac{i}{4} \{\gamma^a, \,
\gamma^b\}_{-}$) with the family quantum numbers (determined by $\tilde{S}^{ab}=
\frac{i}{4} \{\tilde{\gamma}^a, \,\tilde{\gamma}^b\}_{-}$).

%

The even Clifford algebra objects --- which are the superposition of even products of 
$\gamma^a$'s --- offer after the reduction of the two subalgebras  to only
one the description of the second quantized boson fields as the gauge fields of the 
second quantized fermion fields,  the internal space of which are described by the 
odd Clifford algebra objects. The reduction enables to define the generators of the
Lorentz transformations in the internal space of bosons as  ${\bf {\cal S}}^{ab}= S^{ab}
 + \tilde{S}^{ab}$, manifesting the adjoint representation properties of the Clifford 
 even ''basis vectors''.

Both, Clifford odd and Clifford even ''basis vectors'', are discussed in what follows. The
Clifford odd ''basis vectors'' form $2^{\frac{d}{2}-1}$ families, each family has 
$2^{\frac{d}{2}-1}$ members. Their Hermitian conjugated partners form another group 
of $2^{\frac{d}{2}-1}\times 2^{\frac{d}{2}-1}$ members. The Clifford  even ''basis 
vectors'' form two groups of $2^{\frac{d}{2}-1}\times 2^{\frac{d}{2}-1}$ members.
Members of any of the two groups have the Hermitian conjugated partners within the
same group or are self adjoint.

\subsection{''Basis vectors'' are superposition of either odd or even products of 
Clifford objects $\gamma^{a}$'s.}
\label{basisvectors}

In this subsection the properties of the Clifford odd and the Clifford even ''basis vectors''
are discussed, the first ones offering the description of the internal space of fermions, 
the second ones offering the description of the internal space of bosons.
 
 Let us choose the ''basis vectors'' of either the Clifford odd or of the Clifford even character  
 to be eigenvectors of all the Cartan subalgebra members of the  Lorentz algebra. There 
 are $\frac{d}{2}$ members of the Cartan subalgebra in  even dimensional spaces.
 One can choose
\begin{eqnarray}
{\cal {\bf S}}^{03}, {\cal {\bf S}}^{12}, {\cal {\bf S}}^{56}, \cdots, 
{\cal {\bf S}}^{d-1 \;d}\,, \nonumber\\
S^{03}, S^{12}, S^{56}, \cdots, S^{d-1 \;d}\,, \nonumber\\
\tilde{S}^{03}, \tilde{S}^{12}, \tilde{S}^{56}, \cdots,  \tilde{S}^{d-1\; d}\,, 
\nonumber\\
{\cal {\bf S}}^{ab} = S^{ab} +\tilde{S}^{ab}\,.
\label{cartangrasscliff}
\end{eqnarray}

 Let us look for the eigenvectors of each of the Cartan subalgebra members, 
Eq.~(\ref{cartangrasscliff}), for the Grassmann algebra (with ${\cal {\bf S}}^{a b} = 
i \, (\theta^{a} \frac{\partial}{\partial \theta_{b}} - \theta^{b} \frac{\partial}{\partial \theta_{a}})$) and for each of the two kinds of the Clifford subalgebras  separately, and
let ${\cal {\bf S}}^{ab}, S^{ab}, \tilde{S}^{ab} $,   be one of the $\frac{d}{2}$ possibilities of $(ab=03,12,56,\dots, d-1\,d)$.
\begin{eqnarray}
{\cal {\bf S}}^{ab} \,\frac{1}{\sqrt{2}}\, (\theta^a + \frac{\eta^{aa}}{i k} \theta^b) &=&
k\,\frac{1}{\sqrt{2}} (\theta^a + \frac{\eta^{aa}}{ik} \theta^b) \,, 
{\cal {\bf S}}^{ab} \,\frac{1}{\sqrt{2}}\, (1+ \frac{i}{k}  \theta^a \theta^b) =0\,,\;\;
{\rm  or} \;\;
{\cal {\bf S}}^{ab} \,\frac{1}{\sqrt{2}}\,  \frac{i}{k}  \theta^a \theta^b 
=0\,,\nonumber\\
S^{ab} \frac{1}{2} (\gamma^a + \frac{\eta^{aa}}{ik} \gamma^b) &=& \frac{k}{2}  \,
\frac{1}{2} (\gamma^a + \frac{\eta^{aa}}{ik} \gamma^b)\,,\quad
S^{ab} \frac{1}{2} (1 +  \frac{i}{k}  \gamma^a \gamma^b) = \frac{k}{2}  \,
 \frac{1}{2} (1 +  \frac{i}{k}  \gamma^a \gamma^b)\,,\nonumber\\
\tilde{S}^{ab} \frac{1}{2} (\tilde{\gamma}^a + \frac{\eta^{aa}}{ik} \tilde{\gamma}^b) &=& 
\frac{k}{2}  \,\frac{1}{2} (\tilde{\gamma}^a + \frac{\eta^{aa}}{ik} \tilde{\gamma}^b)\,,
\quad
\tilde{S}^{ab} \frac{1}{2} (1 +  \frac{i}{k}  \tilde{\gamma}^a \tilde{\gamma}^b) = 
 \frac{k}{2}  \, \frac{1}{2} (1 +  \frac{i}{k} \tilde{\gamma}^a \tilde{\gamma}^b)\,,
\label{eigencliffcartan}
\end{eqnarray}

with  $k^2=\eta^{aa} \eta^{bb}$.
 The proof of Eq.~(\ref{eigencliffcartan}) is presented in Ref.~\cite{nh2021RPPNP}, 
 App.~(I).
 
 Let us use for the two Clifford subalgebras for the nilpotents $\frac{1}{2} (\gamma^a + \frac{\eta^{aa}}{ik} \gamma^b), (\frac{1}{2} (\gamma^a + \frac{\eta^{aa}}{ik} \gamma^b))^2=0$ and projectors $ \frac{1}{2} (1 +  \frac{i}{k}  \gamma^a \gamma^b),
( \frac{1}{2} (1 +  \frac{i}{k}  \gamma^a \gamma^b))^2 =
 \frac{1}{2} (1 +  \frac{i}{k}  \gamma^a \gamma^b)$ the notation 
%
\begin{eqnarray}
\label{graficcliff}
\stackrel{ab}{(k)}:&=& 
\frac{1}{2}(\gamma^a + \frac{\eta^{aa}}{ik} \gamma^b)\,,\quad 
\stackrel{ab}{[k]}:=\frac{1}{2}(1+ \frac{i}{k} \gamma^a \gamma^b)\,,\nonumber\\
\stackrel{ab}{\tilde{(k)}}:&=& 
\frac{1}{2}(\tilde{\gamma}^a + \frac{\eta^{aa}}{ik} \tilde{\gamma}^b)\,,\quad 
\stackrel{ab}{\tilde{[k]}}:
\frac{1}{2}(1+ \frac{i}{k} \tilde{\gamma}^a \tilde{\gamma}^b)\,. 
\end{eqnarray}
One can derive after taking into account Eq.~(\ref{gammatildeantiher}) the following 
useful relations
\begin{small}
\begin{eqnarray}
%
\gamma^a \stackrel{ab}{(k)}&=& \eta^{aa}\stackrel{ab}{[-k]},\; \quad
\gamma^b \stackrel{ab}{(k)}= -ik \stackrel{ab}{[-k]}, \; \quad 
\gamma^a \stackrel{ab}{[k]}= \stackrel{ab}{(-k)},\;\quad \;\;
\gamma^b \stackrel{ab}{[k]}= -ik \eta^{aa} \stackrel{ab}{(-k)}\,,\nonumber\\
%
\stackrel{ab}{(k)}^{\dagger} &=& \eta^{aa}\stackrel{ab}{(-k)}\,,\quad 
(\stackrel{ab}{(k)})^2 =0\,, \quad \stackrel{ab}{(k)}\stackrel{ab}{(-k)}
=\eta^{aa}\stackrel{ab}{[k]}\,,\nonumber\\
\stackrel{ab}{[k]}^{\dagger} &=& \,\stackrel{ab}{[k]}\,, \quad \quad \quad \quad
(\stackrel{ab}{[k]})^2 = \stackrel{ab}{[k]}\,, 
\quad \stackrel{ab}{[k]}\stackrel{ab}{[-k]}=0\,,
\nonumber\\
\stackrel{ab}{(k)}\stackrel{ab}{[k]}& =& 0\,,\qquad \qquad \qquad 
\stackrel{ab}{[k]}\stackrel{ab}{(k)}=  \stackrel{ab}{(k)}\,, \quad \quad \quad
  \stackrel{ab}{(k)}\stackrel{ab}{[-k]} =  \stackrel{ab}{(k)}\,,
\quad \, \stackrel{ab}{[k]}\stackrel{ab}{(-k)} =0\,,
\nonumber\\
%
\stackrel{ab}{\tilde{(k)}}^{\dagger} &=& \eta^{aa}\stackrel{ab}{\tilde{(-k)}}\,,\quad
(\stackrel{ab}{\tilde{(k)}})^2=0\,, \quad \stackrel{ab}{\tilde{(k)}}\stackrel{ab}{\tilde{(-k)}}
=\eta^{aa}\stackrel{ab}{\tilde{[k]}}\,,\nonumber\\
\stackrel{ab}{\tilde{[k]}}^{\dagger} &=& \,\stackrel{ab}{\tilde{[k]}}\,,
\quad \quad \quad \quad
(\stackrel{ab}{\tilde{[k]}})^2=\stackrel{ab}{\tilde{[k]}}\,,
\quad \stackrel{ab}{\tilde{[k]}}\stackrel{ab}{\tilde{[-k]}}=0\,,\nonumber\\
\stackrel{ab}{\tilde{(k)}}\stackrel{ab}{\tilde{[k]}}& =& 0\,,\qquad \qquad \qquad 
\stackrel{ab}{\tilde{[k]}}\stackrel{ab}{\tilde{(k)}}=  \stackrel{ab}{\tilde{(k)}}\,, 
\quad \quad \quad
  \stackrel{ab}{\tilde{(k)}}\stackrel{ab}{\tilde{[-k]}} =  \stackrel{ab}{\tilde{(k)}}\,,
\quad \, \stackrel{ab}{\tilde{[k]}}\stackrel{ab}{\tilde{(-k)}} =0\,,
\label{graficcliff1}
\end{eqnarray}
\end{small}

 One can define ''basis vectors'' to be eigenvectors of all the members of the Cartan 
 subalgebras as either odd or even products of nilpotents 
and of projectors in any even dimensional space.\\

\vspace{3mm}

{\bf A.} $\;\;$ {\bf Clifford odd ''basis vectors''} members

\vspace{3mm}

The Clifford odd  ''basis vectors'' must be products of an odd products of nilpotents,
the Clifford even  ''basis vectors'' must be products of an even number of nilpotents.
In both cases the rest factors of the ''basis vectors'' form projectors.

Let us define in $d$-dimensional space, $d=2(2n+1)$ and $d=4n$, first the Clifford 
{\it odd} ''basis vectors'', which are eigenvectors of the Cartan subalgebra members, 
Eq.~(\ref{cartangrasscliff}), and are superposition of Clifford odd products of 
$\gamma^a$'s. 

Naming the Clifford odd ''basis vectors'' with $\hat{b}^{m \dagger}_{f}$, where $m$ determines membership in one irreducible representation and $f$ an irreducible representation it follows 
\begin{eqnarray}
\label{allcartaneigenvecodd}
d&=&2(2n+1)\nonumber\\
\hat{b}^{1 \dagger}_{1}&=&\stackrel{03}{(+i)}\stackrel{12}{(+)}\cdots 
\stackrel{d-1 \, d}{(+)}\,,\nonumber\\
\hat{b}^{2 \dagger}_{1}&=&\stackrel{03}{[-i]}\stackrel{12}{[-]} 
\stackrel{56}{(+)} \cdots \stackrel{d-1 \, d}{(+)}\,,  \nonumber\\ 
\hat{b}^{3 \dagger}_{1}&=&\stackrel{03}{(+i)} \stackrel{12}{(+)} 
\stackrel{56}{(+)} \cdots \stackrel{d-3\,d-2}{[-]}\;\stackrel{d-1\,d}{[-]}\,, \nonumber\\
d&=&4n\nonumber\\
\hat{b}^{1 \dagger}_{1}&=&\stackrel{03}{(+i)}\stackrel{12}{(+)}\cdots 
\stackrel{d-1 \, d}{[+]}\,,\nonumber\\
\hat{b}^{2 \dagger}_{1}&=& \stackrel{03}{[-i]}\stackrel{12}{[-i]} 
\stackrel{56}{(+)} \cdots \stackrel{d-1 \, d}{[+]}\,,  \nonumber\\ 
\hat{b}^{3 \dagger}_{1}&=&\stackrel{03}{(+i)} \stackrel{12}{(+)} 
\stackrel{56}{(+)} \cdots \stackrel{d-3\,d-2}{[-]}\;\stackrel{d-1\,d}{(-)}\,, \nonumber\\
\end{eqnarray}

The first line  in $d=2(2n+1)$  case is the product of  nilpotents only and correspondingly 
a superposition of an odd products of $\gamma^a$'s. The second one belongs to the same 
irreducible representation as the first one, since it follows from the first one by the 
application of $S^{01}$, for example, the third one follows from the first one by
the application of $S^{d-3\,d-1}$. One can continue in this way to generate all the 
$2^{\frac{d}{2}-1}$ members of this irreducible representation.

In $(d=4n)$-dimensional space we start with the maximum odd number of nilpotents, 
what requires one projector. Then we repeat the same procedure as in the case of
$d=2(2n+1)$.
 
Since $S^{ab}$ changes two nilpotents into projectors (or one nilpotent and one projector
into projector and nilpotent) the number of nilpotents in the whole irreducible representation remains odd. 

Let $\hat{b}^{m}_{f}$ $= (\hat{b}^{m \dagger}_{f})^{\dagger}$ denotes
the Hermitian conjugated partner of the ''basis vector''  $\hat{b}^{m \dagger}_{f}$. 

Hermitian conjugated partners $\hat{b}^m_f=(\hat{b}^{m \dagger}_f)^{\dagger}$ 
follow by taking into account that projectors are self adjoint, while nilpotents  change sign 
according to Eq.~(\ref{graficcliff1}).

Since $S^{ab}$ transform two nilpotents into projectors it becomes clear that the group of 
Hermitian conjugated partners, which have an odd number of nilpotents $\stackrel{ab}{(k)} $
transformed into  $\stackrel{ab}{(-k)} $, is separated from the group of ''basis vectors''.

It is not difficult to prove, just by the algebraic multiplication, that all the Clifford 
odd  ''basis vectors'' are orthogonal\\
\begin{eqnarray}
\hat{b}^{m \dagger}_f *_{A} \hat{b}^{m`\dagger }_{f `}&=& 0\,, 
\quad \hat{b}^{m}_f *_{A} \hat{b}^{m `}_{f `}= 0\,,\quad \forall (m,m',f,f `)\,, 
\label{orthogonalodd}
\end{eqnarray}
while the relation $\hat{b}^{m}_f *_{A} \hat{b}^{m ` \dagger}_{f `}= 
\delta^{m m'} \delta_{f f `}$ becomes true after  each irreducible representation 
is equipped with the family quantum number.

The orthogonal relations among the  ''basis vectors''  
$\hat{\tilde{b}}^{m \dagger}_{f}$ and among their Hermitian conjugated partners 
$\hat{\tilde{b}}^{m}_{f}$ which are superposition of the Clifford odd products  of 
$\tilde{\gamma}^a$'s, follow by 
replacing in Eq.~(\ref{orthogonalodd}) $\hat{b}^{m \dagger}_f $ by 
$\hat{\tilde{b}}^{m \dagger}_{f}$ and $\hat{b}^{m}_f$ by  $\hat{\tilde{b}}^{m}_{f}$. \\

It is not difficult to prove the anticommutation relations of the Clifford odd ''basis vectors''  
and their Hermitian conjugated partners for both Clifford subalgebras~(\cite{norma92,nh2021RPPNP} and references therein). Let us  here 
present the anticommutation relations of only the one 
of the Clifford subalgebras --- $\gamma^a$'s~\footnote{
Since all the ''basis vectors'' are orthogonal and so are their Hermitian conjugated partners, Eq.~(\ref{orthogonalodd}), one could conclude that the anticommutation relations could 
as well be replaced by commutation relations.  
However,  the properties of the Clifford odd and of the Clifford even ''basis vectors'' 
(we present the properties of the Clifford even ''basis vectors'' in what follows)~\cite{prd2018,n2019PIPII,2020PartIPartII,nh2021RPPNP} are very different:
{\bf a.} The Clifford odd ''basis vectors'' have their Hermitian conjugated partners in 
another group of $2^{\frac{d}{2}-1}$ members of each of $2^{\frac{d}{2}-1}$ 
irreducible representations.
{\bf b.} The Clifford odd ''basis vectors'' have with respect to  the eigenvectors of the 
Cartan 
subalgebra members, Eq.~(\ref{cartangrasscliff}), 
as well as with respect to the superposition of the Cartan subalgebra members, 
properties of fermions, that is they manifest the fundamental 
representations of the $SO(d-1,1)$ group and of subgroups of this group. 
{\bf c.} When  the Clifford even ''basis vectors'' apply on the Clifford odd ''basis vectors'' 
(in an algebraic product $*_{A}$) the Clifford even ''basis vectors'' transform the Clifford odd ''basis vectors'' into another Clifford odd ''basis vectors'', conserving the eigenvalues of the 
Cartan subalgebras of the internal spaces of Clifford odd and Clifford even ''basis vectors''
under the recognition that Clifford even ''basis vectors'' carry integer spins.
{\bf d.} The algebraic products of  the Clifford even ''basis vectors'' make new Clifford even 
''basis vectors'', conserving bosonic quantum numbers (that is integer spins of the Clifford 
even ''basis vectors'').
{\bf e.} The application of the anticommutation relations on the Hilbert space constructed
out of the Clifford odd ''basis vectors'' in the tensor product with the ordinary basis, have 
the properties of the second quantized fermions~(\cite{nh2021RPPNP}, Sect. 5).} 
\begin{eqnarray}
\label{almostDirac}
\hat{b}^{m}_{f} {}_{*_{A}}|\psi_{oc}>&=& 0.\, |\psi_{oc}>\,,\nonumber\\
\hat{b}^{m \dagger}_{f}{}_{*_{A}}|\psi_{oc}>&=&  |\psi^m_{f}>\,,\nonumber\\
\{\hat{b}^{m}_{f}, \hat{b}^{m'}_{f `}\}_{*_{A}+}|\psi_{oc}>&=&
 0.\,|\psi_{oc}>\,, \nonumber\\
\{\hat{b}^{m \dagger}_{f}, \hat{b}^{m' \dagger}_{f  `}\}_{*_{A}+}|\psi_{oc}>
&=& 0. \,|\psi_{oc}>\,,\nonumber\\
\{\hat{b}^{m}_{f}, \hat{b}^{m' \dagger}_{f }\}_{*_{A}+}|\psi_{oc}>
&=& \delta^{m m'}|\psi_{oc}>\,,
\end{eqnarray}
where  ${*_{A}}$ represents the algebraic multiplication of 
$\hat{b}^{m \dagger}_{f}$  and $ \hat{b}^{m'}_{f'} $  among themselves and  
with the vacuum state  $|\psi_{oc}>$ of Eq.(\ref{vaccliffodd}), which takes into account 
Eq.~(\ref{gammatildeantiher0}), 
\begin{eqnarray}
\label{vaccliffodd}
|\psi_{oc}>= \sum_{f=1}^{2^{\frac{d}{2}-1}}\,\hat{b}^{m}_{f}{}_{*_A}
\hat{b}^{m \dagger}_{f} \,|\,1\,>\,,
\end{eqnarray}
for one of the members $m$, anyone, of the odd irreducible representation $f$,
with $|\,1\,>$, which is the vacuum without any structure --- the identity.
It follows that $\hat{b}^{m}_{f}{} |\psi_{oc}>=0$.

The relations are valid for both kinds of the odd Clifford subalgebras. To obtain the
equivalent relations for the ''basis vectors'' $\hat{\tilde{b}}^{m \dagger}_{f}$ we 
only have to replace $\hat{b}^{m \dagger}_{f}$ by $\hat{\tilde{b}}^{m \dagger}_{f}$
and equivalently for the Hermitian conjugated partners.

The Clifford odd ''basis vectors''~\footnote{The Clifford odd ''basis vectors'' in a tensor product 
with the ordinary basis form creation operators for any of the fermion states.  The Hilbert space is constructed from creation operators of single fermion states of all possible ''basis 
vectors'' of all possible momentum $\vec{p}$ as discussed in Sect.~5 of Ref.~\cite{nh2021RPPNP}.
} 
{\it almost} fulfil the second quantization postulates 
for fermions. There is, namely, the  property which the  second quantized fermions 
must fulfil in addition to the relations of Eq.~(\ref{almostDirac}) --- the following property
%
\begin{eqnarray} 
\label{should}
\{\hat{b}^{m}_{f}, \hat{b}^{m'\dagger}_{f'}\}_{*_{A}+}|\psi_{oc}>&=&
\delta^{m m'} \delta_{f f'} |\psi_{oc}>\,,
\end{eqnarray}
for either $\gamma^a$ or $\tilde{\gamma^a}$.
For  any $\hat{b}^{m}_{f}$ and any $\hat{b}^{m'\dagger}_{f'}$ this is not the case. 
It turns out that besides $\hat{b}^{m=1}_{f=1} = \stackrel{d-1 \, d}{(-)}
\cdots  \stackrel{56}{(-)} \stackrel{12}{(-)}\stackrel{03}{(-i)}$, for example, also 
$\hat{b}^{m'}_{f'} = \, \stackrel{d-1 \, d}{(-)} \cdots  \stackrel{56}{(-)}
 \stackrel{12}{[+]}\stackrel{03}{[+i]} $ 
%
and several others give, when applied on $\hat{b}^{m=1\dagger}_{f=1}$,
non zero contributions.
There are namely $2^{\frac{d}{2}-1}-1$  annihilation operators for each 
creation operator which give, applied on the creation operator, non zero contribution. \\


The problem is solvable by the reduction of the two Clifford odd algebras to only one~\cite{prd2018,n2019PIPII,2020PartIPartII,nh2021RPPNP} as it is presented 
in Subsect.~\ref{reduction}:$\;\;$
 If $\gamma^a$'s {\it are chosen to determine internal space of fermions the 
 remaining ones}, $\tilde{\gamma}^a$'s, {\it determine quantum numbers of} 
{\it each family} (described by the eigenvalues of $\tilde{S}^{ab}$ of the Cartan 
subalgebra members). \\ 

Correspondingly the creation operators expressible as tensor products, $*_{T}$, 
of the Clifford odd ''basis vectors'', $\hat{b}^{m \dagger }_{f}$,  and the basis in 
ordinary (momentum or coordinate) space and their Hermitian conjugated partners
annihilation operators fulfil the anticommutation relations for the second quantized 
fermion fields, explaining the postulates of Dirac for the second 
quantized fermion fields. \\

\vspace{3mm}

{\bf B.} $\;\;$ {\bf Clifford even ''basis vectors''}

\vspace{3mm}

Let us denote the Clifford even ''basis vectors'' as ${}^{i}\hat{\cal A}^m_{f}$, where 
$i=(I,II)$ points out that there are two groups of Clifford even ''basis vectors'' which can 
not be transformed into one another by the generators of the Lorentz transformation in 
the space  of ''basis vectors'', since each member of the group  $I$ has a member in the 
group $II$ which differ from the member of the group $I$ only in the sign of either one  
nilpotent or of one projector. All the members within the group, either $i=I$  or $i=II$, 
are reachable by the generators of the Lorentz transformations in the space of ''basis 
vectors''. 
 We make a choice correspondingly
\begin{eqnarray}
\label{allcartaneigenvecevenI}
d&=&2(2n+1)\nonumber\\
{}^I\hat{{\cal A}}^{1 \dagger}_{1}=\stackrel{03}{(+i)}\stackrel{12}{(+)}\cdots 
\stackrel{d-1 \, d}{[+]}\,,\qquad &&
{}^{II}\hat{{\cal A}}^{1 \dagger}_{1}=\stackrel{03}{(-i)}\stackrel{12}{(+)}\cdots 
\stackrel{d-1 \, d}{[+]}\,,\nonumber\\
{}^I\hat{{\cal A}}^{2 \dagger}_{1}=\stackrel{03}{[-i]}\stackrel{12}{[-]} 
\stackrel{56}{(+)} \cdots \stackrel{d-1 \, d}{[+]}\,, \qquad  && 
{}^{II}\hat{{\cal A}}^{2 \dagger}_{1}=\stackrel{03}{[+i]}\stackrel{12}{[-]} 
\stackrel{56}{(+)} \cdots \stackrel{d-1 \, d}{[+]}\,,
\nonumber\\ 
{}^I\hat{{\cal A}}^{3 \dagger}_{1}=\stackrel{03}{(+i)} \stackrel{12}{(+)} 
\stackrel{56}{(+)} \cdots \stackrel{d-3\,d-2}{[-]}\;\stackrel{d-1\,d}{(-)}\,, \qquad &&
{}^{II}\hat{{\cal A}}^{3 \dagger}_{1}=\stackrel{03}{(-i)} \stackrel{12}{(+)} 
\stackrel{56}{(+)} \cdots \stackrel{d-3\,d-2}{[-]}\;\stackrel{d-1\,d}{(-)}\,,  \nonumber\\
d&=&4n\nonumber\\
{}^I\hat{{\cal A}}^{1 \dagger}_{1}=\stackrel{03}{(+i)}\stackrel{12}{(+)}\cdots 
\stackrel{d-1 \, d}{(+)}\,,\qquad &&
{}^{II}\hat{{\cal A}}^{1 \dagger}_{1}=\stackrel{03}{(-i)}\stackrel{12}{(+)}\cdots 
\stackrel{d-1 \, d}{(+)}\,,
\nonumber\\
{}^I\hat{{\cal A}}^{2 \dagger}_{1}= \stackrel{03}{[-i]}\stackrel{12}{[-i]} 
\stackrel{56}{(+)} \cdots \stackrel{d-1 \, d}{(+)}\,, \qquad &&
{}^{II}\hat{{\cal A}}^{2 \dagger}_{1}= \stackrel{03}{[+i]}\stackrel{12}{[-i]} 
\stackrel{56}{(+)} \cdots \stackrel{d-1 \, d}{(+)}\,,  \nonumber\\ 
{}^I\hat{{\cal A}}^{3 \dagger}_{1}=\stackrel{03}{(+i)} \stackrel{12}{(+)} 
\stackrel{56}{(+)} \cdots \stackrel{d-3\,d-2}{[-]}\;\stackrel{d-1\,d}{[-]}\,, \qquad &&
{}^{II}\hat{{\cal A}}^{3 \dagger}_{1}=\stackrel{03}{(-i)} \stackrel{12}{(+)} 
\stackrel{56}{(+)} \cdots \stackrel{d-3\,d-2}{[-]}\;\stackrel{d-1\,d}{[-]}\,.
\end{eqnarray}

The members of the group $II$ follow from the members of group $I$, Eq.~(\ref{allcartaneigenvecevenI}),
if we replace in ${}^I\hat{{\cal A}}^{1 \dagger}_{1}$ the nilpotent 
$\stackrel{03}{(+i)}$ by $\stackrel{03}{(-i)}$,  and in 
${}^I\hat{{\cal A}}^{2 \dagger}_{1}$ the projector $\stackrel{03}{[-i]}$ by 
$\stackrel{03}{[+i]}$, and so on.\\

The Clifford even ''basis vectors'' have an even number of nilpotents which change sign 
under the Hermitian conjugation while projectors are self adjoint according to Eq.~(\ref{graficcliff}).
Correspondingly the Hermitian conjugated partners of the Clifford even ''basis vectors'' 
$({}^{i}\hat{{\cal A}}^{m \dagger}_{f})^{\dagger}$ belong to the same  group $i$:
${}^{i}\hat{{\cal A}}^{m \dagger}_{f}$,  with $2^{\frac{d}{2}-1}\times$ 
$2^{\frac{d}{2}-1}$ members. 
(Because of this property the sign $ \dagger$ has no special meaning.)\\

The creation operators, they are tensor products, $*_{T}$, of the Clifford even ''basis
vectors'' (chosen to be the eigenvectors of the Cartan subalgebra, 
Eq.~(\ref{cartangrasscliff})) and the basis in ordinary space (momentum or 
coordinate) carry after the reduction of the two Clifford subalgebras into 
only one, Subsect.~\ref{reduction},  the Cartan subalgebra eigenvalues of 
${\bf {\cal S}}^{ab} (= S^{ab} + \tilde{S}^{ab})$.  They fulfil the commutation 
relations of the second quantized boson  gauge fields of the corresponding fermion 
fields. \\

Let us say: \\
After the reduction of the two Clifford subalgebras to only one --- the 
one of $\gamma^a$'s  ---  are $2^{\frac{d}{2}-1}$  $2^{\frac{d}{2}-1}$ members 
of the Clifford even ''basis vectors'' ${}^{i}\hat{{\cal A}}^{m \,\dagger}_{f}$ 
reachable from any  other ${}^{i}\hat{{\cal A}}^{m' \dagger}_{f `}$ either by 
$S^{ab}$'s or by $\tilde{S}^{ab}$'s or by both, and have their  Hermitian 
conjugated partners within the same group $i$, for ether $i=I$ or $i=II$. \\ 
The eigenvalues of the Cartan subalgebra are for the Clifford even ''basis vectors '', 
after the reduction of the two Clifford subalgebras to only one determined by  
${\bf {\cal S}}^{ab}= S^{ab} + \tilde{s}^{ab}$.\\

 Contrary, the Clifford odd $2^{\frac{d}{2}-1}$ members of each 
of the  $2^{\frac{d}{2}-1}$ irreducible representations of ''basis vectors'' have 
their Hermitian conjugated partners in another set of $2^{\frac{d}{2}-1}$ 
$\cdot 2^{\frac{d}{2}-1}$ ''basis vectors''.

\vspace{2mm}

We shall demonstrate properties of the Clifford odd and Clifford even ''basis vectors'' 
on the toy model in  $d=(5+1)$ in Subsect.~\ref{cliffordoddevenbasis5+1}in details.

\subsection{Reduction of the Clifford space}
\label{reduction}

The creation and annihilation operators of an odd Clifford  algebra of both kinds, of 
either $\gamma^a$'s or  $\tilde{\gamma}^{a}$'s,   turn out to obey the 
anticommutation relations for the second quantized fermions, postulated by Dirac~%
\cite{nh2021RPPNP}, provided that each of the irreducible representations of the
corresponding Lorentz group, describing the internal space of fermions, carry a 
different quantum number.

But we know that a particular member $m$ has for all the irreducible representations 
the same quantum numbers, that is the same "eigenvalues" of the Cartan subalgebra 
(for the vector space of either $\gamma^a$'s or  $\tilde{\gamma}^{a}$'s), 
Eq.~(\ref{graficcliff}).  

\vspace{2mm}

{\it There is a possibility to "dress" each irreducible representation of one kind of the
two independent vector spaces with a new, let us call it the "family"  quantum number, if 
we "sacrifice" one of the two vector spaces.}\\

Let us ''sacrifice''  $\tilde{\gamma}^{a}$'s,  using $\tilde{\gamma}^{a}$'s to define the 
"family" quantum numbers for each irreducible representation of the vector space of 
''basis vectors'' of an odd products of $\gamma^a$'s, while keeping the relations of Eq.~(\ref{gammatildeantiher0}) unchanged:  $\{\gamma^{a}, 
\gamma^{b}\}_{+}=2 \eta^{a b}= \{\tilde{\gamma}^{a}, 
\tilde{\gamma}^{b}\}_{+}$, $\{\gamma^{a}, \tilde{\gamma}^{b}\}_{+}=0$,
  $ (\gamma^{a})^{\dagger} = \eta^{aa}\, \gamma^{a}$, 
$(\tilde{\gamma}^{a})^{\dagger} =  \eta^{a a}\, \tilde{\gamma}^{a}$, 
$(a,b)=(0,1,2,3,5,\cdots,d)$.
The proof that relations of Eq.~(\ref{gammatildeantiher0}) remain valid after the reduction
of the internal space of odd ''basis vectors'' is presented in~\cite{nh2021RPPNP} in 
App.~I, Statements 2. and Statements 3..\\

We therefore require:\\
  Let  $\tilde{\gamma}^{a}$'s operate on $\gamma^a$'s as follows~%
\cite{nh03,norma93,JMP2013,normaJMP2015,nh2018}
\begin{eqnarray}
\{\tilde{\gamma}^a B &=&(-)^B\, i \, B \gamma^a\}\, |\psi_{oc}>\,,
\label{tildegammareduced}
\end{eqnarray}
with $(-)^B = -1$, if $B$ is (a function of) an odd products of $\gamma^a$'s,  otherwise 
$(-)^B = 1$~\cite{nh03}, $|\psi_{oc}>$ is defined in Eq.~(\ref{vaccliffodd}).

\vspace{2mm}

After the postulate of Eq.~(\ref{tildegammareduced})  the ''basis vectors'' which are  superposition of an odd products of $\gamma^a$'s obey all 
the postulates of Dirac for the second quantized fermion fields, presented in
Eqs.~(\ref{should}, \ref{almostDirac}). 

Each irreducible representation of the odd ''basis vectors'' are after the postulate of Eq.~(\ref{tildegammareduced}) equipped by the quantum numbers of the Cartan 
subalgebra members of $\tilde{S}^{ab}$, Eq.~(\ref{cartangrasscliff}).

The eigenvalues of the operators $S^{ab}$  and $\tilde{S}^{ab}$ on nilpotents and 
projectors of $\gamma^a$'s are after the reduction of Clifford space equal to
\begin{eqnarray}
\label{signature0}
S^{ab} \,\stackrel{ab}{(k)} = \frac{k}{2}  \,\stackrel{ab}{(k)}\,,\quad && \quad
\tilde{S}^{ab}\,\stackrel{ab}{(k)} = \frac{k}{2}  \,\stackrel{ab}{(k)}\,,\nonumber\\
S^{ab}\,\stackrel{ab}{[k]} =  \frac{k}{2}  \,\stackrel{ab}{[k]}\,,\quad && \quad 
\tilde{S}^{ab} \,\stackrel{ab}{[k]} = - \frac{k}{2}  \,\,\stackrel{ab}{[k]}\,,
\end{eqnarray}
demonstrating that the eigenvalues of $S^{ab}$ on nilpotents and projectors of 
$\gamma^a$'s differ from the eigenvalues of $\tilde{S}^{ab}$ on $\gamma^a$'s, so 
that $\tilde{S}^{ab}$ can be used to equip irreducible representations of $S^{ab}$
 with the ''family'' quantum number.

After this  postulate  the vector  space of $\tilde{\gamma}^{a}$'s is 
"frozen out". No vector space of $\tilde{\gamma}^{a}$'s needs to be taken into account  
any longer for the description of the internal space of either fermions or bosons, in agreement 
with the observed properties of fermions. The operators  $\tilde{\gamma}^{a}$'s 
determine from now on properties of fermion and boson ''basis vectors'' written in terms of 
odd and even numbers of $\gamma^a$'s, respectively:\\
{\bf i.} The odd products of the Clifford objects $\gamma^a$'s  offer the  description 
of the internal space of fermions,\\
{\bf ii.} The even products of the Clifford objects $\gamma^a$'s  offer the 
 description of the internal space of bosons, which are the gauge fields of the
 fermions. \\

 We shall demonstrate in Sect.~\ref{even5+1} that the Clifford even ''basis vectors'', 
equipped by the sum of both quantum numbers, $S^{ab}$ and $\tilde{S}^{ab}$,  
${\bf {\cal S}}^{ab}= S^{ab} +\tilde{S}^{ab}$, obey the boson second quantized 
postulates.\\


Let us present some useful relations following if using  Eq.~(\ref{tildegammareduced}),
\begin{eqnarray}
 \tilde{\gamma^a} \stackrel{ab}{(k)} &=& - i\eta^{aa}\stackrel{ab}{[k]},\quad
 \tilde{\gamma^b} \stackrel{ab}{(k)} =  - k \stackrel{ab}{[k]},\;\; \qquad  \,
\tilde{\gamma^a} \stackrel{ab}{[k]} =  \;\;i\stackrel{ab}{(k)},\; \quad\;\;
 \tilde{\gamma^b} \stackrel{ab}{[k]} =  -k \eta^{aa} \stackrel{ab}{(k)}\,, 
\nonumber\\
\stackrel{ab}{\tilde{(k)}} \, \stackrel{ab}{(k)}& =& 0\,, \quad
\qquad 
\stackrel{ab}{\tilde{(-k)}} \, \stackrel{ab}{(k)} = -i \,\eta^{aa}\,  
\stackrel{ab}{[k]}\,,\quad  
\stackrel{ab}{\tilde{(k)}} \, \stackrel{ab}{[k]} = i\, \stackrel{ab}{(k)}\,,\quad
\stackrel{ab}{\tilde{(k)}}\, \stackrel{ab}{[-k]} = 0\,, \nonumber\\
%
%
\stackrel{ab}{\tilde{[k]}} \, \stackrel{ab}{(k)}& =& \, \stackrel{ab}{(k)}\,, 
\qquad \,
\stackrel{ab}{\tilde{[-k]}} \, \stackrel{ab}{(k)} = \, 0 \,,  \qquad  \quad 
\quad \;\;\;
\stackrel{ab}{\tilde{[k]}} \, \stackrel{ab}{[k]} =  0\,,\qquad \;\;\;
\stackrel{ab}{\tilde{[- k]}} \, \stackrel{ab}{[k]} =  \, \stackrel{ab}{[k]}\,.
\label{graphbinomsfamilies}
\end{eqnarray}
%
To point out that the anticommuting properties of  $\gamma^a$'s and 
$\tilde{\gamma}^a$'s remain valid also after the reduction of the two Clifford algebras 
to only one,  Eq.~(\ref{tildegammareduced}), the commutation relations of Eq.(\ref{gammatildeantiher}) are repeated here again 
\begin{eqnarray}
\label{gammatildeantiher0}
\{\gamma^{a}, \gamma^{b}\}_{+}&=&2 \eta^{a b}= \{\tilde{\gamma}^{a}, 
\tilde{\gamma}^{b}\}_{+}\,, \nonumber\\
\{\gamma^{a}, \tilde{\gamma}^{b}\}_{+}&=&0\,,\quad
 (a,b)=(0,1,2,3,5,\cdots,d)\,, \nonumber\\
(\gamma^{a})^{\dagger} &=& \eta^{aa}\, \gamma^{a}\, , \quad 
(\tilde{\gamma}^{a})^{\dagger} =  \eta^{a a}\, \tilde{\gamma}^{a}\,,\nonumber\\
\gamma^a \gamma^a &=& \eta^{aa}\,, \quad 
\gamma^a (\gamma^a)^{\dagger} =I\,,\quad
 \tilde{\gamma}^a  \tilde{\gamma}^a = \eta^{aa} \,,\quad
 \tilde{\gamma}^a  (\tilde{\gamma}^a)^{\dagger} =I\,.
\end{eqnarray}
The proof can be found in~(\cite{nh2021RPPNP}, App. I, Statements 2. and 3.). 
Taking into account the anticommuting properties of $\gamma^a$'s and 
$\tilde{\gamma}^a$'s, Eq.~(\ref{gammatildeantiher0}), it is not difficult to prove  
the relations in Eq.~(\ref{graphbinomsfamilies}).

\subsection{Properties of Clifford odd and even ''basis vectors''  in $d=(5+1)$}
\label{cliffordoddevenbasis5+1}

To clear up the properties of the Clifford odd and Clifford even ''basis vectors'' 
their properties in $d=(5+1)$-dimensional space are presented as:\\
 {\bf i.} The odd products of the Clifford objects $\gamma^a$'s,  offering the 
 description of the internal space of fermion fields,\\
{\bf ii.} The even products of the Clifford objects $\gamma^a$'s,  offering the 
 description of the internal space of boson fields, which are the gauge fields of the
 fermion fields, the internal space of which is described by the Clifford odd ''basis 
 vectors''. 
 
 The properties of the Clifford odd and Clifford even ''basis vectors'' are analysed
 not only with respect to the group $SO(5,1)$  but also with respect to the subgroups 
 $SO(3,1) \times U(1)$ and $SU(3)\times U(1)$ of the group $SO(5,1)$.
 
 \vspace{2mm}
 

Choosing the ''basis vectors'' to be eigenvectors of all the members of the Cartan 
subalgebra of the Lorentz algebra and correspondingly writing them as the products 
of nilpotents $\stackrel{ab}{(+i)}$  ($\stackrel{ab}{(+i)}^2=0$) and projectors 
$\stackrel{ab}{[+]}$ ($\stackrel{ab}{[+]}^2=$ $\stackrel{ab}{[+]}$), each of 
nilpotent or projector is chosen to be the eigenvector of one  of the Cartan 
subalgebra members of the Lorentz algebra in the internal space of fermions 
(the Clifford odd ''basis vectors'') or bosons  (the Clifford even ''basis vectors''), 
one finds the Clifford odd and the Clifford even ''basis vectors'' as presented in 
Table~\ref{Table Clifffourplet.}. The table presents besides the ''basis vectors'' 
also their eigenvalues of the Cartan subalgebra members of 
$S^{ab}$ and $\tilde{S}^{ab}$ for the group $SO(5,1)$ and the handedness 
$\Gamma^{(5+1)}$ and $\Gamma^{(3+1)}$,  representing handedness in 
$d=(5+1)$ and $d=(3+1)$, respectively~\footnote{
 The handedness $\Gamma^{(d)}$,  one of the invariants of the group $SO(d)$, 
with the infinitesimal generators of the Lorentz group $S^{ab}$, is
defined as $\Gamma^{(d)}=\alpha\, \varepsilon_{a_1 a_2\dots a_{d-1} a_d}\, 
S^{a_1 a_2} \cdot S^{a_3 a_4} \cdots S^{a_{d-1} a_d}$, summed over 
$(a_1 a_2\dots a_{d-1} a_d)$, with the constant $\alpha$ chosen so that 
$\Gamma^{(d)}=\pm 1$. 
In the case that states are represented by products of $\gamma^{a}$'s , 
$\Gamma^{(d)}$ simplifies to 
$\Gamma^{(d)} = (i)^{d/2} \prod_a \, (\sqrt{\eta^{aa}} \gamma^a)$ for $d=2n$
and to $\Gamma^{(d)} = (i)^{(d-1)/2}\; \prod_{a} \, (\sqrt{\eta^{aa}} \gamma^a)$, 
for $d = 2n +1$\,. }.

The  {\it odd I} group is presenting the ''basis vectors'' which  are products of an odd 
number of nilpotents (three or one) and of projectors (none or two), offering the 
description of the internal space of fermion fields in $d=(5+1)$-dimensional space. 
Their Hermitian conjugated partners appear in the separate group {\it odd II}.  While 
projectors are self adjoint an odd number of nilpotents change sign under Hermitian conjugationsign, Eq.~(\ref{graficcliff}), what can not be achieved by either $S^{ab }$
 or $\tilde{S}^{ab}$ or both. Correspondingly the ''basis vectors'' and their Hermitian
 conjugated partners differ in $\Gamma^{(d=5+1)}$.

The groups {\it even I} and {\it even II} present commuting Clifford even ''basis vectors'',
with an even number of nilpotents (none or two), which follow from the starting one 
either by $S^{ab }$ or by $\tilde{S}^{ab}$ or by both. Correspondingly the Clifford even ''basis vectors'' have  their  Hermitian conjugated partners within the same group (either  within {\it even I} or {\it even II}) or are self adjoint.
We shall see that they offer the description for the internal space of bosons, having 
properties of the gauge fields of the corresponding fermion fields, the internal space of 
which is described by  the Clifford odd ''basis vectors''.

The ''basis  vectors'', and their Hermitian conjugated partners,
 $\hat{b}^{m \dagger}_{f}$ and $\hat{b}^{m}_{f}$, determining the creation and annihilation operators for fermions, appearing in two separate groups, each with $2^{\frac{d}{2}-1} \times 2^{\frac{d}{2}-1}$ members, algebraically 
anticommute, due to the properties of the Clifford algebra  elements $\gamma^{a}$'s, Eq.~(\ref{gammatildeantiher0}).

Correspondingly the ''basis  vectors'', determining the creation operators for bosons 
${}^{I,II}\hat{\cal A}^{m }_{f}$  algebraically  commute. They appear in two groups 
with $2^{\frac{d}{2}-1} \times 2^{\frac{d}{2}-1}$ members each.

\begin{table*}
\begin{small}
\caption{\label{Table Clifffourplet.}  $2^{(d=6)}=64$ "eigenvectors" of the Cartan 
subalgebra 
of the Clifford  odd and even algebras --- the superposition of odd or 
even products of $\gamma^{a}$'s --- in $d=(5+1)$-dimensional space are presented, 
divided into four groups. The first group, $odd \,I$, is chosen to represent "basis vectors", 
named ${\hat b}^{m \dagger}_f$, appearing in $2^{\frac{d}{2}-1}=4$ 
"families"  ($f=1,2,3,4$), each ''family'' with $2^{\frac{d}{2}-1}=4$  
''family'' members ($m=1,2,3,4$). 
The second group, $odd\,II$, contains Hermitian conjugated partners of the first 
group for each  family separately, ${\hat b}^{m}_f=$ 
$({\hat b}^{m \dagger}_f)^{\dagger}$. Either $odd \,I$ or $odd \,II$ are products
of an odd number of nilpotents, the rest are projectors.
The quantum  numbers of $f$, determined by eigenvalues of $(\tilde{S}^{03}, $ 
$\tilde{S}^{12},$  $\tilde{S}^{56})$, of ${\hat b}^{m \dagger}_f$ are for the 
first {\it odd I } and the two last {\it even I} and {\it even II}  groups written   
above each group of four members with the same $f$. The quantum numbers of 
each member, determined by eigenvalues of ($S^{03}, S^{12}, S^{56})$, are 
in these three cases presented in the three columns before the last two columns.
For the Hermitian conjugated  partners of {\it odd I}, presented in the group 
{\it odd II}, the quantum numbers $(S^{03}, S^{12}, S^{56})$ are presented 
above each group of the Hermitian conjugated partners, the three columns before 
the last two  tell eigenvalues of $(\tilde{S}^{03}, \tilde{S}^{12},\tilde{S}^{56})$.
The  two groups with the even number of $\gamma^a$'s, {\it even \,I} and 
{\it even \,II}, have their Hermitian conjugated partners within their own group 
each. $\Gamma^{(5+1)}$ and $\Gamma^{(3+1)}$ represent handedness in 
$d=(3+1)$ and $d=(5+1)$, respectively, defined in the footnote. }
\end{small}
%
%
\begin{tiny}
\begin{center}
  \begin{tabular}{|c|c|c|c|c|c|r|r|r|r|r|}
\hline
$ $&$$&$ $&$ $&$ $&&$$&$$&$$&&\\
$''basis\, vectors'' $&$m$&$ f=1$&$ f=2 $&$ f=3 $&
$ f=4 $&$$&$$&$$&$$&$$\\
$(\tilde{S}^{03}, \tilde{S}^{12}, \tilde{S}^{56})$&$\rightarrow$&$(\frac{i}{2},- \frac{1}{2},-\frac{1}{2})$&$(-\frac{i}{2},-\frac{1}{2},\frac{1}{2})$&
$(-\frac{i}{2},\frac{1}{2},-\frac{1}{2})$&$(\frac{i}{2},\frac{1}{2},\frac{1}{2})$&$S^{03}$
 &$S^{12}$&$S^{56}$&$\Gamma^{(5+1)}$&$\Gamma^{(3+1)}$\\
\hline
$ $&$$&$ $&$ $&$ $&&$$&$$&$$&&\\
$odd \,I\; {\hat b}^{m \dagger}_f$&$1$& 
$\stackrel{03}{(+i)}\stackrel{12}{[+]}\stackrel{56}{[+]}$&
                        $\stackrel{03}{[+i]}\stackrel{12}{[+]}\stackrel{56}{(+)}$ & 
                        $\stackrel{03}{[+i]}\stackrel{12}{(+)}\stackrel{56}{[+]}$ &  
                        $\stackrel{03}{(+i)}\stackrel{12}{(+)}\stackrel{56}{(+)}$ &
                        $\frac{i}{2}$&$\frac{1}{2}$&$\frac{1}{2}$&$1$&$1$\\
$$&$2$&    $[-i](-)[+] $ & $(-i)(-)(+) $ & $(-i)[-][+] $ & $[-i][-](+) $ &$-\frac{i}{2}$&
$-\frac{1}{2}$&$\frac{1}{2}$&$1$&$1$\\
$$&$3$&    $[-i] [+](-)$ & $(-i)[+][-] $ & $(-i)(+)(-) $ & $[-i](+)[-] $&$-\frac{i}{2}$&
$\frac{1}{2}$&$-\frac{1}{2}$&$1$&$-1$ \\
$$&$4$&    $(+i)(-)(-)$ & $[+i](-)[-] $ & $[+i][-](-) $ & $(+i)[-][-]$&$\frac{i}{2}$&
$-\frac{1}{2}$&$-\frac{1}{2}$&$1$&$-1$ \\
\hline
$ $&$$&$ $&$ $&$ $&&$$&$$&$$&&\\
$(S^{03}, S^{12}, S^{56})$&$\rightarrow$&$(-\frac{i}{2}, \frac{1}{2},\frac{1}{2})$&$(\frac{i}{2},\frac{1}{2},-\frac{1}{2})$&
$(\frac{i}{2},- \frac{1}{2},\frac{1}{2})$&$(-\frac{i}{2},-\frac{1}{2},-\frac{1}{2})$&$\tilde{S}^{03}$
&$\tilde{S}^{12}$&$\tilde{S}^{56}$&$\Gamma^{(5+1)}$&$\tilde{\Gamma}^{(3+1)}$\\
&&
$\stackrel{03}{\;\,}\;\;\,\stackrel{12}{\;\,}\;\;\,\stackrel{56}{\;\,}$&
$\stackrel{03}{\;\,}\;\;\,\stackrel{12}{\;\,}\;\;\,\stackrel{56}{\;\,}$&
$\stackrel{03}{\;\,}\;\;\,\stackrel{12}{\;\,}\;\;\,\stackrel{56}{\;\,}$&
$\stackrel{03}{\;\,}\;\;\,\stackrel{12}{\;\,}\;\;\,\stackrel{56}{\;\,}$&
&&&
&$$\\
\hline
$ $&$$&$ $&$ $&$ $&&$$&$$&$$&&\\
$odd\,II\; {\hat b}^{m}_f$&$1$ &$(-i)[+][+]$ & $[+i][+](-)$ & $[+i](-)[+]$ & $(-i)(-)(-)$&
$-\frac{i}{2}$&$-\frac{1}{2}$&$-\frac{1}{2}$&$-1$&$1$ \\
$$&$2$&$[-i](+)[+]$ & $(+i)(+)(-)$ & $(+i)[-][+]$ & $[-i][-](-)$&
$\frac{i}{2}$&$\frac{1}{2}$&$-\frac{1}{2}$&$-1$&$1$ \\
$$&$3$&$[-i][+](+)$ & $(+i)[+][-]$ & $(+i)(-)(+)$ & $[-i](-)[-]$&
$\frac{i}{2}$&$-\frac{1}{2}$&$\frac{1}{2}$&$-1$&$-1$ \\
$$&$4$&$(-i)(+)(+)$ & $[+i](+)[-]$ & $[+i][-](+)$ & $(-i)[-][-]$&
$-\frac{i}{2}$&$\frac{1}{2}$&$\frac{1}{2}$&$-1$&$-1$ \\
\hline
&&&&&&&&&&\\
\hline
$ $&$$&$ $&$ $&$ $&&$$&$$&$$&&\\
$(\tilde{S}^{03}, \tilde{S}^{12}, \tilde{S}^{56})$&$\rightarrow$&
$(-\frac{i}{2},\frac{1}{2},\frac{1}{2})$&$(\frac{i}{2},-\frac{1}{2},\frac{1}{2})$&
$(-\frac{i}{2},-\frac{1}{2},-\frac{1}{2})$&$(\frac{i}{2},\frac{1}{2},-\frac{1}{2})$&
$S^{03}$&$S^{12}$&$S^{56}$&$\Gamma^{(5+1)}$&
$\Gamma^{(3+1)}$\\ 
&& 
$\stackrel{03}{\;\,}\;\;\,\stackrel{12}{\;\,}\;\;\,\stackrel{56}{\;\,}$&
$\stackrel{03}{\;\,}\;\;\,\stackrel{12}{\;\,}\;\;\,\stackrel{56}{\;\,}$&
$\stackrel{03}{\;\,}\;\;\,\stackrel{12}{\;\,}\;\;\,\stackrel{56}{\;\,}$&
$\stackrel{03}{\;\,}\;\;\,\stackrel{12}{\;\,}\;\;\,\stackrel{56}{\;\,}$&
&&&&\\
\hline
$ $&$$&$ $&$ $&$ $&&$$&$$&$$&&\\
$even\,I \; {}^{I}{\cal A}^{m}_f$&$1$&$[+i](+)(+) $ & $(+i)[+](+) $ & $[+i][+][+] $ & $(+i)(+)[+] $ &$\frac{i}{2}$&
$\frac{1}{2}$&$\frac{1}{2}$&$1$&$1$ \\
$$&$2$&$(-i)[-](+) $ & $[-i](-)(+) $ & $(-i)(-)[+] $ & $[-i][-][+] $ &$-\frac{i}{2}$&
$-\frac{1}{2}$&$\frac{1}{2}$&$1$&$1$ \\
$$&$3$&$(-i)(+)[-] $ & $[-i][+][-] $ & $(-i)[+](-) $ & $[-i](+)(-) $&$-\frac{i}{2}$&
$\frac{1}{2}$&$-\frac{1}{2}$&$1$&$-1$ \\
$$&$4$&$[+i][-][-] $ & $(+i)(-)[-] $ & $[+i](-)(-) $ & $(+i)[-](-) $&$\frac{i}{2}$&
$-\frac{1}{2}$&$-\frac{1}{2}$&$1$&$-1$ \\ 
\hline
$ $&$$&$ $&$ $&$ $&&$$&$$&$$&&\\
$(\tilde{S}^{03}, \tilde{S}^{12}, \tilde{S}^{56})$&$\rightarrow$&
$(\frac{i}{2},\frac{1}{2},\frac{1}{2})$&$(-\frac{i}{2},-\frac{1}{2},\frac{1}{2})$&
$(\frac{i}{2},-\frac{1}{2},-\frac{1}{2})$&$(-\frac{i}{2},\frac{1}{2},-\frac{1}{2})$&
$S^{03}$&$S^{12}$&$S^{56}$&$\Gamma^{(5+1)}$&
$\Gamma^{(3+1)}$\\
&& 
$\stackrel{03}{\;\,}\;\;\,\stackrel{12}{\;\,}\;\;\,\stackrel{56}{\;\,}$&
$\stackrel{03}{\;\,}\;\;\,\stackrel{12}{\;\,}\;\;\,\stackrel{56}{\;\,}$&
$\stackrel{03}{\;\,}\;\;\,\stackrel{12}{\;\,}\;\;\,\stackrel{56}{\;\,}$&
$\stackrel{03}{\;\,}\;\;\,\stackrel{12}{\;\,}\;\;\,\stackrel{56}{\;\,}$&
&&&&\\
\hline
$ $&$$&$ $&$ $&$ $&&$$&$$&$$&&\\
$even\,II \; {}^{II}{\cal A}^{m}_f$&$1$& $[-i](+)(+) $ & $(-i)[+](+) $ & $[-i][+][+] $ & 
$(-i)(+)[+] $ &$-\frac{i}{2}$&
$\frac{1}{2}$&$\frac{1}{2}$&$-1$&$-1$ \\ 
$$&$2$&    $(+i)[-](+) $ & $[+i](-)(+) $ & $(+i)(-)[+] $ & $[+i][-][+] $ &$\frac{i}{2}$&
$-\frac{1}{2}$&$\frac{1}{2}$&$-1$&$-1$ \\
$$&$3$&    $(+i)(+)[-] $ & $[+i][+][-] $ & $(+i)[+](-) $ & $[+i](+)(-) $&$\frac{i}{2}$&
$\frac{1}{2}$&$-\frac{1}{2}$&$-1$&$1$ \\
$$&$4$&    $[-i][-][-] $ & $(-i)(-)[-] $ & $[-i](-)(-) $ & $(-i)[-](-) $&$-\frac{i}{2}$&
$-\frac{1}{2}$&$-\frac{1}{2}$&$-1$&$1$ \\
\hline
 \end{tabular}
\end{center}
\end{tiny}
\end{table*}

To illustrate the properties of Clifford odd and even ''basis vectors'' we present 
their properties as well with respect to the two subgroups $SO(3,1) \times U(1)$ 
and $SU(3) \times U(1)$ of the group $SO(5,1)$, all with the same number of 
commuting operators as $SO(5,1)$.

We use the superposition of members of Cartan subalgebra, Eq.~(\ref{cartangrasscliff}),  
for the subgroup $SO(3,1) \times U(1)$: ($N^3_{\pm}\,,  \tau$) and (for the corresponding 
operators determining the ''family'' quantum numbers) ($\tilde{N}^3_{\pm}\,, 
\tilde{\tau}$) 
\begin{eqnarray}
\label{so1+3 5+1}
&& N^3_{\pm}(= N^3_{(L,R)}): = \,\frac{1}{2} (%
 S^{12}\pm i S^{03} )\,,\quad \tau =S^{56}\,, \quad
\tilde{N}^3_{\pm}(=\tilde{N}^3_{(L,R)}): = \,\frac{1}{2} (%
\tilde{S}^{12}\pm i \tilde{S}^{03} )\,, \quad \tilde{\tau} =\tilde{S}^{56}\,.
\end{eqnarray}
%


Similarly we use for the subgroup  $SU(3)$ $\times U(1)$: ($\tau',\tau^{3}, \tau^{8}$) 
and (for the corresponding operators determining the ''family'' quantum numbers) 
($\tilde{\tau}', \tilde{\tau}^{3}, \tilde{\tau}^{8}$)
%
 \begin{eqnarray}
 \label{so64 5+1}
 \tau^{3}: = &&\frac{1}{2} \,(%
 -S^{1\,2} - iS^{0\,3})\, , \qquad 
\tau^{8}= \frac{1}{2\sqrt{3}} (-i S^{0\,3} + S^{1\,2} -  2 S^{5\;6})\,,\nonumber\\
 \tau' = &&-\frac{1}{3}(-i S^{0\,3} + S^{1\,2} + S^{5\,6})\,.
%
 \end{eqnarray}
 The corresponding relations for ($\tilde{\tau}^{3}, \tilde{\tau}^{8}$,
 $\tilde{\tau}'$) can be read from Eq.~(\ref{so64 5+1}), if replacing $S^{ab}$  
 by $\tilde{S}^{ab}$. \\

 For the operators ${\bf {\cal S}}^{ab}=S^{ab}+ \tilde{S}^{ab}$ the corresponding 
relations for superposition of the Cartan subalgebra elements ($ N^3_{\pm}, \tau$) and
($\tau', \tau^3, \tau^8$) follow if  in Eqs.~(\ref{so1+3 5+1}, \ref{so64 5+1})  
${\bf {\cal S}}^{ab}$ instead of $S^{ab}$ are used.

\subsubsection{''Basis vectors'' of odd products of $\gamma^a$'s in $d=(5+1)$.}
\label{odd5+1}

This is  a short overview of properties of the Clifford odd ''basis vectors'' following a lot Ref.~\cite{nh2021RPPNP} and the references within this reference. 

To illustrate properties of  the Clifford odd ''basis vectors'' let us analyse them  
besides with respect to the Cartan subalgebra of $SO(5,1)$ also with respect 
to two kinds of the subgroups of $SO(5,1)$: $SO(3,1) \times U(1)$ and
$SU(3) \times U(1)$ of the  group $SO(5,1)$, with the same number of Cartan 
subalgebra members in all three cases ($\frac{d}{2}=3$).  In the case of $SO(5,1)$
three eigenvalues of the Cartan subalgebra $S^{ab}$, Eq.~(\ref{cartangrasscliff}), 
($S^{03}, S^{12}, S^{5,6}$) for each of the four members, the same for all four
families carrying eigenvalues of ($\tilde{S}^{03},\tilde{S}^{12}, \tilde{S}^{5,6}$)
are already presented in Table~\ref{Table Clifffourplet.}. The superposition of
the Cartan subalgebra operators for the subgroups $SO(3,1) \times U(1)$ and
$SU(3) \times U(1)$ of the  group $SO(5,1)$ are defined in Eqs.~(\ref{so1+3 5+1}, 
\ref{so64 5+1}.)
    
 The rest of generators of the two subgroups of the group $SO(5,1)$ can be 
 found in Eqs.~(\ref{so1+3}, \ref{so64}) of App.~\ref{grassmannandcliffordfermions}.\\

In Table~\ref{oddcliff basis5+1.}  the Clifford odd ''basis vectors'' 
$\hat{b}^{m \dagger}_{f}$ are presented (they are odd products of nilpotents and 
(the rest) of projectors) as the eigenvectors of all ($3$) commuting Cartan subalgebra members,  Eq.~(\ref{cartangrasscliff}), of the group: {\bf i.} $SO(5,1)$ (with $15$ generators),  {\bf ii.} $SO(4)\times U(1)$ (with $7$ generators) and {\bf iii.} 
$SU(3)\times U(1)$ (with $9$ generators), together with the eigenvalues of the corresponding commuting generators,  presented in Eqs.~(\ref{so1+3 5+1}, 
\ref{so64 5+1}).
These ''basis vectors'' are already presented as part of Table~\ref{Table Clifffourplet.}. 
They fulfil together with their Hermitian conjugated partners the anticommutation 
relations of Eqs.~(\ref{almostDirac}, \ref{should}).

\begin{table}
\begin{tiny}
\begin{small}
\caption{\label{oddcliff basis5+1.} The 
''basis vectors''  $\hat{b}^{m=(ch,s)\dagger}_{f}$ are presented for $d= (5+1)$-dimensional case. Each $\hat{b}^{m=(ch,s)\dagger}_{f}$ is a product of projectors 
and of an odd number of nilpotents and is  the "eigenvector" of all the Cartan 
subalgebra members, ($S^{03}$, $S^{12}$, $S^{56}$) and ($\tilde{S}^{03}$, 
$\tilde{S}^{12}$, $\tilde{S}^{56}$), Eq.~(\ref{cartangrasscliff}), with $ch$  
(representing the charge,  that is the eigenvalue of $S^{56}$) and $s$ (representing
the eigenvalues of $S^{03}$ and $S^{12}$) explaining the index $m$ while $f$ 
determines the family quantum numbers (the eigenvalues of ($\tilde{S}^{03}$, 
$\tilde{S}^{12}$, $\tilde{S}^{56}$)).   
This table presents also the eigenvalues of the three commuting operators 
$N^3_{L,R}$ $=  (S^{12}\pm i S^{03} )$ and $\tau =S^{56}$ of the subgroups 
$SU(2)\times SU(2)\times U(1)$ and the eigenvalues of the three commuting 
operators $\tau^3, \tau^8$ and $\tau'$ of the subgroups  $SU(3)\times U(1)$  
 ($\tau^{3} = \frac{1}{2} \,( -S^{1\,2} - iS^{0\,3})\, , \tau^{8}= 
 \frac{1}{2\sqrt{3}} (-i S^{0\,3} + S^{1\,2} -  2 S^{13\;14})\,,
 \tau' = -\frac{1}{3}(-i S^{0\,3} + S^{1\,2} + S^{5\,6})$).
In these two last cases index $m$ represents the eigenvalues 
 of the corresponding commuting generators.  
 $\Gamma^{(5+1)}=-\gamma^0 
 \gamma^1 \gamma^2 \gamma^3\gamma^5\gamma^6$, $\Gamma^{(3+1)}
 = i\gamma^0 \gamma^1 \gamma^2 \gamma^3$.
Operators $\hat{b}^{m=(ch,s) \dagger}_{f}$  and $\hat{b}^{m=(ch, s)}_{f}$
 fulfil the anticommutation relations of Eqs.~(\ref{almostDirac}, \ref{should}).
\vspace{2mm}}
\end{small}
\label{cliff basis5+1.}
\begin{center}
 \begin{tabular}{|r|l r|r|r|r|r|r|r|r|r|r|r|r|r|r|r|}
 \hline
$\, f $&$m $&$=(ch,s)$&$\hat{b}^{ m=(ch,s) \dagger}_f$
&$S^{03}$&$ S^{1 2}$&$S^{5 6}$&$\Gamma^{3+1}$ &$N^3_L$&$N^3_R$&
$\tau^3$&$\tau^8$&$\tau'$&
$\tilde{S}^{03}$&$\tilde{S}^{1 2}$& $\tilde{S}^{5 6}$\\
\hline
$I$&$1$&$(\frac{1}{2},\frac{1}{2})$&$
\stackrel{03}{(+i)}\,\stackrel{12}{[+]}| \stackrel{56}{[+]}$&
$\frac{i}{2}$&
$\frac{1}{2}$&$\frac{1}{2}$&$1$&
$0$&$\frac{1}{2}$&$0$&$0$&$-\frac{1}{2}$&$\frac{i}{2}$&$-\frac{1}{2}$&$-\frac{1}{2}$\\
$$ &$2$&$(\frac{1}{2},-\frac{1}{2})$&$
\stackrel{03}{[-i]}\,\stackrel{12}{(-)}|\stackrel{56}{[+]}$&
$-\frac{i}{2}$&$-\frac{1}{2}$&$\frac{1}{2}$&$1$
&$0$&$-\frac{1}{2}$&$0$&$-\frac{1}{\sqrt{3}}$&$\frac{1}{6}$&$\frac{i}{2}$&$-\frac{1}{2}$&$-\frac{1}{2}$\\
$$ &$3$&$(-\frac{1}{2},\frac{1}{2})$&$
\stackrel{03}{[-i]}\,\stackrel{12}{[+]}|\stackrel{56}{(-)}$&
$-\frac{i}{2}$&$ \frac{1}{2}$&$-\frac{1}{2}$&$-1$
&$\frac{1}{2}$&$0$&$-\frac{1}{2}$&$\frac{1}{2\sqrt{3}}$&$\frac{1}{6}$&$\frac{i}{2}$&$-\frac{1}{2}$&$-\frac{1}{2}$\\
$$ &$4$&$(-\frac{1}{2},-\frac{1}{2})$&$
\stackrel{03}{(+i)}\, \stackrel{12}{(-)}|\stackrel{56}{(-)}$&
$\frac{i}{2}$&$- \frac{1}{2}$&$-\frac{1}{2}$&$-1$
&$-\frac{1}{2}$&$0$&$\frac{1}{2}$&$\frac{1}{2\sqrt{3}}$&$\frac{1}{6}$&$\frac{i}{2}$&$-\frac{1}{2}$&$-\frac{1}{2}$\\ 
\hline 
$II$&$1$&$(\frac{1}{2},\frac{1}{2})$&$
\stackrel{03}{[+i]}\,\stackrel{12}{(+)}| \stackrel{56}{[+]}$&
$\frac{i}{2}$&
$\frac{1}{2}$&$\frac{1}{2}$&$1$&
$0$&$\frac{1}{2}$&$0$&$0$&$-\frac{1}{2}$&$-\frac{i}{2}$&$\frac{1}{2}$&$-\frac{1}{2}$\\
$$ &$2$&$(\frac{1}{2},-\frac{1}{2})$&$
\stackrel{03}{(-i)}\,\stackrel{12}{[-]}|\stackrel{56}{[+]}$&
$-\frac{i}{2}$&$-\frac{1}{2}$&$\frac{1}{2}$&$1$
&$0$&$-\frac{1}{2}$&$0$&$-\frac{1}{\sqrt{3}}$&$\frac{1}{6}$&$-\frac{i}{2}$&$\frac{1}{2}$&$-\frac{1}{2}$\\
$$ &$3$&$(-\frac{1}{2},\frac{1}{2})$&$
\stackrel{03}{(-i)}\,\stackrel{12}{(+)}|\stackrel{56}{(-)}$&
$-\frac{i}{2}$&$ \frac{1}{2}$&$-\frac{1}{2}$&$-1$
&$\frac{1}{2}$&$0$&$-\frac{1}{2}$&$\frac{1}{2\sqrt{3}}$&$\frac{1}{6}$&$-\frac{i}{2}$&$\frac{1}{2}$&$-\frac{1}{2}$\\
$$ &$4$&$(-\frac{1}{2},-\frac{1}{2})$&$
\stackrel{03}{[+i]} \stackrel{12}{[-]}|\stackrel{56}{(-)}$&
$\frac{i}{2}$&$- \frac{1}{2}$&$-\frac{1}{2}$&$-1$
&$-\frac{1}{2}$&$0$&$\frac{1}{2}$&$\frac{1}{2\sqrt{3}}$&$\frac{1}{6}$&$-\frac{i}{2}$&$\frac{1}{2}$&$-\frac{1}{2}$\\
\hline
$III$&$1$&$(\frac{1}{2},\frac{1}{2})$&$
\stackrel{03}{[+i]}\,\stackrel{12}{[+]}| \stackrel{56}{(+)}$&
$\frac{i}{2}$&
$\frac{1}{2}$&$\frac{1}{2}$&$1$&
$0$&$\frac{1}{2}$&$0$&$0$&$-\frac{1}{2}$&$-\frac{i}{2}$&$-\frac{1}{2}$&$\frac{1}{2}$\\
$$ &$2$&$(\frac{1}{2},-\frac{1}{2})$&$
\stackrel{03}{(-i)}\,\stackrel{12}{(-)}|\stackrel{56}{(+)}$&
$-\frac{i}{2}$&$-\frac{1}{2}$&$\frac{1}{2}$&$1$
&$0$&$-\frac{1}{2}$&$0$&$-\frac{1}{\sqrt{3}}$&$\frac{1}{6}$&$-\frac{i}{2}$&$-\frac{1}{2}$&$\frac{1}{2}$\\
$$ &$3$&$(-\frac{1}{2},\frac{1}{2})$&$
\stackrel{03}{(-i)}\,\stackrel{12}{[+]}|\stackrel{56}{[-]}$&
$-\frac{i}{2}$&$ \frac{1}{2}$&$-\frac{1}{2}$&$-1$
&$\frac{1}{2}$&$0$&$-\frac{1}{2}$&$\frac{1}{2\sqrt{3}}$&$\frac{1}{6}$&$-\frac{i}{2}$&$-\frac{1}{2}$&$\frac{1}{2}$\\
$$ &$4$&$(-\frac{1}{2},-\frac{1}{2})$&$
\stackrel{03}{[+i]}\, \stackrel{12}{(-)}|\stackrel{56}{[-]}$&
$\frac{i}{2}$&$- \frac{1}{2}$&$-\frac{1}{2}$&$-1$
&$-\frac{1}{2}$&$0$&$\frac{1}{2}$&$\frac{1}{2\sqrt{3}}$&$\frac{1}{6}$&$-\frac{i}{2}$&$-\frac{1}{2}$&$\frac{1}{2}$\\ 
\hline
$IV$&$1$&$(\frac{1}{2},\frac{1}{2})$&$
\stackrel{03}{(+i)}\,\stackrel{12}{(+)}| \stackrel{56}{(+)}$&
$\frac{i}{2}$&$\frac{1}{2}$&$\frac{1}{2}$&$1$
&$0$&$\frac{1}{2}$&$0$&$0$&$-\frac{1}{2}$&$\frac{i}{2}$&$\frac{1}{2}$&
$\frac{1}{2}$\\
$$ &$2$&$(\frac{1}{2},-\frac{1}{2})$&$
\stackrel{03}{[-i]}\,\stackrel{12}{[-]}|\stackrel{56}{(+)}$&
$-\frac{i}{2}$&$-\frac{1}{2}$&$\frac{1}{2}$&$1$
&$0$&$-\frac{1}{2}$&$0$&$-\frac{1}{\sqrt{3}}$&$\frac{1}{6}$&$\frac{i}{2}$&$\frac{1}{2}$&$\frac{1}{2}$\\
$$ &$3$&$(-\frac{1}{2},\frac{1}{2})$&$
\stackrel{03}{[-i]}\,\stackrel{12}{(+)}|\stackrel{56}{[-]}$&
$-\frac{i}{2}$&$ \frac{1}{2}$&$-\frac{1}{2}$&$-1$
&$\frac{1}{2}$&$0$&$-\frac{1}{2}$&$\frac{1}{2\sqrt{3}}$&$\frac{1}{6}$&$\frac{i}{2}$&$\frac{1}{2}$&$\frac{1}{2}$\\
$$ &$4$&$(-\frac{1}{2},-\frac{1}{2})$&$
\stackrel{03}{(+i)}\,\stackrel{12}{[-]}|\stackrel{56}{[-]}$&
$\frac{i}{2}$&$- \frac{1}{2}$&$-\frac{1}{2}$&$-1$
&$-\frac{1}{2}$&$0$&$\frac{1}{2}$&$\frac{1}{2\sqrt{3}}$&$\frac{1}{6}$&$\frac{i}{2}$&$\frac{1}{2}$&$\frac{1}{2}$\\
\hline
 \end{tabular}
 \end{center}
\end{tiny}
\end{table}

\vspace{2mm}

\begin{small}
Let us notice that the right handed,  $\Gamma^{(5+1)}=1$, fourplet of the fourth family of 
 Table~\ref{oddcliff basis5+1.}  can be found in the first  four  lines of 
Table~\ref{Table so13+1.} if only the $d=(5+1)$ part is taken into 
 account. It is repeated three more times (in the four lines from $9$ to $12$, from $17$ 
 to $20$ and from $25$ to $28$). 
 \end{small}

\begin{small}
  In any chosen $d=2n$-dimensional space there is besides the choice that the 
''basis vectors'' are the right handed Clifford odd objects, like in 
Table~\ref{Table Clifffourplet.},  as well the choice that
the Clifford odd objects are left handed. Their Hermitian conjugated  partners are 
then correspondingly the right handed Clifford odd objects. 
%
\end{small} 

\begin{figure}
  \centering
   \includegraphics[width=0.45\textwidth]{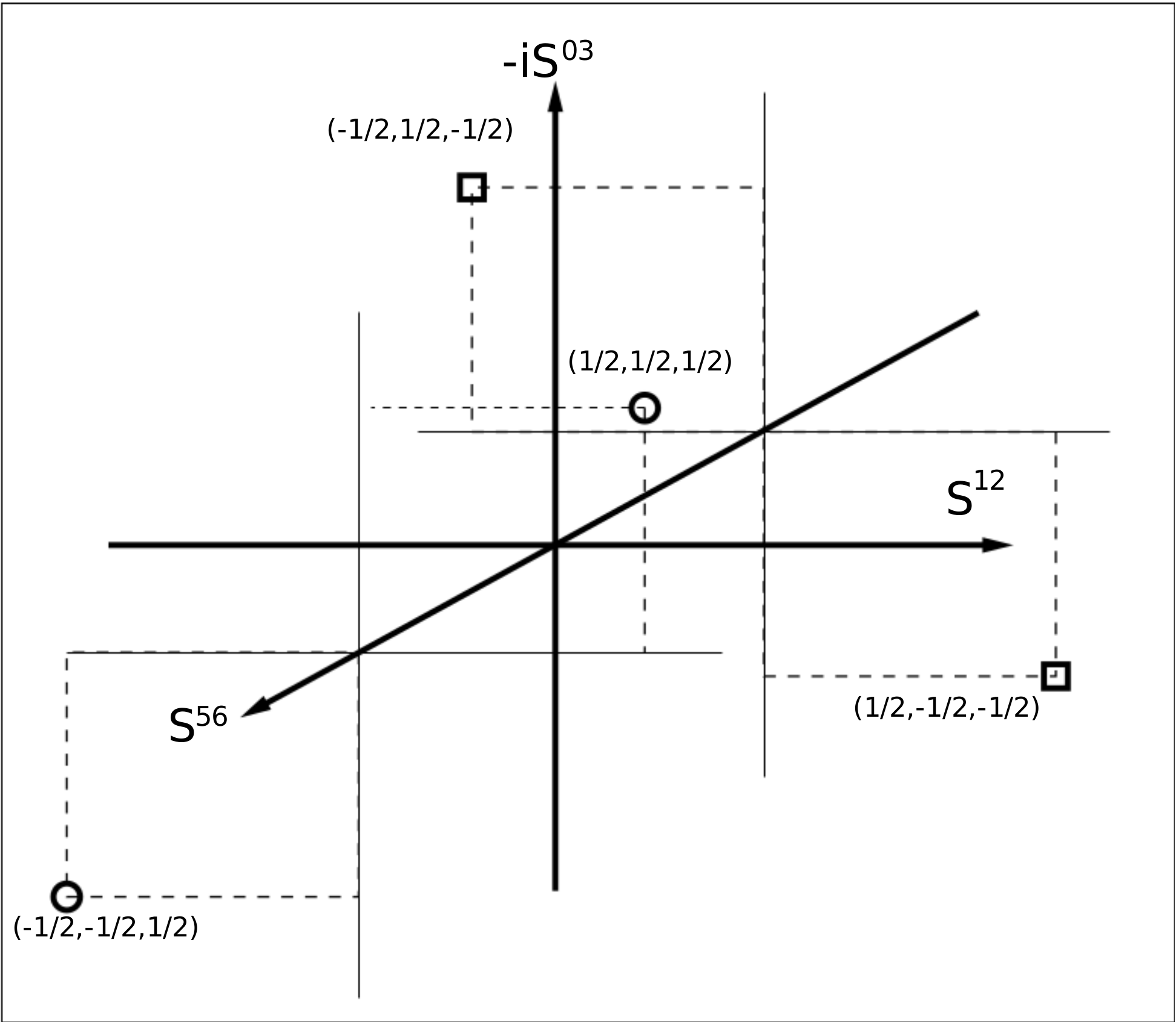}
  \caption{\label{Fig.SO5+1odd} The eigenvalues of the Cartan subalgebra 
  members of the group $SO(5,1)$ are presented for the ''basis vectors'', the properties 
of which are described in  Table~\ref{oddcliff basis5+1.}. 
On the abscissa axis and on the ordinate axis the eigenvalues of the two 
operators, $S^{12}$ and $ -i S^{03}$ are presented, respectively, 
while the third axis presents the eigenvalues of $S^{56}$.
There are two  Clifford odd family members ''basis vectors''  with $S^{56}= 
\frac{1}{2}$, denoted on  the figure
by the circle ${\bf \bigcirc}$, having 
($S^{12}=-\frac{1}{2}$, $ -i S^{03} =-\frac{1}{2}, S^{56} = \frac{1}{2}$) 
and ($S^{12}=\frac{1}{2}$, $ -i S^{03} =\frac{1}{2}, S^{56} = \frac{1}{2}$), 
respectively, and the two  Clifford odd family members ''basis vectors''  with $S^{56}= 
-\frac{1}{2}$, denoted on  the figure 
by the square  $\Box $ 
having ($S^{12}=-\frac{1}{2}$, $ -i S^{03} =\frac{1}{2}, S^{56} = -\frac{1}{2}$)
and ($S^{12}=\frac{1}{2}$, $ -i S^{03} =-\frac{1}{2}, S^{56} = -\frac{1}{2}$), respectively.
They all appear in four families.}
\end{figure}

In Fig.~\ref{Fig.SO5+1odd} 
the four ''basis vectors'' of one of the four families, anyone, are presented, demonstrating 
the eigenvalues of the Cartan subalgebra members ($S^{03}, S^{12}, S^{56}$). We notice
two ''basis vectors'' with $S^{56}=\frac{1}{2}$ and two with $S^{56}=-\frac{1}{2}$.

\vspace{2mm}

\begin{figure}
  \centering
   \includegraphics[width=0.45\textwidth]{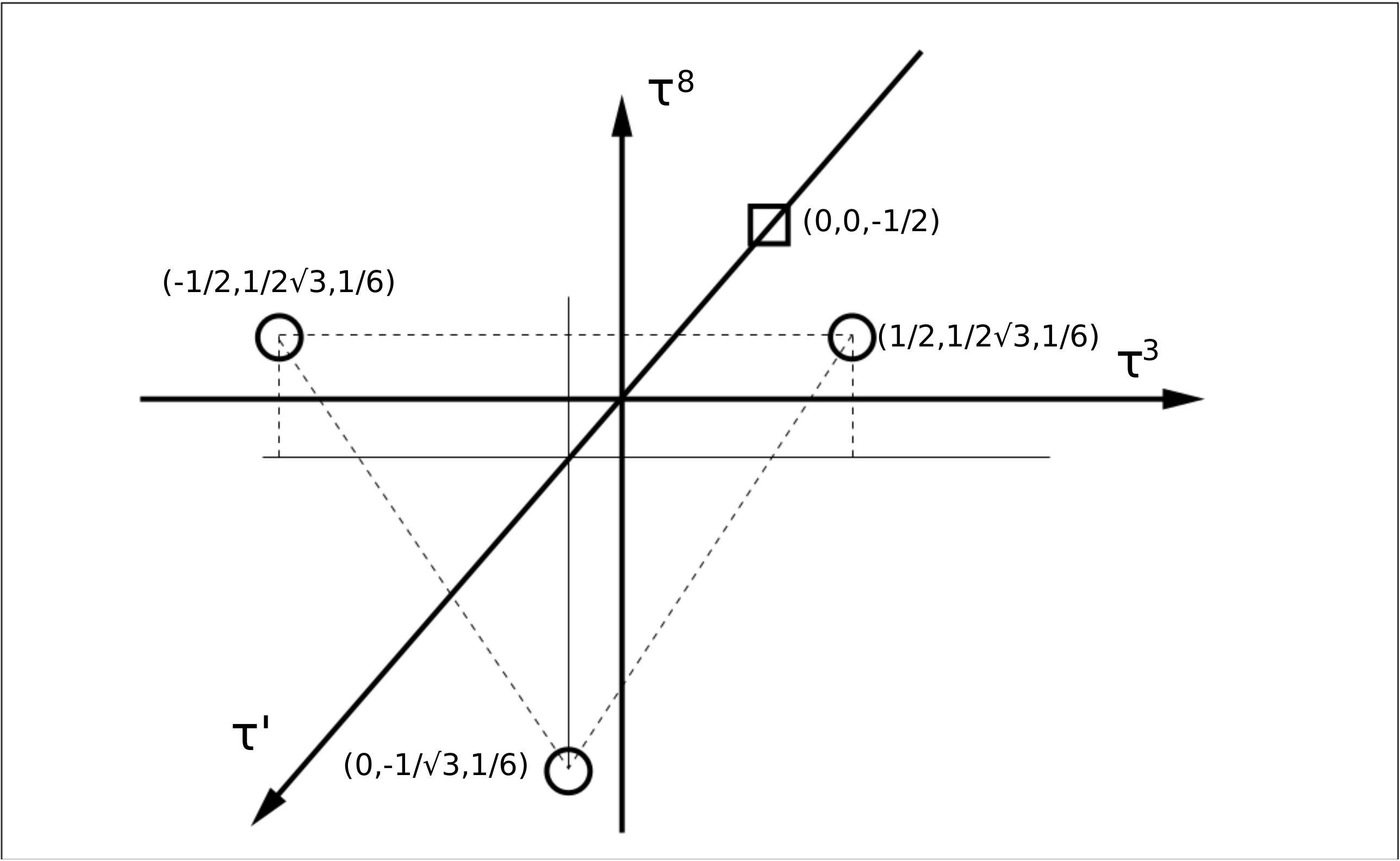}
  \caption{\label{Fig.SU3U1odd} 
  The eigenvalues of the superposition of the Cartan subalgebra eigenvalues of the 
  subgroups $SU(3)$ and $U(1)$ of the group $SO(5,1)$ for the ''basis vectors'', the 
  properties  of which appear in  Table~\ref{oddcliff basis5+1.}, are presented. 
On the abscissa axis, on the ordinate axis and on the third axis the eigenvalues of the  superposition of the three Cartan subalgebra members, $\tau^3=\frac{1}{2} 
(- S^{12} -i S^{03}), \tau^8=\frac{1}{2\sqrt{3}} ( S^{12} -i S^{03}- 2 S^{56}),
\tau'=- \frac{1}{3} (S^{12} -i S^{03} +S^{56})$ are presented. One notices one triplet,
denoted by ${\bf \bigcirc}$, with the values ($\tau'=\frac{1}{6}$, $\tau^3=-\frac{1}{2},
\tau^8=\frac{1}{2\sqrt{3}}), (\tau'=\frac{1}{6}$, $\tau^3=\frac{1}{2},
\tau^8=\frac{1}{2\sqrt{3}}), (\tau'=\frac{1}{6}$, $\tau^3=0,
\tau^8=-\frac{1}{\sqrt{3}}$), 
respectively, and one singlet denoted 
by the square $\Box$ ($\tau.^3=0, \tau^8=0, \tau'=-\frac{1}{2}$). 
The triplet and the singlet appear  in four families.} 
\end{figure}

In Fig.~\ref{Fig.SU3U1odd} the same four family members (the ''basis  vectors'') of any of 
the four families as the ones on Fig.~\ref{Fig.SO5+1odd} are presented, this way with 
respect to the superposition of the Cartan subalgbra members  ($S^{03}, S^{12}, S^{56}$), manifesting the $SU(3) \times U(1)$ subgroups of the $SO(5,1)$ group. We notice one 
''colour'' triplet with the ''fermion'' number $\tau'=\frac{1}{6}$ and one ''colourless''  singlet 
with the ''fermion'' number $\tau'=-\frac{1}{2}$.

\begin{small}
In the case of the group $SO(6)$ manifesting as $SU(3) \times U(1)$ and representing the
$SU(3)$ as the colour subgroup and $U(1)$ as the ''fermion'' number if embedded into 
$SO(13,1)$ the triplet would represent quarks and the singlet leptons, as can be seen in 
Table~\ref{Table so13+1.}.\\
\end{small}

It is not difficult to generalize what we learned from the case of $d=(5+1)$
to any even dimensional space.\\
 
 Let us summarize the properties of the Clifford odd ''basis vectors'' in even 
 dimensional spaces:\\
 
 {\bf i.} The ''basis vectors'', described by superposition of the Clifford odd 
 products of $\gamma^{a}$'s, belonging to either the same or to different 
 families, have  their  Hermitian conjugated partners in another group of 
 superposition of the Clifford odd products of $\gamma^{a}$'s.\\
 {\bf ii.} All the ''basis vectors'' belonging to the same or different families 
 are algebraically orthogonal and so are algebraically orthogonal among 
 themselves  their Hermitian conjugated partners.\\
 {\bf iii.} The Clifford odd ''basis vectors'' carry half integer spin (and from 
 the point of view of the lower dimensional space), the charges in the 
fundamental representations of the corresponding subgroups of the group 
$SO(d-1,1)$), and the half integer family quantum numbers.\\
{\bf iv.} The family members of ''basis vectors'' have the same properties in 
all the families~\footnote{
$S^{ab}$ rotate any ''basis vector'' into any other ''basis vector'' of the 
same family and $\tilde{S}^{ab}$ rotate any family member into the same 
family member of another family.},~ 
independently whether one observes the group $SO(d-1,1)$ or the subgroups 
with the same number of commuting operators. 
 %
In Table~\ref{oddcliff basis5+1.} presented eigenvalues of the commuting operators 
($S^{03}, S^{12}, S^{56}$), describing the properties of the ''basis vectors''  in $d=(5+1)$
(the same for each family carrying different family quantum numbers), 
are drown in  Fig.~\ref{Fig.SO5+1odd}. 
In  Table~\ref{oddcliff basis5+1.} presented the $SU(3)\times U(1)$ superposition of 
($S^{03}, S^{12}$, $S^{56}$),  that is ($\tau^3, \tau^8, \tau' $) appearing in 
Eq.~(\ref{so64 5+1}), (they are the same for each family carrying different family quantum numbers) are drown jn Fig.~\ref{Fig.SU3U1odd}.\\
{\bf v.} The sum of all the eigenvalues of all the commuting operators over 
the  $2^{\frac{d}{2}-1}$ family members is equal to zero for each of
 $2^{\frac{d}{2}-1}$ families, 
 independently whether the group $SO(5,1)$ or the subgroups of this group are 
 considered. This can be checked in Table~\ref{oddcliff basis5+1.} for the particular 
 case with $d=(5+1)$  for ($S^{03}, S^{12}$, $S^{56}$), as well as for 
 ($\tau^3, \tau^8, \tau' $).\\
%
{\bf vi.} The sum of the family quantum numbers over all the families is zero, as  
it can be checked for the $d=(5+1)$  case in Table~\ref{oddcliff basis5+1.}.\\
{\bf vii.} For a chosen even $d$ there is a choice of either right or left handed ''basis
vectors''. The choice of the handedness of  the ''basis vectors'' determines also the 
vacuum state for the ''basis vectors''.  The properties of the left handed ''basis 
vectors'' differ strongly from the right handed ones. 
The $d=(5+1)$ case demonstrates this difference: When looking at the eigenvalues
of the superposition of the Cartan subalgebra members  numbers 
($\tau^3, \tau^8, \tau'$) of the subgroups $SU(3)\times U(1)$ of the $SO(5,1)$ 
group one sees that the right handed realization manifests the ''colour'' properties 
of ''quarks'' and ''leptons'', Fig.~\ref{Fig.SU3U1odd}, while the left handed would 
represent the ''colour'' properties of  ''antiquarks'' and ''antileptons'' (words 
''quarks'' and ''antiquarks'' and ''leptons'' and ''antileptons'' are put into 
quotation marks since in the treated case $SU(3)$ and $U(1)$ are subgroups  of 
$SO(5,1)$ and not of $SO(6)$ as presented in Table~\ref{Table so13+1.}).\\
{\bf viii.} One can define the Hilbert space made of single fermion states which 
are tensor, $*_{T}$, products of $2^{\frac{d}{2}-1}$ ''basis vectors'' for each of
the $2^{\frac{d}{2}-1}$ families and of (continuously) infinite basis in ordinary 
space applying on the vacuum state, starting with no fermion state, one single 
fermion state, continuing with all possible ''basis vectors'' of all possible momenta 
(or coordinates)~(\cite{nh2021RPPNP}, Sect. 5 and in Refs. cited therein).\\
{\bf ix.} The Clifford odd ''basis vectors'',  constructed as superposition of  odd 
products of anticommuting operators $\gamma^a$'s, manifest all the properties 
desired for the internal space of fermion fields with the anticommutation relations 
for fermion fields included: When forming creation operators for fermions as the 
tensor products, $*_T$, of the Clifford odd ''basis vectors'' and the basis in ordinary (momentum or coordinate) space, the single fermion creation operators and (their 
Hermitian conjugated) annihilation operators fulfil all the requirements postulated 
by Dirac for the second quantized fermion fields, explaining correspondingly the 
postulates of  Dirac, Sect.~\ref{fermionsbosons}. 
\begin{eqnarray}
\label{Diracinternal}
\hat{b}^{m}_{f} {}_{*_{A}}|\psi_{oc}>&=& 0.\, |\psi_{oc}>\,,\nonumber\\
\hat{b}^{m \dagger}_{f}{}_{*_{A}}|\psi_{oc}>&=&  |\psi^m_{f}>\,,\nonumber\\
\{\hat{b}^{m}_{f}, \hat{b}^{m'}_{f `}\}_{*_{A}+}|\psi_{oc}>&=&
 0.\,|\psi_{oc}>\,, \nonumber\\
\{\hat{b}^{m \dagger}_{f}, \hat{b}^{m' \dagger}_{f  `}\}_{*_{A}+}|\psi_{oc}>
&=& 0. \,|\psi_{oc}>\,,\nonumber\\
\{\hat{b}^{m}_{f}, \hat{b}^{m' \dagger}_{f `}\}_{*_{A}+}|\psi_{oc}>
&=& \delta^{m m'} \delta_{f f `}|\psi_{oc}>\,.
\end{eqnarray}

So far we have considered the description of the internal space of fermion fields using 
the Clifford odd "basis vectors". They demonstrate the half integer spins (and charges 
from the point of view of $(3+1)$-dimensional space in the fundamental representation 
of the charge groups).  

In the next subsection the properties of the Clifford even ''basis vectors'' are discussed. 
We shall see that  the Clifford even ''basis vectors'' demonstrate the integer spins (and charges from the point of view of $(3+1)$-dimensional space in the adjoint 
representation of the charge groups). The Clifford even ''basis vectors'' obviously 
manifest properties of boson fields which are the gauge fields of the corresponding 
fermion fields the internal space of which is described by the Clifford odd ''basis 
vectors''. 

\subsubsection{''Basis vectors'' of even products of $\gamma^a$'s in $d=(5+1)$}

\label{even5+1}
%

%
The  Clifford even ''basis vectors''  are products of an even number of nilpotents,  
$\stackrel{ab}{(k)}$, with  the rest up to $\frac{d}{2}$ of  projectors, 
$\stackrel{ab}{[k]}$.

Table~\ref{Cliff basis5+1even I.} repeats the   Clifford even ''basis vectors''  ${}^{I}{\hat{\cal A}}^{m \dagger}_{f}$ from Table~\ref{Table Clifffourplet.}, pointing out 
the self adjoint members and the Hermitian conjugated pairs and  presenting the 
eigenvalues of the superposition of the Cartan subalgebra members, manifesting the
subgroups $SO(3,1)\times U(1)$ and  $SU(3)\times U(1)$, Eqs.~(\ref{so1+3, 5+1},
\ref{so64 5+1}).

Each of  ${}^{I}{\hat{\cal A}}^{m \dagger}_{f}$ is the product of projectors and 
an even number of nilpotents, and each is the eigenvector of all the Cartan subalgebra members, (${\cal S}^{03}$,  ${\cal S}^{03}$,  ${\cal S}^{03}$).

\vspace{2mm}

%
\begin{table}
\begin{tiny}
\caption{The  Clifford even ''basis vectors''  ${}^{I}{\hat{\cal A}}^{m \dagger}_{f}$,  
each of them is the product of projectors and an even number of nilpotents, and each is 
the eigenvector of all the Cartan subalgebra members, ${\cal S}^{03}$, 
${\cal S}^{12}$, ${\cal S}^{56}$, Eq.~(\ref{cartangrasscliff}), are presented for 
$d= (5+1)$-dimensional case. Indexes $m$ and $f$ determine  
$2^{\frac{d}{2}-1}\times 2^{\frac{d}{2}-1}$ different members  
${}^{I}{\hat{\cal A}}^{m \dagger}_{f}$. 
In the third column  the  ''basis vectors'' ${}^{I}{\hat{\cal A}}^{m \dagger}_{f}$  
which are  Hermitian conjugated partners to each other 
are pointed out with the same symbol. For example, with $\star \star$ 
are equipped the first member with $m=1$ and $f=1$ and the last member of $f=3$ 
with $m=4$.
The sign $\bigcirc$ denotes the  Clifford even ''basis vectors'' which are self adjoint 
$({}^{I}{\hat{\cal A}}^{m \dagger}_{f})^{\dagger}$ 
$={}^{I}{\hat{\cal A}}^{m \dagger}_{f}$. It is obvious that the sign 
${}^{\dagger}$ looses meaning, since ${}^{I}{\hat{\cal A}}^{m \dagger}_{f}$ 
are self adjoint or are (mutually, in pairs) Hermitian conjugated  to another 
${}^{I}{\hat{\cal A}}^{m' \dagger}_{f `}$.
This table represents also the  eigenvalues of the three commuting operators 
${\cal N}^3_{L,R}$ and ${\cal S}^{56 }$ of the subgroups
 $SU(2)\times SU(2)\times U(1)$ of the group
$SO(5,1)$ and the eigenvalues of the three 
commuting operators ${\tau}^3, {\tau}^8$ and ${ \tau'}$ of the 
subgroups  $SU(3)\times U(1)$.
\vspace{3mm}
}
%
%
\label{Cliff basis5+1even I.} 
 %
 \begin{center}
 \begin{tabular}{|r| r|r|r|r|r|r|r|r|r|r|r|}
 \hline
$\, f $&$m $&$*$&${}^{I}\hat{\cal A}^{m \dagger}_f$
&${\cal S}^{03}$&$ {\cal S}^{1 2}$&${\cal S} ^{5 6}$&
${\cal N}^3_L$&${\cal N}^3_R$&
${\cal \tau}^3$&${\cal \tau}^8$&${\cal \tau}'$
\\
\hline
%
$I$&$1$&$\star \star$&$
\stackrel{03}{[+i]}\,\stackrel{12}{(+)} \stackrel{56}{(+)}$&
$0$&$1$&$1$
&$\frac{1}{2}$&$\frac{1}{2}$&$-\frac{1}{2}$&$-\frac{1}{2\sqrt{3}}$&$-\frac{2}{3}$
\\
$$ &$2$&$\bigtriangleup$&$
\stackrel{03}{(-i)}\,\stackrel{12}{[-]}\,\stackrel{56}{(+)}$&
$- i$&$0$&$1$
&$\frac{1}{2}$&$-\frac{1}{2}$&$-\frac{1}{2}$&$-\frac{3}{2\sqrt{3}}$&$0$
\\
$$ &$3$&$\ddagger$&$
\stackrel{03}{(-i)}\,\stackrel{12}{(+)}\,\stackrel{56}{[-]}$&
$-i$&$ 1$&$0$
&$1 $&$0$&$-1$&$0$&$0$
\\
$$ &$4$&$\bigcirc$&$
\stackrel{03}{[+i]}\,\stackrel{12}{[-]}\,\stackrel{56}{[-]}$&
$0$&$0$&$0$
&$0$&$0$&$0$&$0$&$0$
\\
\hline 
$II$&$1$&$\bullet$&$
\stackrel{03}{(+i)}\,\stackrel{12}{[+]}\, \stackrel{56}{(+)}$&
$i$&$0$&$1$&
$-\frac{1}{2}$&$\frac{1}{2}$&$\frac{1}{2}$&$-\frac{1}{2\sqrt{3}}$&$-\frac{2}{3}$\\
$$ &$2$&$\otimes$&$
\stackrel{03}{[-i]}\,\stackrel{12}{(-)}\,\stackrel{56}{(+)}$&
$0$&$-1$&$1$
&$-\frac{1}{2}$&$-\frac{1}{2}$&$\frac{1}{2}$&$-\frac{3}{2\sqrt{3}}$&$0$
\\
$$ &$3$&$\bigcirc$&$
\stackrel{03}{[-i]}\,\stackrel{12}{[+]}\,\stackrel{56}{[-]}$&
$0$&$ 0$&$0$
&$0$&$0$&$0$&$0$&$0$
\\
$$ &$4$&$\ddagger$&$
\stackrel{03}{(+i)}\, \stackrel{12}{(-)}\,\stackrel{56}{[-]}$&
$i$&$-1$&$0$
&$-1$&$0$&$1$&$0$&$0$
\\ 
%
%
 \hline
$III$&$1$&$\bigcirc$&$
\stackrel{03}{[+i]}\,\stackrel{12}{[+]}\, \stackrel{56}{[+]}$&
$0$&$0$&$0$&
$0$&$0$&$0$&$0$&$0$\\
$$ &$2$&$\odot \odot$&$
\stackrel{03}{(-i)}\,\stackrel{12}{(-)}\,\stackrel{56}{[+]}$&
$-i$&$-1$&$0$
&$0$&$-1$&$0$&$-\frac{1}{\sqrt{3}}$&$\frac{2}{3}$\\
$$ &$3$&$\bullet$&$
\stackrel{03}{(-i)}\,\stackrel{12}{[+]}\,\stackrel{56}{(-)}$&
$-i$&$ 0$&$-1$
&$\frac{1}{2}$&$-\frac{1}{2}$&$-\frac{1}{2}$&$\frac{1}{2\sqrt{3}}$&$\frac{2}{3}$
\\
$$ &$4$&$\star \star$&$
\stackrel{03}{[+i]} \stackrel{12}{(-)}\,\stackrel{56}{(-)}$&
$0$&$- 1$&$- 1$
&$-\frac{1}{2}$&$-\frac{1}{2}$&$\frac{1}{2}$&$\frac{1}{2\sqrt{3}}$&$\frac{2}{3}$
\\
\hline
$IV$&$1$&$\odot \odot $&$
\stackrel{03}{(+i)}\,\stackrel{12}{(+)}\, \stackrel{56}{[+]}$&
$i$&$1$&$0$&
$0$&$1$&$0$&$\frac{1}{\sqrt{3}}$&$-\frac{2}{3}$
\\
$$ &$2$&$\bigcirc$&$
\stackrel{03}{[-i]}\,\stackrel{12}{[-]}\,\stackrel{56}{[+]}$&
$0$&$0$&$0$
&$0$&$0$&$0$&$0$&$0$
\\
$$ &$3$&$\otimes$&$
\stackrel{03}{[-i]}\,\stackrel{12}{(+)}\,\stackrel{56}{(-)}$&
$0$&$ 1$&$-1$
&$\frac{1}{2}$&$\frac{1}{2}$&$-\frac{1}{2}$&$\frac{3}{2\sqrt{3}}$&$0$
\\
$$ &$4$&$\bigtriangleup$&$
\stackrel{03}{(+i)}\, \stackrel{12}{[-]}\,\stackrel{56}{(-)}$&
$i$&$0$&$-1$
&$-\frac{1}{2}$&$\frac{1}{2}$&$\frac{1}{2}$&$\frac{3}{2\sqrt{3}}$&$0$\\ 
\hline 
 \end{tabular}
 \end{center}
\end{tiny}
\end{table}
%

To realize that even Clifford ''basis vectors'' have all the properties  needed to
describe the internal space of boson fields let us study their properties in the case of $d=(5+1)$-dimensional space,
analysing  them  with respect to the eigenvalues of the Cartan subalgebra members 
($S^{03}, S^{12}, S^{56}$), Eq.~(\ref{cartangrasscliff}), of the group $SO(5,1)$, 
as well as of the subgroups $SU(2)\times SU(2)\times U(1)$ and $SU(3)\times U(1)$,  
the commuting operators of which are presented in Egs.~(\ref{so1+3 5+1}, 
\ref{so64 5+1}). 

The  Clifford even ''basis vectors'' can have in the $d=(5+1)$ case none or two 
nilpotents. The rest are projectors (three or one).
In Table~\ref{Table Clifffourplet.} the  Clifford even ''basis vectors'' are denoted 
by $even \,{}^{I}\hat{{\cal A}}^{m \dagger}_{f}$ and  
$even \,{}^{II} \hat{{\cal A}}^{m \dagger}_{f} $. 
They obey commutation relations since even products of (anticommuting)  
$\gamma^a$'s obey commutation relations. 

The  Clifford even ''basis vectors'' ${}^{i}\hat{{\cal A}}^{m \dagger}_{f}, i=I,II$  
have their Hermitian conjugated partners within the same group,or they are self 
adjoint, for $i=I$ or $i=II$.

\vspace{3mm} 

Let us analyse what happens when the  Clifford even ''basis vectors'' apply 
algebraically on the Clifford odd ''basis vectors''.  We shall see that the spin of the
Clifford even ''basis vectors'' have the integer value, determined by ${{\bf \cal S}}^{ab}=
S^{ab} + \tilde{S}^{ab}$. Correspondingly the subgroups of $SO(d-1,d)$ manifest 
 the adjoint representations.
 
 \vspace{1mm}

{\bf a.} $\,\,$  The properties of the algebraic, $*_{A}$, application of the Clifford 
even ''basis vectors'', presented in Table~\ref{Table Clifffourplet.} as 
${}^{i}{\hat{\cal A}}^{m \dagger}_{f}$, $(i=I,II)$, on the Clifford odd ''basis 
vectors'' $\hat{b}^{m \dagger}_{f}$, presented in Table~\ref{oddcliff basis5+1.} 
(as well as in Table~\ref{Table Clifffourplet.} under 
$odd \, I \, \hat{b}^{m \dagger}_{f}$), teaches us that the Clifford even ''basis 
vectors''  describe the internal space of the 
gauge fields of  the corresponding $\hat{b}^{m \dagger}_{f}$. To see this let us 
evaluate:

%
The algebraic application, $*_A$, of ${}^{I}{\hat{\cal A}}^{m \dagger}_{f=3}, 
m=(1,2,3,4)$, presented in Table~\ref{Table Clifffourplet.} in the third column of 
$even\,I$, on $\hat{b}^{m=1 \dagger}_{f=1}$, presented as the first  Clifford 
$odd \,I$ ''basis vector'' on both Tables~(\ref{Table Clifffourplet.}, \ref{oddcliff basis5+1.}),
can easily be evaluated by taking into account Eq.~(\ref{graficcliff1}) for any $m$.
%
%
%
\begin{small}
\begin{eqnarray}
&&{}^{I}{\hat{\cal A}}^{m \dagger}_{3}*_A \hat{b}^{1 \dagger}_{1} 
(\equiv \stackrel{03}{(+i)} \stackrel{12}{[+]} \stackrel{56}{[+]}):\nonumber\\
&&{}^{I}{\hat{\cal A}}^{1 \dagger}_{3} (\equiv \stackrel{03}{[+i]}
\stackrel{12}{[+]} \stackrel{56}{[+]})  *_{A} \hat{b}^{1 \dagger}_{1} 
(\equiv \stackrel{03}{(+i)} \stackrel{12}{[+]} \stackrel{56}{[+]}) \rightarrow
\hat{b}^{1 \dagger}_{1}\,,
\quad ({}^{I}{\hat{\cal A}}^{1 \dagger}_{3})^{\dagger} \equiv
{}^{I}{\hat{\cal A}}^{1 \dagger}_{3}, \quad {\rm  selfadjoint}\nonumber\\
&&{}^{I}{\hat{\cal A}}^{2 \dagger}_{3} (\equiv \stackrel{03}{(-i)}
\stackrel{12}{(-)} \stackrel{56}{[+]}) *_{A} \hat{b}^{1 \dagger}_{1}
\rightarrow \hat{b}^{2 \dagger}_{1} 
(\equiv \stackrel{03}{[-i]} \stackrel{12}{(-)} \stackrel{56}{[+]})\,, \quad  
({}^{I}{\hat{\cal A}}^{2 \dagger}_{3})^{\dagger} \rightarrow 
{}^{I}{\hat{\cal A}}^{1 \dagger}_{4}\,,\; *1*
\nonumber\\ 
&& {}^{I}{\hat{\cal A}}^{3 \dagger}_{3} (\equiv \stackrel{03}{(-i)}
\stackrel{12}{[+]} \stackrel{56}{(-)}) *_{A} \hat{b}^{1 \dagger}_{1}
\rightarrow \hat{b}^{3 \dagger}_{1} 
(\equiv \stackrel{03}{[-i]} \stackrel{12}{[+]} \stackrel{56}{(-)})\,, \quad  
({}^{I}{\hat{\cal A}}^{3 \dagger}_{3})^{\dagger} \rightarrow 
{}^{I}{\hat{\cal A}}^{1 \dagger}_{2}\,,\,*2*
\nonumber\\
&&{}^{I}{\hat{\cal A}}^{4 \dagger}_{3} (\equiv \stackrel{03}{[+i]}
\stackrel{12}{(-)} \stackrel{56}{(-)}) *_{A} \hat{b}^{1 \dagger}_{1}
\rightarrow \hat{b}^{4 \dagger}_{1}
(\equiv \stackrel{03}{(+i)} \stackrel{12}{(-)} \stackrel{56}{(-)})\,, \quad
({}^{I}{\hat{\cal A}}^{4 \dagger}_{3})^{\dagger} \rightarrow 
{}^{I}{\hat{\cal A}}^{1 \dagger}_{1}\,,\;  *3*\,.
\label{calAb1}
\end{eqnarray}
\end{small}
%
%
The arrow sign, $\rightarrow$\,, means that the relation is valid up to the constant.
The signs {\it selfadjoint} and $*1*,*2*,*3*$ denote whether
${}^{I}{\hat{\cal A}}^{m \dagger}_{f}$ is selfadjoint or it has its Hermitian conjugated
partner  within the same group of ${}^{I}{\hat{\cal A}}^{m \dagger}_{f }$ denoted by
the same sign also in Eq.~(\ref{calAb234}).

We conclude that the  algebraic, $*_A$,  application  of
 ${}^{I}{\hat{\cal A}}^{m \dagger}_{3}$ 
on $\hat{b}^{1 \dagger}_{1}$ leads to
the same or another family member of the same family $f=1$, namely to
$\hat{b}^{m \dagger}_{1} $, $m=(1,2,3,4)$.

The reader can calculate the eigenvalues  of the Cartan subalgebra members, Eq.~(\ref{cartangrasscliff}), before and after the algebraic multiplication, $*_A$.
Since the  Clifford odd ''basis vectors''  appearing in the algebraic, $*_A$, multiplication
(''interacting'') with ${}^{I}{\hat{\cal A}}^{m \dagger}_{3}$ carry the half integer 
eigenvalues of ($S^{03}, S^{12}, S^{56}$), and so do $\hat{b}^{m \dagger}_{1}$ 
after the algebraic multiplication, it follows that 
${}^{I}{\hat{\cal A}}^{m \dagger}_{3}$ carry for all $m$ the integer eigenvalues 
of the Cartan subalgebra members, what means that  their Lorentz generators 
${\bf {\cal S}}^{ab}$  are the sum ${\bf {\cal S}}^{ab}= S^{ab} + \tilde{S}^{ab} $.

We conclude (we shall demonstrate in what follows  new examples) that the 
eigenvalues of the Cartan subalgebra members of all the Clifford even basis vectors 
${}^{I,II}{\hat{\cal A}}^{m \dagger}_{f}$ are equal to 
 ${\bf {\cal S}}^{ab}$ $= S^{ab} + \tilde{S}^{ab} $.
 
Let us calculate the eigenvalues of the Cartan subalgebra members of 
$\hat{b}^{m \dagger}_{1}$, for $m=(1,2,3,4)$. 
Since each  nilpotent and projector  determining the ''basis vector'' 
$\hat{b}^{m \dagger}_{1}$ is the eigenvector of the Cartan subalgebras of 
$S^{ab}$ and $\tilde{S}^{ab}$, one can read the corresponding eigenvalues from 
Eqs.~(\ref{eigencliffcartan}, \ref{signature0}).  These values are written also in 
Table~\ref{oddcliff basis5+1.}. Since all the members $\hat{b}^{m \dagger}_{1}$ of
the family ($f=1$  in this case) have the same eigenvalues of ($\tilde{S}^{03}, 
\tilde{S}^{12}, \tilde{S}^{56}$), we can conclude that
${}^{I}{\hat{\cal A}}^{m \dagger}_{3}$ contributes for $m=1$ the Cartan subalgbera
eigenvalues $(0,0,0)$, transforming $\hat{b}^{1 \dagger}_{1}$  back into 
$\hat{b}^{1 \dagger}_{1}$, ${}^{I}{\hat{\cal A}}^{2 \dagger}_{3}$  contributes the
eigenvalues $(-i.-1,0)$, transforming $\hat{b}^{1 \dagger}_{1}$  into  
$\hat{b}^{2 \dagger}_{1}$,   ${}^{I}{\hat{\cal A}}^{3 \dagger}_{3}$ contributes  
the Cartan subalgebra eigenvalues $(-i,0,-1)$, transforming
$\hat{b}^{1 \dagger}_{1}$ into $\hat{b}^{3 \dagger}_{1}$, and 
${}^{I}{\hat{\cal A}}^{4 \dagger}_{3}$ contributes the Cartan subalgbera eigenvalues 
$(0,-1,-1)$, transforming $\hat{b}^{1 \dagger}_{1}$ into $\hat{b}^{4 \dagger}_{1}$.\\

Proceeding like in Eq.~(\ref{calAb1}), while taking into account the fourth line of Eq.~(\ref{graficcliff1}), one finds
\begin{small}
\begin{eqnarray}
&&{}^{I}{\hat{\cal A}}^{m\dagger}_{4}*_A \hat{b}^{2 \dagger}_{1} 
(\equiv \stackrel{03}{[-i]} \stackrel{12}{(-)} \stackrel{56}{[+]}):\nonumber\\
&&{}^{I}{\hat{\cal A}}^{1 \dagger}_{4} (\equiv \stackrel{03}{(+i)}
\stackrel{12}{(+)} \stackrel{56}{[+]})  *_{A} \hat{b}^{2 \dagger}_{1} 
(\equiv \stackrel{03}{[-i]} \stackrel{12}{(-)} \stackrel{56}{[+]}) \rightarrow
\hat{b}^{1 \dagger}_{1}\,,\quad ({}^{I}{\hat{\cal A}}^{1 \dagger}_{4})^{\dagger} 
\rightarrow  {}^{I}{\hat{\cal A}}^{1 \dagger}_{3}\,,\; *1* \nonumber\\
&&{}^{I}{\hat{\cal A}}^{2 \dagger}_{4} (\equiv \stackrel{03}{[-i]}
\stackrel{12}{[-]} \stackrel{56}{[+]}) *_{A} \hat{b}^{2 \dagger}_{1}
\rightarrow \hat{b}^{2 \dagger}_{1} 
(\equiv \stackrel{03}{[-i]} \stackrel{12}{(-)} \stackrel{56}{[+]})\,, \quad
({}^{I}{\hat{\cal A}}^{2 \dagger}_{4})^{\dagger} \rightarrow 
{}^{I}{\hat{\cal A}}^{2 \dagger}_{4}\,,\;\, {\rm selfadjoint} \nonumber\\ 
&&{}^{I}{\hat{\cal A}}^{3 \dagger}_{4} (\equiv \stackrel{03}{[-i]}
\stackrel{12}{(+)} \stackrel{56}{(-)}) *_{A} \hat{b}^{2 \dagger}_{1}
\rightarrow \hat{b}^{3 \dagger}_{1} 
(\equiv \stackrel{03}{[-i]} \stackrel{12}{[+]} \stackrel{56}{(-)})\,, \quad 
({}^{I}{\hat{\cal A}}^{3 \dagger}_{4})^{\dagger} \rightarrow 
{}^{I}{\hat{\cal A}}^{2 \dagger}_{2}\,,\;  *4*\nonumber\\
&&{}^{I}{\hat{\cal A}}^{4 \dagger}_{4} (\equiv \stackrel{03}{(+i)}
\stackrel{12}{[-]} \stackrel{56}{(-)}) *_{A} \hat{b}^{2 \dagger}_{1}
\rightarrow \hat{b}^{4 \dagger}_{1}
(\equiv \stackrel{03}{(+i)} \stackrel{12}{(-)} \stackrel{56}{(-)})\,, \quad
({}^{I}{\hat{\cal A}}^{4 \dagger}_{4})^{\dagger} \rightarrow 
{}^{I}{\hat{\cal A}}^{2 \dagger}_{1}\,,\; *5*\nonumber\\
&&{}^{I} {\hat{\cal A}}^{m \dagger}_{2}*_A \hat{b}^{3 \dagger}_{1} 
(\equiv \stackrel{03}{[-i]} \stackrel{12}{[+]} \stackrel{56}{(-)}):\nonumber\\
&&{}^{I}{\hat{\cal A}}^{1 \dagger}_{2} (\equiv \stackrel{03}{(+i)}
\stackrel{12}{[+]} \stackrel{56}{(+)} ) *_{A} \hat{b}^{3 \dagger}_{1} 
(\equiv \stackrel{03}{[-i]} \stackrel{12}{[+]} \stackrel{56}{(-)]}) \rightarrow
\hat{b}^{1 \dagger}_{1}\,,\quad
({}^{I}{\hat{\cal A}}^{1 \dagger}_{2})^{\dagger} \rightarrow 
{}^{I}{\hat{\cal A}}^{3 \dagger}_{3}\,,\; *2*\nonumber\\
&&{}^{I}{\hat{\cal A}}^{2 \dagger}_{2} (\equiv \stackrel{03}{[-i]}
\stackrel{12}{(-)} \stackrel{56}{(+) } ) *_{A} \hat{b}^{3 \dagger}_{1}
\rightarrow \hat{b}^{2 \dagger}_{1} 
(\equiv \stackrel{03}{[-i]} \stackrel{12}{(-)} \stackrel{56}{[+]})\,,\quad
({}^{I}{\hat{\cal A}}^{2 \dagger}_{2})^{\dagger} \rightarrow 
{}^{I}{\hat{\cal A}}^{3 \dagger}_{4}\,,\;  *4* \nonumber\\ 
&& {}^{I}{\hat{\cal A}}^{3 \dagger}_{2} (\equiv \stackrel{03}{[-i]}
\stackrel{12}{[+] } \stackrel{56}{[-]}) *_{A} \hat{b}^{3 \dagger}_{1}
\rightarrow \hat{b}^{3 \dagger}_{1} 
(\equiv \stackrel{03}{[-i]} \stackrel{12}{[+]} \stackrel{56}{(-)})\,, \quad
({}^{I}{\hat{\cal A}}^{3 \dagger}_{2})^{\dagger} \rightarrow 
{}^{I}{\hat{\cal A}}^{3 \dagger}_{2}\,,\;  \,{\rm  selfadjoint}\nonumber\\
&&{}^{I}{\hat{\cal A}}^{4 \dagger}_{2} (\equiv \stackrel{03}{(+i)}
\stackrel{12}{(-)} \stackrel{56}{[-]}) *_{A} \hat{b}^{1 \dagger}_{1}
\rightarrow \hat{b}^{4 \dagger}_{1}
(\equiv \stackrel{03}{(+i)} \stackrel{12}{(-)} \stackrel{56}{(-)})\,,\quad
 ({}^{I}{\hat{\cal A}}^{4 \dagger}_{2})^{\dagger} \rightarrow 
{}^{I}{\hat{\cal A}}^{3 \dagger}_{1}\,,\;  *6*\nonumber\\
&& {}^{I}{\hat{\cal A}}^{m \dagger}_{1}*_A \hat{b}^{4 \dagger}_{1} 
(\equiv \stackrel{03}{(+i)} \stackrel{12}{(-)} \stackrel{56}{(-)}):\nonumber\\
&&{}^{I}{\hat{\cal A}}^{1\dagger}_{1} (\equiv \stackrel{03}{[+i]}
\stackrel{12}{(+)} \stackrel{56}{(+)})  *_{A} \hat{b}^{4 \dagger}_{1} 
(\equiv \stackrel{03}{(+i)} \stackrel{12}{(-)} \stackrel{56}{(-)}) \rightarrow
\hat{b}^{1 \dagger}_{1}\,, \quad 
({}^{I}{\hat{\cal A}}^{1 \dagger}_{1})^{\dagger} \rightarrow 
{}^{I}{\hat{\cal A}}^{4 \dagger}_{3}\,,\;  *3*
\nonumber\\
&&{}^{I}{\hat{\cal A}}^{2\dagger}_{1} (\equiv \stackrel{03}{(-i)}
\stackrel{12}{[-]} \stackrel{56}{(+)}) *_{A} \hat{b}^{4 \dagger}_{1}
\rightarrow \hat{b}^{2 \dagger}_{1} 
(\equiv \stackrel{03}{[-i]} \stackrel{12}{(-)} \stackrel{56}{[+]})\,, \quad
({}^{I}{\hat{\cal A}}^{2 \dagger}_{1})^{\dagger} \rightarrow 
{}^{I}{\hat{\cal A}}^{4 \dagger}_{4}\,,\;  *5*\nonumber\\ 
&& {}^{I}{\hat{\cal A}}^{3 \dagger}_{1} (\equiv \stackrel{03}{(-i)}
\stackrel{12}{(+)} \stackrel{56}{[-]}) *_{A} \hat{b}^{4 \dagger}_{1}
\rightarrow \hat{b}^{3 \dagger}_{1} 
(\equiv \stackrel{03}{[-i]} \stackrel{12}{[+]} \stackrel{56}{(-)})\,, \quad
({}^{I}{\hat{\cal A}}^{3 \dagger}_{1})^{\dagger} \rightarrow 
{}^{I}{\hat{\cal A}}^{4 \dagger}_{2}\,,\; *6*\nonumber\\
&&{}^{I}{\hat{\cal A}}^{4 \dagger}_{1} (\equiv \stackrel{03}{[+i]}
\stackrel{12}{[-]} \stackrel{56}{[-]}) *_{A} \hat{b}^{4 \dagger}_{1}
\rightarrow \hat{b}^{4 \dagger}_{1}
(\equiv \stackrel{03}{(+i)} \stackrel{12}{(-)} \stackrel{56}{(-)})\,,\quad
({}^{I}{\hat{\cal A}}^{4 \dagger}_{1})^{\dagger} \rightarrow 
{}^{I}{\hat{\cal A}}^{4 \dagger}_{1}\,,\; {\rm  selfadjoint}\,.
\label{calAb234}
\end{eqnarray}
\end{small}
Eqs.~(\ref{calAb1}, \ref{calAb234}) include on the very right hand side of any equation 
the information whether ${}^{I}{\hat{\cal A}}^{m \dagger}_{f}$ is self adjoint 
or is Hermitian conjugated to another ${}^{I}{\hat{\cal A}}^{m' \dagger}_{f'}$. The
pairs, which are Hermitian conjugated to each other, carry the same number: *1*, *2*,  
and so on up to *6*.

All the rest of ${}^{I}{\hat {\cal A}}^{m \dagger}_f$, applying on  $\hat{b}^{n \dagger}_{1}$,
give zero for any other  $f$ except the one presented in Eqs.~(\ref{calAb1}, 
\ref{calAb234}). 

We can repeat this calculation for all four family members 
$\hat{b}^{m ` \dagger}_{f `}$ of any of families $f `=(1,2,3,4)$, concluding
\begin{eqnarray}
&&{}^{I}{\hat{\cal A}}^{m \dagger}_{3} *_{A} \hat{b}^{1 \dagger}_{f}
\rightarrow \hat{b}^{m \dagger}_{f}\,,\nonumber\\
&&{}^{I}{\hat{\cal A}}^{m \dagger}_{4} *_{A} \hat{b}^{2 \dagger}_{f}
\rightarrow \hat{b}^{m \dagger}_{f}\,,\nonumber\\
&&{}^{I}{\hat{\cal A}}^{m \dagger}_{2} *_{A} \hat{b}^{3 \dagger}_{f}
\rightarrow \hat{b}^{m \dagger}_{f}\,,\nonumber\\
&&{}^{I}{\hat{\cal A}}^{m \dagger}_{1} *_{A} \hat{b}^{4 \dagger}_{f}
\rightarrow \hat{b}^{m \dagger}_{f}\,.
\label{calAb1234}
\end{eqnarray}
%


Let us add what we just learned  about the Clifford even ''basis vectors'' 
${}^{I}{\hat{\cal A}}^{m \dagger}_{f}$ the properties presented in
Table~\ref{Cliff basis5+1even I.}:

\noindent
{\bf i.} The Clifford even ''basis vectors'' ${}^{I}{\hat{\cal A}}^{m \dagger}_{f}$ 
are products of an even number of nilpotents,  $\stackrel{ab}{(k)}$, 
with the rest up to $\frac{d}{2}$ of projectors, $\stackrel{ab}{[k]}$.\\
{\bf ii.} Nilpotents and projectors are eigenvectors of the Cartan 
subalgebra members ${\bf {\cal S}}^{ab}$ $= S^{ab} + \tilde{S}^{ab} $, Eq.~(\ref{cartangrasscliff}), carrying correspondingly the integer eigenvalues 
of the Cartan subalgebra members.\\
{\bf iii.} They have their Hetmitian conjugated partners within the same group  of
${}^{I}{\hat{\cal A}}^{m \dagger}_{f}$ with $2^{\frac{d}{2}-1}$ 
$\times$ $2^{\frac{d}{2}-1}$ members.\\
{\bf iv.} They have properties of the boson gauge fields as  we recognized for the 
case of $d=(5+1)$-dimensional space in Eqs.~(\ref{calAb1}, \ref{calAb234},
\ref{calAb1234});  When applying on the Clifford odd ''basis vectors'' (offering the description of the fermion fields) they transform the Clifford odd ''basis vectors'' into
another Clifford odd ''basis vectors'', transferring to the Clifford odd ''basis vectors'' 
the integer spins with respect to the  $SO(d-1,1)$ group, 
while with respect to subgroups of the  $SO(d-1,1)$ group they transfer appropriate
superposition of the eigenvalues (manifesting the properties of the adjoint representations of the corresponding groups).\\
%
%
{\bf v.} The sum of all the eigenvalues of each of the Cartan subalgebra members over 
the $16$ members of ${}^{I}{\hat{\cal A}}^{m \dagger}_{f}$ is equal to zero, regardless of which subgroups of $SO(5,1)$ group we are dealing with: $SU(2)\times SU(2) \times U(1)$ 
or  $SU(3)\times U(1)$, what can easily generalize to $2^{\frac{d}{2}-1}$ 
$\times$ $2^{\frac{d}{2}-1}$ case.\\


When looking at the eigenvalues of (${\bf \cal{S}}^{03}, {\bf \cal{S}}^{12},
 {\bf \cal{S}}^{56}$), we read from Table~\ref{Cliff basis5+1even I.} that 
 four of  $16$ ''basis vectors'' ${}^{I}{\hat{\cal A}}^{m \dagger}_{f}$  are singlets, 
 all with  (${\bf \cal{S}}^{03}=0,  {\bf \cal{S}}^{12}=0, {\bf \cal{S}}^{56}=0$).\\ 
 Four ''basis vectors'' form the  fourplet with $\{$($-i{\bf \cal{S}}^{03}=1, 
 {\bf \cal{S}}^{12}=1,  {\bf \cal{S}}^{56}=0$) and ($-i{\bf \cal{S}}^{03}=-1, 
 {\bf \cal{S}}^{12}=-1,  {\bf \cal{S}}^{56}=0$) (these two, Hermitian conjugated 
 to each other, are denoted by $\otimes \otimes$) and  
 ($-i{\bf \cal{S}}^{03}=-1, {\bf \cal{S}}^{12}=1, {\bf \cal{S}}^{56}=0$) 
 and ($-i{\bf \cal{S}}^{03}=1, {\bf \cal{S}}^{12}=-1,
 {\bf \cal{S}}^{56}=0$) (these last two, Hermitian conjugated to each other, are 
 denoted by $\ddagger$)$\}$.\\
Four   ''basis vectors'' form the fourplet with  $\{$($-i{\bf \cal{S}}^{03}=0, {\bf \cal{S}}^{12}=1, {\bf \cal{S}}^{56}=1$) (denoted by $\star\star$),  
($-i{\bf \cal{S}}^{03}=0,  {\bf \cal{S}}^{12}=-1, {\bf \cal{S}}^{56}=1$) 
(denoted by $\otimes$),   ($-i{\bf \cal{S}}^{03}=-1, {\bf \cal{S}}^{12}=0, 
{\bf \cal{S}}^{56}=1$)  (denoted by $\bigtriangleup$) and 
($-i{\bf \cal{S}}^{03}=1, {\bf \cal{S}}^{12}=0,  {\bf \cal{S}}^{56}=1$) 
(denoted by $\bullet$)$\}$.\\  
 The Hermitian conjugated partners  of this last fourplet appear in the third fourplet 
 with  $\{$($-i{\bf \cal{S}}^{03}=0, {\bf \cal{S}}^{12}=-1,
 {\bf \cal{S}}^{56}=-1$) (denoted by $\star\star$),  ($-i{\bf \cal{S}}^{03}=0, 
 {\bf \cal{S}}^{12}= 1, {\bf \cal{S}}^{56}=-1$) (denoted by $\otimes$),  
 ($-i{\bf \cal{S}}^{03}=1, {\bf \cal{S}}^{12}=0, {\bf \cal{S}}^{56}=-1$) 
 (denoted by $\bigtriangleup$) and ($-i{\bf \cal{S}}^{03}=-1, 
 {\bf \cal{S}}^{12}=0, {\bf \cal{S}}^{56}=-1$) (denoted by $\bullet$)$\}$.\\ 

These 16 Clifford even ''basis vectors'' are offering the description of the internal 
space of the gauge fields of four families of the Clifford odd ''basis vectors'' 
presented in Fig.~\ref{Fig.SO5+1odd}.

\begin{figure}
  \centering
   \includegraphics[width=0.45\textwidth]{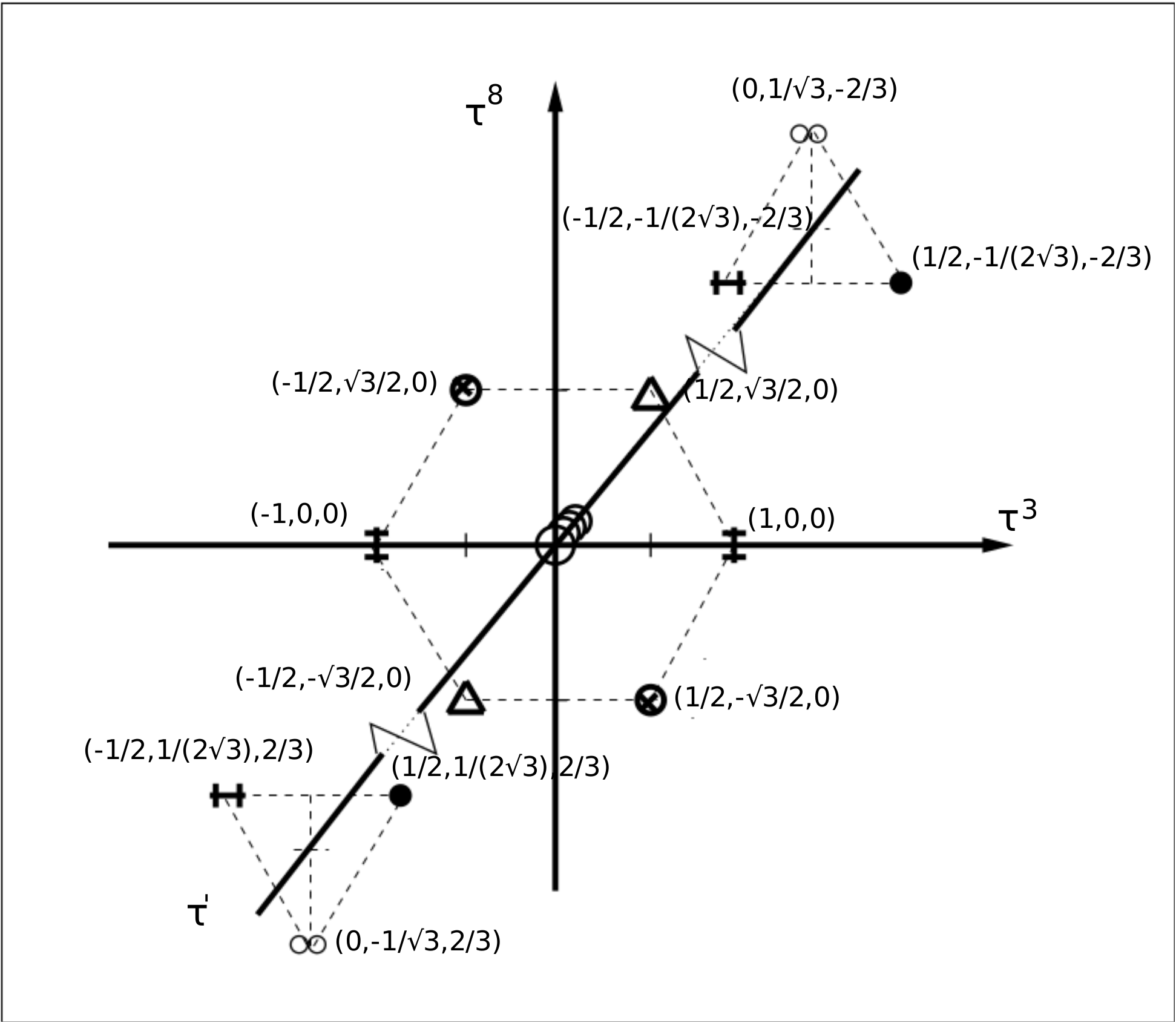}
  \caption{\label{FigSU3U1even} 
The Clifford even ''basis vectors'' ${}^{I}{\hat{\cal A}}^{m\dagger}_{f}$ in the case that 
$d=(5+1)$ are presented with respect to the eigenvalues of the commuting operators 
of the subgroups $SU(3)$ and $U(1)$ of the group $SO(5,1)$: ($\tau^3=\frac{1}{2}$
$(- {\bf {\cal S}}^{12} -i {\bf {\cal S}}^{03})$, $\tau^8=\frac{1}{2\sqrt{3}} 
({\bf {\cal S}}^{12} -i {\bf {\cal S}}^{03}- 2 {\bf {\cal S}}^{56})$, 
$\tau'=- \frac{1}{3} ({\bf {\cal S}}^{12} -i {\bf {\cal S}}^{03} + 
{\bf {\cal S}}^{56})$). 
Their properties appear  also in  Table~\ref{Cliff basis5+1even I.}. 
The abscissa axis carries the eigenvalues of $\tau^3$, the ordinate axis carries the 
eigenvalues of $\tau^8$ and the third axis carries the eigenvalues of $\tau'$.
One notices {\bf i.} four singlets with ($\tau^3=0, \tau^8=0, \tau'=0$), denoted by $\bigcirc$, representing four self adjoint  Clifford even ''basis vectors'' 
${}^{I}{\hat{\cal A}}^{m \dagger}_{f}$, with ($f=1, m=4$), ($f=2, m=3$), 
($f=3, m=1$), ($f=4, m=2$)\,, 
{\bf ii.} one sextet of three pairs, Hermitian conjugated to each other, with $\tau'=0$, denoted by $\bigtriangleup$ (${}^{I}{\hat{\cal A}}^{2\dagger}_{1}$ with  
($\tau'=0, \tau^3=-\frac{1}{2}, \tau^8=-\frac{3}{2\sqrt{3}}$) 
and ${}^{I}{\hat{\cal A}}^{4\dagger}_{4}$ with ($\tau'=0, \tau^3=\frac{1}{2}, \tau^8=\frac{3}{2\sqrt{3}}$)),  
by $\ddagger$ (${}^{I}{\hat{\cal A}}^{3\dagger}_{1}$ with  ($\tau'=0, \tau^3=-1, \tau^8=0$) and ${}^{I}{\hat{\cal A}}^{4\dagger}_{2}$ with $\tau'=0, \tau^3=1, \tau^8=0$)),  and by $\otimes$ 
(${}^{I}{\hat{\cal A}}^{2\dagger}_{2}$ with  ($\tau'=0, \tau^3=\frac{1}{2}, \tau^8=-\frac{3}{2\sqrt{3}}$) 
and ${}^{I}{\hat{\cal A}}^{3\dagger}_{4}$ with ($\tau'=0, \tau^3=- \frac{1}{2}, 
\tau^8=\frac{3}{2\sqrt{3}}$)), 
{\bf iii.} and one triplet, denoted by $\star \star$ (${}^{I}{\hat{\cal A}}^{4\dagger}_{3}$ with 
($\tau'=\frac{2}{3}, \tau^3=\frac{1}{2}, \tau^8=\frac{1}{2\sqrt{3}}$)), 
by $\bullet$ (${}^{I}{\hat{\cal A}}^{3\dagger}_{3}$ with  ($\tau'=\frac{2}{3}, 
\tau^3= -\frac{1}{2}, \tau^8=\frac{1}{2\sqrt{3}}$)), and by $\odot \odot$ 
(${}^{I}{\hat{\cal A}}^{2\dagger}_{3}$ with
 ($\tau'=\frac{2}{3}, \tau^3=0, \tau^8=-\frac{1}{\sqrt{3}}$)), 
 {\bf iv.} as well as one  antitriplet, Hermitian conjugated to triplet, denoted by $\star \star$ 
 (${}^{I}{\hat{\cal A}}^{1 \dagger}_{1}$ with  ($\tau'=-\frac{2}{3}, 
 \tau^3=-\frac{1}{2}, \tau^8=-\frac{1}{2\sqrt{3}}$)), 
 by $\bullet$ (${}^{I}{\hat{\cal A}}^{1\dagger}_{2}$ with  ($\tau'=-\frac{2}{3}, \tau^3= \frac{1}{2}, \tau^8=- \frac{1}{2\sqrt{3}}$)), and by $\odot \odot$ (${}^{I}{\hat{\cal A}}^{4 \dagger}_{1}$ with  ($\tau'=-\frac{2}{3},  \tau^3=0, \tau^8=\frac{1}{\sqrt{3}}$)).} 
\end{figure}

Fig.~\ref{FigSU3U1even} represents the $2^{\frac{d}{2}-1}\times 2^{\frac{d}{2}-1}$ 
members  ${}^{I}{\hat{\cal A}}^{m}_{f}$ of the Clifford even ''basis vectors'' for the
case that $d=(5+1)$. The properties of ${}^{I}{\hat{\cal A}}^{m}_{f}$ are presented 
also in Table~\ref{Cliff basis5+1even I.}. There are in this case  again $16$ members. Manifesting the structure of subgroups $SU(3) \times U(1)$ of the group $SO(5,1)$ they 
are represented as eigenvectors of the superposition of the Cartan subalgebra members 
(${\bf {\cal S}}^{03}, {\bf {\cal S}}^{12}, {\bf {\cal S}}^{56}$), that is with 
$\tau^3=\frac{1}{2} (- {\bf {\cal S}}^{12} -i {\bf {\cal S}}^{03})$, 
$\tau^8=\frac{1}{2\sqrt{3}} ( {\bf {\cal S}}^{12} -i {\bf {\cal S}}^{03}- 
2 {\bf {\cal S}}^{56})$, and $\tau'=- \frac{1}{3} 
({\bf {\cal S}}^{12} -i {\bf {\cal S}}^{03} + {\bf {\cal S}}^{56})$. 
There are four self adjoint Clifford even ''basis vectors'' with ($\tau^3=0, \tau^8=0, \tau'=0$),
one sextet of three pairs Hermitian conjugated to each other, one triplet and one 
antitriplet with the members of the triplet Hermitian conjugated to the corresponding 
members of the antitriplet and opposite. These $16$ members of the Clifford even 
''basis vectors'' ${}^{I}{\hat{\cal A}}^{m}_{f}$ are the boson ''partners'' of the 
Clifford odd ''basis vectors'' $\hat{b}^{m \dagger }_{f}$, presented in Fig.~\ref{Fig.SU3U1odd} for one of four families, anyone.  The reader can check 
that the algebraic application of ${}^{I}{\hat{\cal A}}^{m}_{f}$,  belonging to 
the triplet, transforms the Clifford odd singlet, denoted on Fig.~\ref{Fig.SU3U1odd} 
by $\Box$, to one of the members of the triplet, denoted  on Fig.~\ref{Fig.SU3U1odd} 
by  the circle ${\bf \bigcirc}$.

\begin{small}
Let us, for example, algebraically apply ${}^{I}{\hat{\cal A}}^{2}_{3}$ 
($\equiv \stackrel{03}{(-i)}\,\stackrel{12}{(-)}\,\stackrel{56}{[+]}$), denoted 
by $\odot\odot$, carrying $(\tau^3=0, \tau^8= - \frac{1}{\sqrt 3}, \tau'=\frac{2}{3})$
on the Clifford odd ''basis vector''  $\hat{b}^{1\dagger}_{1}$, with  $(\tau^3=0, \tau^8= 0, 
\tau'=-\frac{1}{2})$ presented in Table~\ref{oddcliff basis5+1.} and represented on Fig.~\ref{Fig.SU3U1odd} by $\Box$ as a singlet. ${}^{I}{\hat{\cal A}}^{2}_{3}$ 
transforms $\hat{b}^{1\dagger}_{1}$  (by transferring to $\hat{b}^{1\dagger}_{1}$ 
$(\tau^3=0, \tau^8= - \frac{1}{\sqrt 3}, \tau'=\frac{2}{3})$)  to $\hat{b}^{1\dagger}_{2}$ with $(\tau^3=0, \tau^8= - \frac{1}{\sqrt{3}}, \tau'=\frac{1}{6})$, belonging 
on Fig.~\ref{Fig.SU3U1odd} to the triplet, denoted by $\bigcirc$.

We can see that ${}^{I}{\hat{\cal A}}^{m \dagger}_{3}$  with $(m=2,3,4)$,
if applied on the $SU(3)$ singlet $\hat{b}^{1 \dagger}_{1}$ with ($\tau'= -\frac{1}{2}, \tau^3=0,\tau^8=0$), transforms it to  $\hat{b}^{m=2,3,4) \dagger}_{1}$, respectively, which are members of the $SU(3 )$ triplet. All these Clifford even 
''basis vectors'' have $\tau'$ equal to $\frac{2}{3}$, changing correspondingly 
$\tau'= -\frac{1}{2}$ into 
$\tau'=\frac{1}{6}$ and  bringing the needed values of $\tau^3$ and $\tau^8$. 
 \end{small}

We conclude:\\
Algebraic, $*_A$, application  of ${}^{I}{\hat{\cal A}}^{m \dagger}_{3} $, with 
$m=1,2,3,4$,
on $\hat{b}^{1 \dagger}_{1}$ transforms $\hat{b}^{1 \dagger}_{1}$ into
$\hat{b}^{m \dagger}_{1}$, $m=(1,2,3,4)$.\\                  
Algebraic, $*_A$, application of ${}^{I}{\hat {\cal A}}^{m}_4,$ with $m=(1,2,3,4)$ on   
$\hat{b}^{2 \dagger}_{1}$ transforms $\hat{b}^{2 \dagger}_{1}$ into
$\hat{b}^{m \dagger}_{1}, m=(1,2,3,4)$.\\
Algebraic, $*_A$, application of ${}^{I}{\hat {\cal A}}^{m}_2,$ with $m=(1,2,3,4)$ on   
$\hat{b}^{3 \dagger}_{1}$ transforms $\hat{b}^{3 \dagger}_{1}$ into
$\hat{b}^{m \dagger}_{1}, m=(1,2,3,4)$.\\
Algebraic, $*_A$, application of ${}^{I}{\hat {\cal A}}^{m}_1,$ with $m=(1,2,3,4)$ on   
$\hat{b}^{4 \dagger}_{1}$ transforms $\hat{b}^{4 \dagger}_{1}$ into
$\hat{b}^{m \dagger}_{1}, m=(1,2,3,4)$~\footnote{
We can comment the above events, concerning the internal space of  fermions and 
bosons, as:\\
 If the fermion, the internal space of which is described by Clifford odd
''basis vector'' $\hat{b}^{1 \dagger}_{1}$, absorbs the boson with the ''basis vector'' 
${}^{I}{\hat{\cal A}}^{1}_{3}$ (with ${\cal{S}}^{03}=0, {\cal{S}}^{12}=0, 
{\cal{S}}^{56}=0$), its  ''basis vector'' $\hat{b}^{1 \dagger}_{1}$ remains unchanged.\\
The fermion with the ''basis vector'' $\hat{b}^{1 \dagger}_{1}$, absorbing the boson 
with ${\hat{\cal A}}^{2}_{3}$ (with ${\cal{S}}^{03}=-i, {\cal{S}}^{12}=-1, 
{\cal{S}}^{56}=0$), 
changes into fermion with the ''basis vector'' $\hat{b}^{2 \dagger}_{1}$ (which 
carries  $S^{03}=-\frac{i}{2},
S^{12}=-\frac{1}{2}$, and the same $S^{56}=\frac{1}{2}$ as before). \\
The fermion with ''basis vector'' $\hat{b}^{1 \dagger}_{1}$,  absorbing the boson with 
the ''basis vector'' ${\hat{\cal A}}^{3}_{3}$ (carrying ${\cal{S}}^{03}=-i, 
{\cal{S}}^{12}=0, {\cal{S}}^{56}=-1$) changes to $\hat{b}^{3 \dagger}_ {1}$, 
(with $S^{03}=-\frac{i}{2},S^{12}=\frac{1}{2}$, $S^{56}=-\frac{1}{2}$).\\
While  the  fermion with the ''basis vector'' $\hat{b}^{1 \dagger}_{1}$  absorbing 
the boson with the ''basis vector'' ${\hat{\cal A}}^{4}_{3}$  (with ${\cal{S}}^{03}=0, 
{\cal{S}}^{12}=-1, {\cal{S}}^{56}=-1$) changes to $\hat{b}^{4 \dagger}_ {1}$ (with $S^{03}=\frac{i}{2}, S^{12}=-\frac{1}{2}$, $S^{56}=-\frac{1}{2}$).

These  comments are meaningful when bosons and fermions start to
interact, that is when the interaction between fermions and bosons is included.

The {\it spin-charge-family} assumes the simple starting action as presented in Eq.~(\ref{wholeaction}), in which fermions interact in $d=(13 +1)$ with gravity only.
}.

\vspace{3mm} 

Let us analyse what happens when the  Clifford even ''basis vectors'' apply algebraically, 
$*_{A}$, on  the Clifford even ''basis vectors''. \\

{\bf b.} $\,\,$ There are  $16$ ($2^{\frac{d=6}{2}-1}\times 2^{\frac{d=6}{2}-1}$) 
Clifford  even basis vectors ${}^{I}{\hat{\cal A}}^{m \dagger}_{f}$, presented in 
Table~\ref{Cliff basis5+1even I.} and on Fig.~\ref{FigSU3U1even}.

There are six pairs of the Clifford even ''basis vectors'' 
$ {}^{I}{\hat{\cal A}}^{m \dagger}_{f}$, which are Hermitian conjugated to each other, denoted in Table~\ref{Cliff basis5+1even I.} by
($\star \star$, $\bigtriangleup$, $\ddagger$, $\bullet$, $\otimes$, $\odot \odot$). 
The algebraic multiplication, $*_A$, of such pairs leads to one of four self adjoint 
operators,  denoted in Table~\ref{Cliff basis5+1even I.} by $\bigcirc$. 

One recognizes that $ {}^{I}{\hat{\cal A}}^{m \dagger}_{f}  *_{A} 
{}^{I}{\hat{\cal A}}^{m' \dagger}_{f} =0 \, \;\forall f, \; {\rm if}\, m'\ne m_{of}\,$
and $m\ne m'$, 
where  $ {}^{I}{\hat{\cal A}}^{m_{o f} \dagger}_{f} $ represents the self adjoint
member for each $f$ separately, denoted on Fig.~\ref{FigSU3U1even} by $\bigcirc$.

Each  $ {}^{I}{\hat{\cal A}}^{m \dagger}_{f} $ has in each group with $f '=f$
as well as $f `\ne f$ only one member $ {}^{I}{\hat{\cal A}}^{m' \dagger}_{f `} $
for which is $ {}^{I}{\hat{\cal A}}^{m \dagger}_{f} *_{A}
 {}^{I}{\hat{\cal A}}^{m' \dagger}_{f '}= {}^{I}{\hat{\cal A}}^{m \dagger}_{f `} $, which is not equal zero. For $f `= f $ the member is the self adjoint 
 $ {}^{I}{\hat{\cal A}}^{m_{of} \dagger}_{f} $ one. All other algebraic 
 multiplication give zero.
 
 Two ''basis vectors'' ${}^{I}{\hat{\cal A}}^{m \dagger}_{f}$  and 
${}^{I}{\hat{\cal A}}^{m ' \dagger}_{f}$ of the same $f$ and of 
$(m, m')\ne m_{of}$ are in algebraic multiplication orthogonal (giving zero).

We summarize the above findings in
\begin{eqnarray}
\label{ruleAAI}
{}^{I}{\hat{\cal A}}^{m \dagger}_{f} \,*_A\, {}^{I}{\hat{\cal A}}^{m' \dagger}_{f `}
\rightarrow \left \{ \begin{array} {r} {}^{I}{\hat{\cal A}}^{m \dagger}_{f `}\,, 
{\rm only\; one\; for }\;
\forall f `\,,\\
{\rm or \,zero}\,.
\end{array} \right.
\end{eqnarray}

Two ''basis vectors'' ${}^{I}{\hat{\cal A}}^{m \dagger}_{f}$  and 
${}^{I}{\hat{\cal A}}^{m' \dagger}_{f '}$, the algebraic product, $*_{A}$, of which 
 gives nonzero contribution, ''scatter'' into the third one  
 ${}^{I}{\hat{\cal A}}^{m \dagger}_{f `}$, like  ${}^{I}{\hat{\cal A}}^{1 \dagger}_{1}$ 
  $*_A \,{}^{I}{\hat{\cal A}}^{4 \dagger}_{2}\rightarrow$  
  ${}^{I}{\hat{\cal A}}^{1 \dagger}_{2}$ and  
 ${}^{I}{\hat{\cal A}}^{2 \dagger}_{2}$   
 $*_A \,{}^{I}{\hat{\cal A}}^{3 \dagger}_{4}\rightarrow$  
 ${}^{I}{\hat{\cal A}}^{2 \dagger}_{4}$~
  %
\footnote{The word ''scatter''  is used in quotation marks since the  ''basis vectors'' 
${}^{I}{\hat{\cal A}}^{m \dagger}_{f}$ determine only the internal space of bosons,
as also the ''basis vectors'' $\hat{b}^{m \dagger}_{f}$ determine only the internal 
space of fermions.}.

Looking at the ''basis vectors'' of boson fields 
${}^{I}{\hat{\cal A}}^{m \dagger}_{f}$ from the point 
of view of subgroups  $SU(3)\times U(1)$ of the group $SO(5+1)$ we  recognize 
in the part of fields forming the octet the colour gauge fields of quarks and leptons 
and antiquarks and antileptons.  \\
$U(1)$ fields carry   no $U(1)$ charges.\\

Let us write the commutation relations for Clifford even ''basis vectors'' taking into account Eq.~(\ref{ruleAAI}).\\
$\;\;$ {\bf i.} $\;\;$ In the case that ${}^{I}{\hat{\cal A}}^{m \dagger}_{f} \,*_{A} \, 
{}^{I}{\hat{\cal A}}^{m' \dagger}_{f `} \rightarrow $ 
$ {}^{I}{\hat{\cal A}}^{m \dagger}_{f `}$ and 
${}^{I}{\hat{\cal A}}^{m' \dagger}_{f `} \,*_{A} \,
{}^{I}{\hat{\cal A}}^{m \dagger}_{f}= 0$ it follows
 %
%
\begin{eqnarray}
\label{ruleAAI1}
\{{}^{I}{\hat{\cal A}}^{m \dagger}_{f} \,, \, 
{}^{I}{\hat{\cal A}}^{m' \dagger}_{f `}\}_{*_A \,-}
\rightarrow \left \{ \begin{array} {r} {}^{I}{\hat{\cal A}}^{m \dagger}_{f `}\,, \;\;\;\;
({\rm if\; }\quad \;{}^{I}{\hat{\cal A}}^{m \dagger}_{f} \,*_{A} \, 
{}^{I}{\hat{\cal A}}^{m' \dagger}_{f `} 
\rightarrow  {}^{I}{\hat{\cal A}}^{m \dagger}_{f `}
\\
{\rm and }\;\;{}^{I}{\hat{\cal A}}^{m' \dagger}_{f `} \,*_{A} \,
{}^{I}{\hat{\cal A}}^{m \dagger}_{f}= 0)\,,
\end{array} \right.
\end{eqnarray}\\
 $\;\;$ {\bf ii.} $\;\;$ In  the case that 
 ${}^{I}{\hat{\cal A}}^{m \dagger}_{f} \,*_{A} \, 
{}^{I}{\hat{\cal A}}^{m' \dagger}_{f `} \rightarrow $ 
$ {}^{I}{\hat{\cal A}}^{m \dagger}_{f `}$ and 
${}^{I}{\hat{\cal A}}^{m' \dagger}_{f `} \,*_{A} \,
{}^{I}{\hat{\cal A}}^{m \dagger}_{f}\rightarrow $ 
${}^{I}{\hat{\cal A}}^{m' \dagger}_{f } \,$ it follows
\begin{eqnarray}
\label{ruleAAI2}
\{{}^{I}{\hat{\cal A}}^{m \dagger}_{f} \,, \, 
{}^{I}{\hat{\cal A}}^{m' \dagger}_{f `} \}_{*_A \,-}
\rightarrow \left \{ \begin{array} {r} {}^{I}{\hat{\cal A}}^{m \dagger}_{f `}
- {}^{I}{\hat{\cal A}}^{m' \dagger}_{f }\,, \;\;\;
({\rm if\; } \quad \;{}^{I}{\hat{\cal A}}^{m \dagger}_{f} \,*_{A} \, 
{}^{I}{\hat{\cal A}}^{m' \dagger}_{f `} 
\rightarrow  {}^{I}{\hat{\cal A}}^{m \dagger}_{f `}
\\
{\rm and }\;\;{}^{I}{\hat{\cal A}}^{m' \dagger}_{f `} \,*_{A} \,
{}^{I}{\hat{\cal A}}^{m \dagger}_{f}\rightarrow 
{}^{I}{\hat{\cal A}}^{m' \dagger}_{f})\,,
\end{array} \right.
\end{eqnarray}\\
%

 $\;\;$ {\bf iii.} $\;\;$ In all other cases we have
\begin{eqnarray}
\label{ruleAAI3}
\{{}^{I}{\hat{\cal A}}^{m \dagger}_{f} \,,\,
{}^{I}{\hat{\cal A}}^{m' \dagger}_{f `}\}_{*_A \,-} =0\;.
\end{eqnarray}
$\{{}^{I}{\hat{\cal A}}^{m \dagger}_{f} \,, \, 
{}^{I}{\hat{\cal A}}^{m' \dagger}_{f `} \}_{*_A \,-}$ means
${}^{I}{\hat{\cal A}}^{m \dagger}_{f} \, *_A \, 
{}^{I}{\hat{\cal A}}^{m' \dagger}_{f `}  - {}^{I}{\hat{\cal A}}^{m' \dagger}_{f `} \, *_A \, 
{}^{I}{\hat{\cal A}}^{m \dagger}_{f }$.\\

\vspace{2mm}

There are besides the $2^{\frac{d(=6)}{2}-1}\times 2^{\frac{d(=6)}{2}-1}$ 
Clifford  even ''basis vectors'' of ${}^{I}{\hat{\cal A}}^{m \dagger}_{f}$, 
presented as $even \,I$ in Table~\ref{Table Clifffourplet.}, the same number, 
$2^{\frac{d(=6)}{2}-1}\times 2^{\frac{d(=6)}{2}-1}$, of the Clifford even 
''basis vectors'' ${}^{II}{\hat{\cal A}}^{m \dagger}_{f}$, presented in 
Table~\ref{Table Clifffourplet.} as $even \,II$. \\

Let be pointed out that the choice of the Clifford odd ''basis vectors'', $odd\, I$, 
describing  the internal space of fermions, and consequently  the choice  of 
$odd \, II$ as their Hermitian conjugated partners,  and consequently also 
the choice of the Clifford even ''basis vectors'', $even \, I$, describing  the 
internal space of their gauge fields in the way presented in Eqs.~(\ref{calAb1}, 
\ref{calAb234}, \ref{calAb1234}), is ours. If  we choose in 
Table~\ref{Table Clifffourplet.} $odd \,II $ to represent  the ''basis vectors''
describing the internal space of fermions, then the corresponding ''basis vectors''
representing the internal space of boson gauge fields are those of $even\, II$. \\

\vspace{3mm}

{\bf c.} $\,\,$ The above findings  for the ''basis vectors'' 
${}^{I}{\hat{\cal A}}^{m \dagger}_{f}$ in Eq.~(\ref{ruleAAI}) are valid as 
well for ${}^{II}{\hat{\cal A}}^{m \dagger}_{f}$
\begin{eqnarray}
\label{ruleAAII}
{}^{II}{\hat{\cal A}}^{m \dagger}_{f} \,*_A\, 
{}^{II}{\hat{\cal A}}^{m' \dagger}_{f `}
\rightarrow \left \{ \begin{array} {r} {}^{II}{\hat{\cal A}}^{m \dagger}_{f `}\,, 
{\rm only\; one\; for \,}
\forall f `\,,\\
{\rm or \,zero}\,.
\end{array} \right.
\end{eqnarray}
Equivalent figure to Fig.~\ref{FigSU3U1even} can be drown also for 
${}^{II}{\hat{\cal A}}^{m \dagger}_{f}$, as well as the relations presented in 
Eqs.~(\ref{ruleAAI1}, \ref{ruleAAI2}, \ref{ruleAAI3}). One only has to replace 
in these equations ${}^{I}{\hat{\cal A}}^{m \dagger}_{f}$ by 
${}^{II}{\hat{\cal A}}^{m \dagger}_{f} $. 
But we recognize in Table~\ref{Table Clifffourplet.} the differences among 
${}^{I}{\hat{\cal A}}^{m \dagger}_{f}$ and ${}^{II}{\hat{\cal A}}^{m \dagger}_{f} $,
which cause the orthogonality relation presented in Eq.~(\ref{AIAIIorth}).\\


It remains therefore to see how do the Clifford even ''basis vectors'' 
${}^{I}{\hat{\cal A}}^{m \dagger}_{f}$ and 
${}^{II}{\hat{\cal A}}^{m' \dagger}_{f `}$  algebraically apply (''interact'') on each other.\\

\vspace{3mm}
{\bf d.} $\,\,$  
Taking into account Eqs.~(\ref{gammatildeantiher}, \ref{tildegammareduced}) and 
the third and the fourth line of Eq.~(\ref{graficcliff1}) one recognizes  that in 
Table~\ref{Table Clifffourplet.} any member of $even\, II$ follows from $even\, I$ 
by the application of $\gamma^a \tilde{\gamma}^a$ for any index $a$. We correspondingly find, due to the fact that the two projectors $\stackrel{ab}{[k]}$ 
and $\stackrel{ab}{[-k]}$ give zero in the algebraic application $\stackrel{ab}{[k]} *_{A} \stackrel{ab}{[-k]}=0$, the orthogonality relation
\begin{eqnarray}
\label{AIAIIorth}
{}^{I}{\hat{\cal A}}^{m \dagger}_{f} *_A {}^{II}{\hat{\cal A}}^{m \dagger}_{f} 
&=&0={}^{II}{\hat{\cal A}}^{m \dagger}_{f} *_A 
{}^{I}{\hat{\cal A}}^{m \dagger}_{f}\,.
\end{eqnarray}
Correspondingly the application  of ${}^{II}{\hat{\cal A}}^{m \dagger}_{f}$
on the Clifford odd ''basis vectors'' ${\hat b}^{m \dagger}_f$  and 
their Hermitian conjugated partners ${\hat b}^{m}_f$,  differ from the 
application  of ${}^{I}{\hat{\cal A}}^{m \dagger}_{f}$
on the Clifford odd ''basis vectors'' ${\hat b}^{m \dagger}_f$ studied so far.\\

The  properties of the algebraic application, $*_{A}$, of 
${}^{II}{\hat{\cal A}}^{m \dagger}_{f}$ on the Clifford odd ''basis vectors'' 
${\hat b}^{m \dagger}_f$  and their Hermitian conjugated partners 
${\hat b}^{m}_f$ will be studied separately. 

The main message of this article to convince the reader that
 the Clifford algebra provides a description of the inner space of {\bf fermion and 
 boson fields}, {\bf thereby offering a new understanding of the second 
quantization postulates for  fermion and bosonic fields}, is fulfilled.\\

\vspace{3mm}


%
\subsection{''Basis vectors'' describing internal space of fermions and bosons in any
even dimensional space}
\label{generalbasisinternal}

In  Subsect.~\ref{cliffordoddevenbasis5+1}  the properties of "basic vectors", which
describe the internal space of fermions  (with the Clifford odd ''basis vectors'') and 
bosons (with the Clifford even ''basis vectors'') in $d=(5+1)$=dimensional space, 
are presented in an illustrative way.

The properties of the Clifford odd ''basis vectors'' describing the internal space 
of fermions are discussed in~\ref{odd5+1}. In~\ref{even5+1} the properties 
of the Clifford even ''basis vectors'' are described, representing the internal space 
of bosons, manifesting in the algebraic application among themselves and with 
the Clifford odd ''basis vectors'' the properties of the gauge fields of the fermions, 
the internal space of which is described by the Clifford odd ''basis vectors''.

Generalization to any even $d$ is not difficult. The description of the 
internal space of fermions follows Ref.~\cite{nh2021RPPNP}, of the 
internal space of bosons started in Ref.~\cite{n2021SQ}. \\
 
 {\bf a.} The  ''basis vectors'' offering the description of the internal space of 
fermions, $\hat{b}^{m \dagger}_f$, must be superposition of  odd  products of 
nilpotents $\stackrel{ab}{(k)}$, $2n'+1$, in $d=2(2n +1)$,  
$n'=(0,1,2,\dots, \frac{1}{2}(\frac{d}{2}-1)$ (the minimum number of nilpotents is 
one and the maximum $2n+1$), and the rest is the product of $n''$ 
projectors $\stackrel{ab}{[k]}$, $n''=\frac{d}{2}-(2n'+1)$. (For $d=4n$ the 
minimum number of nilpotents is one and the maximum $2n-1$.) 

In even dimensional spaces the nilpotents and projectors are chosen to be 
''eigenvectors'' of the $\frac{d}{2}$ members of the Cartan subalgebra of the 
Lorentz algebra describing the internal  space of of fermions. 
 
 After reducing both types of Clifford subalgebras ($\gamma^a$'s and 
 $\tilde{\gamma}^a$'s) to just one ($\gamma^a$'s are chosen), the 
generators   $S^{ab}$ of the Lorentz transformations in the internal space of 
fermions described by $\gamma^a$'s determine the $2^{\frac{d}{2}-1}$ 
family members for each  of  $2^{\frac{d}{2}-1}$ families,  
while $\tilde{S}^{ab}$'s determine the $\frac{d}{2}$ numbers (the eigenvalues 
of the Cartan subalgebra members) for the $2^{\frac{d}{2}-1}$ families. 
 
The Clifford odd ''basis vectors'' $\hat{b}^{m \dagger}_{f}$ and their Hermitian 
conjugated partners $\hat{b}^{m}_{f}(=(\hat{b}^{m \dagger}_{f})^{\dagger}$ ) 
obey the postulates of Dirac for the second quantized fermion fields
\begin{eqnarray}
\{ \hat{b}^{m}_{f}, \hat{b}^{m' \dagger}_{f'} \}_{*_{A}+}\, |\psi_{oc}> 
&=& \delta^{m m'} \, \delta_{ff'} \,  |\psi_{oc}>\,,\nonumber\\
\{ \hat{b}^{m}_{f}, \hat{b}^{m'}_{f'} \}_{*_{A}+}  \,  |\psi_{oc}>
&=& 0 \,\cdot\,  |\psi_{oc}>\,,\nonumber\\
\{\hat{b}^{m  \dagger}_{f},\hat{b}^{m' \dagger}_{f'}\}_{*_{A}+} \, |\psi_{oc}>
&=& 0 \, \cdot\, |\psi_{oc}>\,,\nonumber\\
 \hat{b}^{m \dagger}_{f} \,{}_{*_{A}} |\psi_{oc}>&=& |\psi^{m}_{f}>\,, \nonumber\\
 \hat{b}^{m}_{f}   \,{*_{A}}  |\psi_{oc}>&=& 0 \,\cdot\,  |\psi_{oc}>\,,
\label{alphagammatildeprod}
\end{eqnarray}
with ($m,m'$) denoting the "family" members and ($f,f '$) denoting the "families" 
of ''basis vectors'', ${*_{A}}$, represents the algebraic multiplication of 
$ \hat{b}^{m \dagger}_{f} $ among themselves and with their Hermitian 
conjugated objects $ \hat{b}^{m}_{f } $.

The vacuum state is $|\psi_{oc}>$,  Eq.~(\ref{vaccliffodd}). It is not difficult to 
prove the above relations if taking into account Eq.~(\ref{gammatildeantiher0}). 

The Clifford odd ''basis vectors'' $ \hat{b}^{m \dagger}_{f} $'s and their Hermitian 
conjugated partners  $ \hat{b}^{m}_{f} $'s appear in two independent groups, 
each with $2^{\frac{d}{2}-1}\times$  $2^{\frac{d}{2}-1}$ members, the members 
of one group have their Hermitian conjugated partners in another group.

It is our choice which one of these two groups with $2^{\frac{d}{2}-1}\times$  
$2^{\frac{d}{2}-1}$ members to take  as  ''basis vectors'' 
$ \hat{b}^{\dagger m}_{f} $'s. Making the opposite choice the  ''basis vectors'' 
change handedness.\\

{\bf b.} $\,\,$ The  ''basis vectors'' for bosons, 
${}^{I}{\hat{\cal A}}^{m \dagger}_{f }$  and  
${}^{II}{\hat{\cal A}}^{m \dagger}_{f }$,
must contain in even dimensional space an  even number of nilpotents  
$\stackrel{ab}{(k)}$, $2n'$. In $d=2(2n+1)$, 
$n'=(0,1,2,\dots, \frac{1}{2} (\frac{d}{2}-1$)), the rest,  $n''$, are projectors   
$\stackrel{ab}{[k]}$,  $n''=(\frac{d}{2}-(2n'))$. (In $d=4n$ one can have 
maximally $\frac{d}{2}$ nilpotents and minimally zero.)

The generators of the Lorentz transformations of 
${}^{I}{\hat{\cal A}}^{m \dagger}_{f} $  and of 
${}^{II}{\hat{\cal A}}^{m \dagger}_{f} $ are determined by 
${\bf {\cal S}}^{ab}=S^{ab} + \tilde{S}^{ab}$. Their properties are (chosen) 
to be denoted by  the Cartan subalgebra members of the Lorentz group with the
infinitesimal generators- ${\bf {\cal S}}^{ab}$. 
 %

The  ''basis vectors'' are either self adjoint or have the Hermitian conjugated 
partners within the same group of $2^{\frac{d}{2}-1}\times$  
$2^{\frac{d}{2}-1}$ members.
 
They do not form families, $m$ and $f$ only note a particular ''basis vector''
${}^{I,II}{\hat{\cal A}}^{m \dagger}_{f}$. One of the members of particular  
$f $ is self adjoint. 

%
The algebraic application, $*_{A}$, of the Clifford even ''basis vectors'' 
${}^{I}{\hat{\cal A}}^{m \dagger}_{f }$ on the Clifford odd ''basis vectors'' 
$ \hat{b}^{m' \dagger}_{f `} $  
can be in general case represented as 
\begin{eqnarray}
\label{calAb1234gen}
{}^{I}{\hat{\cal A}}^{m \dagger}_{f `} \,*_A \, \hat{b}^{m' \dagger }_{f}
\rightarrow \left \{ \begin{array} {r} \hat{b }^{m \dagger}_{f }\,, \\
{\rm or \,zero}\,.
\end{array} \right.
\end{eqnarray}
For each ${}^{I}{\hat{\cal A}}^{m \dagger}_{f}$  there are among 
$2^{\frac{d}{2}-1}\times 2^{\frac{d}{2}-1}$ members of the Clifford odd 
''basis vectors'' (describing the internal space of fermion fields) 
$2^{\frac{d}{2}-1}$ members, $\hat{b}^{m' \dagger}_{f `}$, fulfilling the
relation of Eq.~(\ref{calAb1234gen}). All the rest ($2^{\frac{d}{2}-1}\times 
(2^{\frac{d}{2}-1}-1)$, give zero contributions.\\

 The algebraic application, $*_{A}$, of the Clifford even ''basis vectors'' 
 ${}^{I}{\hat{\cal A}}^{m \dagger}_{f}$ among themselves fulfil the relation 
\begin{eqnarray}
\label{ruleAAIgen0}
{}^{I}{\hat{\cal A}}^{m \dagger}_{f} \,*_A\, {}^{I}{\hat{\cal A}}^{m' \dagger}_{f `}
\rightarrow \left \{ \begin{array} {r} {}^{I}{\hat{\cal A}}^{m \dagger}_{f `}\,, 
{\rm only\; one\; for }\;
\forall f `\,,\\
{\rm or \,zero}\,.
\end{array} \right.
\end{eqnarray}
Eq.~(\ref{ruleAAIgen0}) means that for each 
${}^{I}{\hat{\cal A}}^{m \dagger}_{f}$  there are among 
$2^{\frac{d}{2}-1}\times 2^{\frac{d}{2}-1}$ members of the Clifford even
''basis vectors'' (describing the internal space of boson fields) 
$2^{\frac{d}{2}-1}$ members, ${}^{I}{\hat{\cal A}}^{m' \dagger}_{f `}$, 
which lead to ${}^{I}{\hat{\cal A}}^{m \dagger}_{f `}$. All the rest 
($2^{\frac{d}{2}-1}\times (2^{\frac{d}{2}-1}-1)$ of
${}^{I}{\hat{\cal A}}^{m' \dagger}_{f `}$ give zero contributions.~\footnote{ 
''Basis vectors'' ${}^{I}{\hat{\cal A}}^{m \dagger}_{f}$  and 
${}^{I}{\hat{\cal A}}^{m' \dagger}_{f '}$, the algebraic products, $*_{A}$, of which 
 lead nonzero contributions, ''scatter'' into the third one  
 ${}^{I}{\hat{\cal A}}^{m \dagger}_{f `}$, like  it is the case of $d=(5+1)$ in which
 ${}^{I}{\hat{\cal A}}^{1 \dagger}_{1}$ 
  $*_A \,{}^{I}{\hat{\cal A}}^{4 \dagger}_{2}\rightarrow$  
  ${}^{I}{\hat{\cal A}}^{1 \dagger}_{2}$ and  the case 
 ${}^{I}{\hat{\cal A}}^{2 \dagger}_{2}$   
 $*_A \,{}^{I}{\hat{\cal A}}^{3 \dagger}_{4}\rightarrow$  
 ${}^{I}{\hat{\cal A}}^{2 \dagger}_{4}$.
  %
The world ''scatter''  is used in quotation marks since the  ''basis vectors'' 
${}^{I}{\hat{\cal A}}^{m \dagger}_{f}$ determine only the internal space of bosons,
as also the ''basis vectors'' $\hat{b}^{m \dagger}_{f}$ determine only the internal 
space of fermions.}\\

Let us write the commutation relations for Clifford even ''basis vectors'' taking into account Eq.~(\ref{ruleAAIgen0}). \\ 

$\;\;$ {\bf i.} $\;\;$ For the case that ${}^{I}{\hat{\cal A}}^{m \dagger}_{f}
 \,*_{A} \, {}^{I}{\hat{\cal A}}^{m' \dagger}_{f `} \rightarrow $ 
$ {}^{I}{\hat{\cal A}}^{m \dagger}_{f `}$ and 
${}^{I}{\hat{\cal A}}^{m' \dagger}_{f `} \,*_{A} \,
{}^{I}{\hat{\cal A}}^{m \dagger}_{f}= 0$ then
\begin{eqnarray}
\label{ruleAAIgen1}
\{{}^{I}{\hat{\cal A}}^{m \dagger}_{f} \,, \, 
{}^{I}{\hat{\cal A}}^{m' \dagger}_{f `}\}_{*_A \,-}
\rightarrow 
{}^{I}{\hat{\cal A}}^{m \dagger}_{f `}\,. \;\;\;\;
\end{eqnarray}\\

 $\;\;$ {\bf ii.} $\;\;$  For the case that 
 ${}^{I}{\hat{\cal A}}^{m \dagger}_{f} \,*_{A} \, 
 {}^{I}{\hat{\cal A}}^{m' \dagger}_{f `} \rightarrow $ 
$ {}^{I}{\hat{\cal A}}^{m \dagger}_{f `}$ and 
${}^{I}{\hat{\cal A}}^{m' \dagger}_{f `} \,*_{A} \,
{}^{I}{\hat{\cal A}}^{m \dagger}_{f}\rightarrow $ 
${}^{I}{\hat{\cal A}}^{m' \dagger}_{f } \,$ then
\begin{eqnarray}
\label{ruleAAIgen2}
\{{}^{I}{\hat{\cal A}}^{m \dagger}_{f} \,, \, 
{}^{I}{\hat{\cal A}}^{m' \dagger}_{f `} \}_{*_A \,-}
\rightarrow  
{}^{I}{\hat{\cal A}}^{m \dagger}_{f `}
- {}^{I}{\hat{\cal A}}^{m' \dagger}_{f }\,. \;\;\;
\end{eqnarray}\\
%

 $\;\;$ {\bf iii.} $\;\;$ In all other cases we have
\begin{eqnarray}
\label{ruleAAIgen3}
\{{}^{I}{\hat{\cal A}}^{m \dagger}_{f} \,,\,
{}^{I}{\hat{\cal A}}^{m' \dagger}_{f `}\}_{*_A \,-} =0\;.
\end{eqnarray}
Let us remind the reader that the note $\rightarrow$ means that 
relations are fulfilled up to a sign, while the relation
$\{{}^{I}{\hat{\cal A}}^{m \dagger}_{f} \,, \, 
{}^{I}{\hat{\cal A}}^{m' \dagger}_{f `} \}_{*_A \,-}$ means
${}^{I}{\hat{\cal A}}^{m \dagger}_{f} \, *_A \, 
{}^{I}{\hat{\cal A}}^{m' \dagger}_{f `}  - 
{}^{I}{\hat{\cal A}}^{m' \dagger}_{f `} \, *_A \, 
{}^{I}{\hat{\cal A}}^{m \dagger}_{f }$.\\

\vspace{2mm}

The equivalent relations, presented in Eqs.~(\ref{ruleAAIgen1}, \ref{ruleAAIgen2}, \ref{ruleAAIgen3}), are valid also for the Clifford even ''basis vectors''
${}^{II}{\hat{\cal A}}^{m \dagger}_{f}$.

 Let be repeated that exchanging  the role of the Clifford odd ''basis vector'' 
 $\hat{b}^{m \dagger}_{f}$ and their Hermitian conjugated partners 
 $\hat{b}^{m}_{f}$ (what means in the case of $d=(5+1)$ the exchange of 
 $odd \,I$, which is right handed,  with $odd \,II$, which is left handed, 
 Table~\ref{Table Clifffourplet.}) not only causes the change of the handedness  
 of the new $\hat{b}^{n \dagger}_{f}$, but also the exchange of 
the role of  the Clifford even ''basis vectors'' (what means in the case 
of $d=(5+1)$ the exchange of $even \,II$ with $even \,I$).

Let be added that looking at the ''basis vectors'' of boson fields 
${}^{I}{\hat{\cal A}}^{m \dagger}_{f}$ from the point of view of subgroups  
$SU(3)\times U(1)$ of the group $SO(5+1)$ we  recognize in the part of ''basis 
vectors'' forming the octet, Table~\ref{Cliff basis5+1even I.} and 
Fig.~\ref{FigSU3U1even}, the colour gauge fields of quarks and leptons 
and antiquarks and antileptons.  \\
$U(1)$ fields carry   no $U(1)$ charges.\\

\section{Second quantized fermion  and boson fields with internal space described 
by Clifford algebra}
\label{fermionsbosons}

We learned in the previous section that in  even dimensional spaces ($d=2(2n+1)$ 
or $d=4n$) the Clifford odd and the Clifford even ''basis vectors'', which are the 
superposition of the Clifford odd and the Clifford even products of $\gamma^a$'s, 
respectively, offer the description of the internal spaces of fermion and boson fields,
after the reduction of the Clifford space of $\gamma^a$'s and 
$\tilde{\gamma}^a$'s~\footnote{
$\tilde{\gamma}^a$'s keep the anticommutation relations among themselves and 
with $\gamma^a$'s, Eqs.~(\ref{gammatildeantiher}, \ref{gammatildeantiher0}), as  
they did before the reduction of the Clifford space, Eq.~(\ref{gammatildeantiher}). 
The proof can be found in Ref.~(\cite{nh2021RPPNP}, in App. I).}
to only the part determined by $\gamma^a$'s. 

The Clifford odd algebra offers $2^{\frac{d}{2}-1}$ ''basis vectors''  $\hat{b}^{m \dagger}_{f}$, appearing in $2^{\frac{d}{2}-1}$ families (with the family quantum numbers determined by  $\tilde{S}^{ab}= \frac{i}{2} \{ \tilde{\gamma}^a, \tilde{\gamma}^b\}_{-}$),  which together with their $2^{\frac{d}{2}-1}\times$ 
$2^{\frac{d}{2}-1}$ Hermitian conjugated partners $\hat{b}^{m}_{f}$ fulfil the 
postulates for the second quantized fermion fields,  Eq.~(\ref{alphagammatildeprod}),~\cite{nh2021RPPNP}, explaining the second 
quantization postulates of Dirac.

The Clifford even algebra offers $2^{\frac{d}{2}-1}\times$ $2^{\frac{d}{2}-1}$ 
''basis vectors'' of ${}^{I}{\hat{\cal A}}^{m \dagger}_{f}$ (and the same number
of ${}^{II}{\hat{\cal A}}^{m \dagger}_{f}$)  with the properties of the second 
quantized boson fields manifesting as the gauge fields of fermion fields described 
by the Clifford odd ''basis vectors'' $\hat{b}^{m \dagger}_{f}$ (and their Hermitian 
conjugated partners $\hat{b}^{m}_{f}$), Eqs.~(\ref{ruleAAIgen1}, 
\ref{ruleAAIgen2}, \ref{ruleAAIgen3}, \ref{calAb1234gen}).

The Clifford odd and the Clifford even ''basis vectors'' are chosen to be products of 
nilpotents, $\stackrel{ab}{(k)}$ (with the odd number of nilpotents if describing 
fermions and the even number of nilpotents if describing bosons), and projectors,  
$\stackrel{ab}{[k]}$. Nilpotents and projectors are (chosen to be) eigenvectors 
of the Cartan subalgebra members of the Lorentz algebra in the internal space of 
$S^{ab}$ for the Clifford odd ''basis vectors''  and of ${\bf {\cal S}}^{ab} (=
S^{ab}+ \tilde{S}^{ab}$) for  the Clifford even ''basis vectors''.

\vspace{3mm}

To define the creation operators, either for fermions or for bosons 
besides the ''basis vectors'' defining the internal space of fermions and bosons 
also the basis in ordinary space in momentum or coordinate representation is needed.
 Here  Ref.~(\cite{nh2021RPPNP}, Subsect.~3.3 and App. J) is overviewed. \\

Let us introduce the momentum  part of the  single particle states. 
The longer version is presented in Ref.~(\cite{nh2021RPPNP} in Subsect.~3.3 
and in App. J).
\begin{eqnarray}
\label{creatorp}
|\vec{p}>&=& \hat{b}^{\dagger}_{\vec{p}} \,|\,0_{p}\,>\,,\quad 
<\vec{p}\,| = <\,0_{p}\,|\,\hat{b}_{\vec{p}}\,, \nonumber\\
<\vec{p}\,|\,\vec{p }'>&=&\delta(\vec{p}-\vec{p}')=
<\,0_{p}\,|\hat{b}_{\vec{p}}\; \hat{b}^{\dagger}_{\vec{p}'} |\,0_{p}\,>\,, 
\nonumber\\
&&{\rm leading \;to\;} \nonumber\\
\hat{b}_{\vec{p}}\, \hat{b}^{\dagger}_{\vec{p}'} &=&\delta(\vec{p}-\vec{p}')\,,
\end{eqnarray}
where the normalization to identity is assumed,  $<\,0_{p}\, |\,0_{p}\,>=1$. 
While the quantized operators $\hat{\vec{p}}$ and  $\hat{\vec{x}}$ commute
 $\{\hat{p}^i\,, \hat{p}^j \}_{-}=0$ and  $\{\hat{x}^k\,, \hat{x}^l \}_{-}=0$, 
it follows for  $\{\hat{p}^i\,, \hat{x}^j \}_{-}=i \eta^{ij}$. One correspondingly 
finds 
\begin{small}
 \begin{eqnarray}
 \label{eigenvalue10}
 <\vec{p}\,| \,\vec{x}>&=&<0_{\vec{p}}\,|\,\hat{b}_{\vec{p}}\;
\hat{b}^{\dagger}_{\vec{x}} 
 |0_{\vec{x}}\,>=(<0_{\vec{x}}\,|\,\hat{b}_{\vec{x}}\;
\hat{b}^{\dagger}_{\vec{p}} \,
 |0_{\vec{p}}\,>)^{\dagger}\, \nonumber\\
 \{\hat{b}^{\dagger}_{\vec{p}}\,,  \,
\hat{b}^{\dagger}_{\vec{p}\,'}\}_{-}&=&0\,,\qquad 
\{\hat{b}_{\vec{p}},  \,\hat{b}_{\vec{p}\,'}\}_{-}=0\,,\qquad
\{\hat{b}_{\vec{p}},  \,\hat{b}^{\dagger}_{\vec{p}\,'}\}_{-}=0\,,
\nonumber\\
\{\hat{b}^{\dagger}_{\vec{x}},  \,\hat{b}^{\dagger}_{\vec{x}\,'}\}_{-}&=&0\,,
\qquad 
\{\hat{b}_{\vec{x}},  \,\hat{b}_{\vec{x}\,'}\}_{-}=0\,,\qquad
\{\hat{b}_{\vec{x}},  \,\hat{b}^{\dagger}_{\vec{x}\,'}\}_{-}=0\,,
\nonumber\\
\{\hat{b}_{\vec{p}},  \,\hat{b}^{\dagger}_{\vec{x}}\}_{-}&=&
 e^{i \vec{p} \cdot \vec{x}} \frac{1}{\sqrt{(2 \pi)^{d-1}}}\,,\qquad,
\{\hat{b}_{\vec{x}},  \,\hat{b}^{\dagger}_{\vec{p}}\}_{-}=
 e^{-i \vec{p} \cdot \vec{x}} \frac{1}{\sqrt{(2 \pi)^{d-1}}}\,,
\end{eqnarray}
\end{small}

 While the internal space of either fermion or boson fields has the finite number of
''basis vectors'' (finite number of degrees of freedom), 
$2^{\frac{d}{2}-1}\times 2^{\frac{d}{2}-1}$,
 the momentum basis is obviously continuously infinite  (has continuously many 
 degrees of freedom).\\

The creation operators for either fermions or bosons must be a tensor product, 
$*_{T}$, of both contributions, the ''basis vectors'' describing the internal space of 
fermions or bosons and the basis in ordinary, momentum or coordinate, space. 

The creation operators for a free massless fermion of the energy 
$p^0 =|\vec{p}|$, belonging to a family $f$ and  to a superposition of 
family members $m$  applying on the vacuum state  
$|\psi_{oc}>\,*_{T}\, |0_{\vec{p}}>$ 
can be written as~(\cite{nh2021RPPNP}, Subsect.3.3.2, and the references therein)
 \begin{eqnarray}
\label{wholespacefermions}
{\bf \hat{b}}^{s \dagger}_{f} (\vec{p}) \,&=& \,
\sum_{m} c^{sm}{}_f  (\vec{p}) \,\hat{b}^{\dagger}_{\vec{p}}\,*_{T}\,
\hat{b}^{m \dagger}_{f} \, \,,   
 \end{eqnarray}
where the vacuum state for fermions $|\psi_{oc}>\,*_{T}\, |0_{\vec{p}}> $ 
includes both spaces, the internal part, Eq.(\ref{vaccliffodd}), and the momentum 
part, Eq.~(\ref{creatorp}) (in a tensor product for a starting  single particle state 
with zero momentum, from which one obtains the other single fermion states of the
same ''basis vector'' by the operator  $\hat{b}^{\dagger}_{\vec{p}}$ which pushes 
the momentum by an amount $\vec{p}$~\footnote{
The creation operators and their Hermitian conjugated partners annihilation 
operators in the coordinate representation can be
read in~\cite{nh2021RPPNP} and the references therein:
$\hat{\bf b}^{s \dagger}_{f }(\vec{x},x^0)=
\sum_{m} \,\hat{b}^{ m \dagger}_{f} \,  \int_{- \infty}^{+ \infty} \,
\frac{d^{d-1}p}{(\sqrt{2 \pi})^{d-1}} \, c^{m s }{}_{f}\; 
(\vec{p}) \;  \hat{b}^{\dagger}_{\vec{p}}\;
e^{-i (p^0 x^0- \varepsilon \vec{p}\cdot \vec{x})}
$
~(\cite{nh2021RPPNP}, subsect. 3.3.2., Eqs.~(55.57.64) and the references therein).}). 
\\
The creation operators fulfil the anticommutation relations for the second quantized 
fermion fields
\begin{small}
\begin{eqnarray}
\{  \hat{\bf b}^{s' }_{f `}(\vec{p'})\,,\, 
\hat{\bf b}^{s \dagger}_{f }(\vec{p}) \}_{+} \,|\psi_{oc}> |0_{\vec{p}}>&=&
\delta^{s s'} \delta_{f f'}\,\delta(\vec{p}' - \vec{p})\, |\psi_{oc}> |0_{\vec{p}}>
\,,\nonumber\\
\{  \hat{\bf b}^{s' }_{f `}(\vec{p'})\,,\, 
\hat{\bf b}^{s}_{f }(\vec{p}) \}_{+} \,|\psi_{oc}> |0_{\vec{p}}>&=&0\, . \,
 |\psi_{oc}> |0_{\vec{p}}>
\,,\nonumber\\
\{  \hat{\bf b}^{s' \dagger}_{f '}(\vec{p'})\,,\, 
\hat{\bf b}^{s \dagger}_{f }(\vec{p}) \}_{+}\, |\psi_{oc}> |0_{\vec{p}}>&=&0\, . \,
\,|\psi_{oc}> |0_{\vec{p}}>
\,,\nonumber\\
 \hat{\bf b}^{s \dagger}_{f }(\vec{p}) \,|\psi_{oc}> |0_{\vec{p}}>&=&
|\psi^{s}_{f}(\vec{p})>\,\nonumber\\
 \hat{\bf b}^{s}_{f }(\vec{p}) \, |\psi_{oc}> |0_{\vec{p}}>&=&0\, . \,
 \,|\psi_{oc}> |0_{\vec{p}}>\nonumber\\
 |p^0| &=&|\vec{p}|\,.
\label{Weylpp'comrel}
\end{eqnarray}
\end{small}
The creation operators $  \hat{\bf b}^{s\dagger}_{f }(\vec{p}) )$  and their 
Hermitian conjugated partners annihilation operators  
$\hat{\bf b}^{s}_{f }(\vec{p}) $, creating and annihilating the single fermion 
states, respectively, fulfil when applying on the vacuum state,  
$|\psi_{oc}>|0_{\vec{p}}>$, the anticommutation relations for the second quantized 
fermions, postulated by Dirac (Ref.~\cite{nh2021RPPNP}, Subsect.~3.3.1, 
Sect.~5).~\footnote{
 The anticommutation relations of Eq.~(\ref{Weylpp'comrel}) are valid also if we 
 replace  the vacuum state,  $|\psi_{oc}>|0_{\vec{p}}>$, by the Hilbert space of 
Clifford fermions generated by the tensor product multiplication, $*_{T_{H}}$, of 
any number of the Clifford odd fermion states of all possible internal quantum 
numbers and all possible momenta (that is of any number of 
$ \hat {\bf b}^{s\, \dagger}_{f} (\vec{p})$ of any
 $(s,f, \vec{p})$), Ref.~(\cite{nh2021RPPNP}, Sect. 5.).}\\

To write the creation operators for boson fields we must take into account that 
boson gauge fields have the space index $\alpha$, describing the $\alpha$
component of the boson field in the ordinary space~\footnote{
In the  {\it  spin-charge-family} theory also the Higgs's scalars origin 
in the boson gauge fields with the vector index $(7,8)$, Ref.~(\cite{nh2021RPPNP}, Sect.~7.4.1, and the references therein).}.
We therefore write
 \begin{eqnarray}
\label{wholespacebosons}
{\bf {}^{I}{\hat{\cal A}}^{m \dagger}_{f \alpha}} (\vec{p}) \,&=& 
\hat{b}^{\dagger}_{\vec{p}}\,*_{T}\, 
{\cal C}^{ m}{}_{f \alpha}\, {}^{I}{\hat{\cal A}}^{m \dagger}_{f} \, \,.                                                                              
 \end{eqnarray}
We treat free massless bosons of momentum $\vec{p}$ and energy $p^0=|\vec{p}|$ 
and of particular ''basis vectors'' ${}^{I}{\hat{\cal A}}^{m \dagger}_{f}$ which are eigenvectors of  all the Cartan subalgebra members~\footnote{
In general the energy eigenstates of bosons  are in superposition of 
${\bf {}^{I}{\hat{\cal A}}^{m \dagger}_{f} }$. One example, which uses the 
superposition of the Cartan subalgebra eigenstates manifesting the $SU(3)\times U(1)$ subgroups of the group $SO(6)$,  is presented  in Fig.~\ref{FigSU3U1even}.},
${\cal C}^{ m}{}_{f \alpha}$ determines the vector component of the boson 
field for a particular $(m,f)$. Creation operators operate on the vacuum state 
$|\psi_{oc_{ev}}>\,*_{T}\, |0_{\vec{p}}> $ with the internal space part
just a constant, $|\psi_{oc_{ev}}>=$ $|\,1>$, and for 
a starting  single boson state with a zero momentum from which one obtains 
the other single boson states with the same ''basis vector'' by the operator 
$\hat{b}^{\dagger}_{\vec{p}}$ which push the momentum by an amount 
$\vec{p}$.


For the creation operators for boson fields in  a coordinate
representation we find using  Eqs.~(\ref{creatorp}, \ref{eigenvalue10})
 \begin{eqnarray}
{\bf {}^{I}{\hat{\cal A}}^{m \dagger}_{f \alpha}} 
(\vec{x}, x^0)& =&  \int_{- \infty}^{+ \infty} \,
\frac{d^{d-1}p}{(\sqrt{2 \pi})^{d-1}} \, 
{}^{I}{\hat{\cal A}}^{m \dagger}_{f \alpha}  (\vec{p})\, 
e^{-i (p^0 x^0- \varepsilon \vec{p}\cdot \vec{x})}|_{p^0=|\vec{p}|}\,.
\label{Weylbosonx}
\end{eqnarray}

To understand what new does the Clifford algebra description of the internal space 
of fermion and boson fields, Eqs.~(\ref{wholespacebosons}, \ref{Weylbosonx}, 
\ref{wholespacefermions}), bring to our understanding of the second quantized 
fermion and boson fields and what new can we learn from this offer, 
we need to relate $\sum_{ab} c^{ab} \omega_{ab \alpha}$ and 
$ \sum_{m f} {}^{I}{\hat{\cal A}}^{m \dagger}_{f} {\cal C}^{m f}_{\alpha}$,
recognizing that ${}^{I}{\hat{\cal A}}^{m \dagger}_{f} {\cal C}^{m f}_{\alpha}$
are eigenstates of the Cartan subalgebra members, while  $\omega_{ab \alpha}$
are not.

The gravity fields, the vielbeins and the two kinds of the spin connection fields,
$f^{a}{}_{\alpha}$, $\omega_{ab \alpha}$, $\tilde{\omega}_{ab \alpha}$, 
respectively, are in the {\it spin-charge-family} theory~\footnote{
This is the case for most of the Kaluza-Klein-like theories.} 
(unifying spins, charges and families of fermions and offering not only the 
explanation for all the assumptions of the {\it standard model} but also for the 
increasing number of phenomena observed so far) the only boson fields in 
$d=(13+1)$, observed in $d=(3+1)$ besides as  gravity also as all the other 
boson fields with the Higgs's scalars included~\cite{nd2017}.

We therefore need to relate
\begin{eqnarray}
\label{relationomegaAmf0}
\{\frac{1}{2}  \sum_{ab} S^{ab}\, \omega_{ab \alpha} \} 
\sum_{m } \beta^{m f}\, \hat{\bf b}^{m \dagger}_{f }(\vec{p}) &{\rm relate\, \,to}&
\{ \sum_{m' f '} {}^{I}{\hat{\cal A}}^{m' \dagger}_{f '} \,
{\cal C}^{m' f '}_{\alpha} \}
\sum_{m } \beta^{m f} \, \hat{\bf b}^{m \dagger}_{f }(\vec{p}) \,, \nonumber\\
 &&\forall f \,{\rm and}\,\forall \, \beta^{m f}\,, \nonumber\\
{\bf \cal S}^{cd} \,\sum_{ab} (c^{ab}{}_{mf}\, \omega_{ab \alpha})  &{\rm relate\, \,to}& 
{\bf \cal S}^{cd}\, ({}^{I}{\hat{\cal A}}^{m \dagger}_{f}\, {\cal C}^{m f}_{\alpha})\,, \nonumber\\
&& \forall \,(m,f), \nonumber\\
&&\forall \,\,{\rm Cartan\,\,subalgebra\, \, \, member}  \,{\bf \cal S}^{cd} \,.
\end{eqnarray}
Let be repeated that ${}^{I}{\hat{\cal A}}^{m \dagger}_{f } $ are chosen to be
the eigenvectors of the Cartan subalgebra members, Eq.~(\ref{cartangrasscliff}).
Correspondingly we can relate  a particular ${}^{I}{\hat{\cal A}}^{m \dagger}_{f } 
{\cal C}^{m f }_{\alpha}$ with such a superposition of $\omega_{ab \alpha}$'s
which  is   the eigenvector with  the same values of the Cartan subalgebra members as 
there is a particular ${}^{I}{\hat{\cal A}}^{m \dagger}_{f } {\cal C}^{m f }_{\alpha}$. 
We can do this in two ways:\\
 {\bf i.} $\;\;$ Using the first relation in Eq.~(\ref{relationomegaAmf0}).  On the left 
hand side of this relation ${S}^{ab}$'s apply  on $ \hat{b}^{m \dagger}_{f} $  part of 
 $ \hat{\bf b}^{m \dagger}_{f }(\vec{p}) $.
On the right hand side ${}^{I}{\hat{\cal A}}^{m \dagger}_{f }$ apply as well on  the
same ''basis vector''  $ \hat{b}^{m \dagger}_{f} $. \\
  {\bf ii.} $\;\;$ Using  the second relation, in which  ${\bf \cal S}^{cd}$ apply  on 
   the left hand side on  $\omega_{ab \alpha}$'s
\begin{eqnarray}
\label{sonomega}
 \, {\bf \cal S}^{cd} \,\sum_{ab}\, c^{ab}{}_{mf}\, \omega_{ab \alpha}
 &=& \sum_{ab}\, c^{ab}{}_{mf}\, i \,(\omega_{cb \alpha} \eta^{ad}- 
\omega_{db \alpha} \eta^{ac}+ \omega_{ac \alpha} \eta^{bd}-
\omega_{ad \alpha} \eta^{bc}),
\end{eqnarray}
on  each $ \omega_{ab \alpha}$ separately; $c^{ab}{}_{mf}$ are constants to be 
determined from the second relation, where  on the right hand side of this relation
${\bf \cal S}^{cd} (= S^{cd}+ \tilde{S}^{cd})$ apply on the ''basis vector'' 
${}^{I}{\hat{\cal A}}^{m \dagger}_{f }$ of the corresponding gauge field. 

Let us demonstrate the first of the two relations on the toy model case of $d=(3+1)$.

\vspace{2mm}

\begin{small}
The ''basis vectors'' $\hat{b}^{m \dagger}_{f=1}$ --- $\hat{b}^{1 \dagger}_{1}= 
\stackrel{03}{(+i)}\stackrel{12}{[+]}$, $\hat{b}^{2 \dagger}_{1}= 
\stackrel{03}{[-i]}\stackrel{12}{(-)}$ --- can be found in 
Table~\ref{Table Clifffourplet.},  
as the first and the second ''basis vectors'' of the first family of {\it odd I} if 
only the first and the second factors are taken into account and the third one  
$\stackrel{56}{[+]}$ is neglected. 
$\hat{b}^{1 \dagger}_{2}= 
\stackrel{03}{[+i]}\stackrel{12}{(+)}$, $\hat{b}^{2 \dagger}_{2}= 
\stackrel{03}{(-i)}\stackrel{12}{[-]}$ can be found in the same 
Table~\ref{Table Clifffourplet.} as the first and the second ''basis vectors'' of the third 
family if $\stackrel{56}{[+]}$ is neglected. 

The corresponding even ''basis vectors'' can be found in the same 
table~\ref{Table Clifffourplet.} in the third and the fourth column  under {\it even I} 
in the first two lines, if neglecting $\stackrel{56}{[+]}$; ${}^{I}{\hat{\cal A}}^{1 \dagger}_{1}= \stackrel{03}{[+i]}\stackrel{12}{[+]}$, ${}^{I}{\hat{\cal A}}^{2 \dagger}_{1}= \stackrel{03}{(-i)}
\stackrel{12}{(-i)}$, ${}^{I}{\hat{\cal A}}^{1 \dagger}_{2}= \stackrel{03}{(+i)}
\stackrel{12}{(+)}$, ${}^{I}{\hat{\cal A}}^{2 \dagger}_{2}= \stackrel{03}{[-i]}
\stackrel{12}{[-i]}$.  

Applying ($S^{01} \omega_{01 \alpha}+S^{02} \omega_{02 \alpha}+ 
S^{13} \omega_{13 \alpha}+ S^{23} \omega_{23 \alpha}+ 
S^{03} \omega_{03  \alpha}+ S^{12} \omega_{12 \alpha}$) on 
$\hat{b}^{1 \dagger}_{1}$,   and relating this
 to  $({}^{I}{\hat{\cal A}}^{1 \dagger}_{1} {\cal C}^{1 1}_{\alpha}+
 {}^{I}{\hat{\cal A}}^{2 \dagger}_{1} {\cal C}^{2 1}_{\alpha} + 
 {}^{I}{\hat{\cal A}}^{1 \dagger}_{2} {\cal C}^{1 2}_{\alpha} +
 {}^{I}{\hat{\cal A}}^{2 \dagger}_{2} {\cal C}^{2 2}_{\alpha})$
 $\hat{b}^{1 \dagger}_{1}$ we end up, after taking into account 
 Eqs.~(\ref{gammatildeantiher}, \ref{graficcliff1}, \ref{sonomega}), with the
 first relation of the first line of Eq.~(\ref{calCtoomega})
 \begin{eqnarray}
 \label{calCtoomega}
 {\cal C}^{11}_{\alpha} &=&\frac{1}{2} (i  \omega_{03  \alpha} + 
  \omega_{12  \alpha})\,, \quad  (-)\, {\cal C}^{21}_{\alpha} =\frac{1}{2} 
  (- i \omega_{01  \alpha} +   \omega_{02  \alpha} + i  \omega_{13  \alpha} - 
  \omega_{23  \alpha}))\,, \nonumber\\
{\cal C}^{22}_{\alpha} &=&\frac{1}{2} (-i  \omega_{03  \alpha} - 
  \omega_{12  \alpha})\,, \quad  (-)\, {\cal C}^{12}_{\alpha} =\frac{1}{2} 
  (- i \omega_{01  \alpha} -  \omega_{02  \alpha} - i  \omega_{13  \alpha} - 
  \omega_{23  \alpha}))\,. \nonumber\\  
 \end{eqnarray}
The last three relations, of the above equation follow from the equivalent application of ($S^{01} \omega_{01 \alpha}+S^{02} \omega_{02 \alpha}+ 
S^{13} \omega_{13 \alpha}+ S^{23} \omega_{23 \alpha}+ 
S^{03} \omega_{03  \alpha}+ S^{12} \omega_{12 \alpha}$) on 
($\hat{b}^{2 \dagger}_{1}$,  $\hat{b}^{1 \dagger}_{2}$, 
$\hat{b}^{2 \dagger}_{2}$), respectively,   after relating them 
 to  $({}^{I}{\hat{\cal A}}^{1 \dagger}_{1} {\cal C}^{1 1}_{\alpha}+
 {}^{I}{\hat{\cal A}}^{2 \dagger}_{1} {\cal C}^{2 1}_{\alpha} + 
 {}^{I}{\hat{\cal A}}^{1 \dagger}_{2} {\cal C}^{1 2}_{\alpha} +
 {}^{I}{\hat{\cal A}}^{2 \dagger}_{2} {\cal C}^{2 2}_{\alpha})$ applying on
 ($\hat{b}^{2 \dagger}_{1}$, $\hat{b}^{1 \dagger}_{2}$, $\hat{b}^{2 \dagger}_{2}$), respectively. 
\end{small}

\vspace{2mm}

Let us conclude this section by pointing out that either the Clifford odd ''basis vectors''
$\hat{b}^{m \dagger}_{f}$ or the Clifford even ''basis vectors'' 
${}^{i}{\hat{\cal A}}^{m \dagger}_{f}, i=(I,II) $ have in any even $d$-dimensional 
space $2^{\frac{d}{2}-1}$ $\times 2^{\frac{d}{2}-1}$ members, while 
$\omega_{ab \alpha}$ as well as $\tilde{\omega}_{ab \alpha}$ have each for each 
$\alpha$ $\frac{d}{2}(d-1)$ members. It is needed to find out what new can this 
difference bring into the application of the Kaluza-Klein-like theories on elementary 
physics and cosmology.

%
\section{Conclusions}
\label{conclusions}

In the {\it spin-charge-family} theory~(\cite{nh2021RPPNP} and references therein) 
the Clifford odd algebra is used to describe the internal space of fermion fields. The 
Clifford odd ''basis vectors'' in the tensor product with a basis in ordinary space 
form the creation and annihilation operators, in which the anticommutativity of the
''basis vectors'' is transferred  to the creation and annihilation operators for fermions, 
offering the explanation for the second quantization postulates for fermion fields~(\cite{nh2021RPPNP} and references therein). The  Clifford odd ''basis vectors'' 
have all the properties of fermions: Half integer spins with respect to the Cartan 
subalgebra members of the Lorentz algebra in the internal space of fermions in even dimensional spaces ($d=2(2n+1)$ or $d=4n$), as discussed in  Subsects.~\ref{generalbasisinternal} and~\ref{basisvectors} and in 
Sect.~\ref{fermionsbosons}.\\
With respect to the subgroups of the $SO(d-1, 1)$ group the Clifford odd ''basis 
vectors'' appear in the fundamental representations, as illustrated in 
Subsects.~\ref{cliffordoddevenbasis5+1} and~\ref{odd5+1}.

In this article, it is demonstrated that the Clifford even algebra is offering the description 
of the internal space of boson fields. The Clifford even ''basis vectors'' in the tensor product 
with a basis in ordinary space form the creation and annihilation operators which
manifest the commuting properties of the second quantized boson fields, offering 
explanation for the second quantization postulates for boson fields~\cite{n2021SQ}.
The Clifford even ''basis vectors'' have all the properties of bosons: Integer spins with 
respect to the Cartan subalgebra members of the Lorentz algebra in the internal space 
of bosons, as discussed in  Subsects.~\ref{generalbasisinternal} and~\ref{basisvectors}
and in Sect.~\ref{fermionsbosons}.\\
With respect to the subgroups of the $SO(d-1, 1)$ group the Clifford even ''basis vectors'' 
manifest the  adjoint representations, as illustrated in
Subsect.~\ref{cliffordoddevenbasis5+1} and~\ref{even5+1}. 

There are two kinds of the anticommuting algebras~\cite{norma93}: The Grassmann
algebra, offering in $d$-dimensional space $2\,.\, 2^d$ operators ($2^d$ $\,\theta^a$'s 
and $2^d$ $\frac{\partial}{\partial \theta_a}$'s, Hermitian conjugated to each other, Eq.~(\ref{thetaderher0})),  and the two Clifford subalgebras, each with $2^d$ operators 
called $\gamma^a$'s and $\tilde{\gamma}^a$'s, respectively, \cite{norma93,nh02,nh03}, Eqs.~(\ref{thetaderanti0}-\ref{gammatildeantiher}), Subsect.~\ref{GrassmannClifford}
of this article. 

The operators in each of the two  Clifford subalgebras  appear in two groups of 
$2^{\frac{d}{2}-1}\times $ $2^{\frac{d}{2}-1}$ of  the Clifford odd  operators 
(the odd products of either $\gamma^a$'s in one subalgebra or of 
$\tilde{\gamma}^a$'s in the other subalgebra),  which are Hermitian conjugated 
to each other. In each Clifford odd group of any of the two subalgebras there appear 
$2^{\frac{d}{2}-1}$ irreducible representation each with the $2^{\frac{d}{2}-1}$
members.

There are as well the Clifford even operators (the even products of either 
$\gamma^a$'s in one subalgebra or of $\tilde{\gamma}^a$'s in the another 
subalgebra) which  again appear in two groups of $2^{\frac{d}{2}-1}\times $ 
$2^{\frac{d}{2}-1}$ members each. In the case of the Clifford even objects the 
members of each group of $2^{\frac{d}{2}-1}\times $ $2^{\frac{d}{2}-1}$ 
members have the Hermitian conjugated partners within the same group, ~Subsect.~\ref{basisvectors}, Table~\ref{Table Clifffourplet.}.

The Grassmann algebra operators are  expressible with the operators of the two Clifford subalgebras and opposite, Eq.~~(\ref{clifftheta1}). The two Clifford subalgebras are 
independent of each other, Eq.~(\ref{gammatildeantiher}), forming two independent 
spaces.

Either the Grassmann algebra~\cite{ND2018Grass,n2019PIPII} or the two Clifford 
subalgebras can be used to describe 
the internal space of anticommuting objects, if the odd products of operators 
($\theta^a$'s or $\gamma^a$'s, or $ \tilde{\gamma}^a$'s) are used to describe the 
internal space of these objects. Describing the commuting objects the even products of 
operators  ($\theta^a$'s or $ \gamma^a$'s or $\tilde{\gamma}^a$'s) have to be 
used.

\vspace{2mm}

No integer spin anticommuting objects have been observed so far, and to describe the 
internal space of the so far observed fermions only one of the two Clifford odd 
subalgebras are needed. \\

The problem can be solved by reducing  the two Clifford subalgebras to only one, the one 
(chosen to be) determined by $\gamma^{ab}$'s, Subsect.~\ref{reduction}. The decision  
that $ \tilde{\gamma}^a$  apply  on $ \gamma^a$ as follows: 
$\{ \tilde{\gamma}^a B =(-)^B\, i \, B \gamma^a\}\, |\psi_{oc}>$, 
Eq.~(\ref{tildegammareduced}), 
(with $(-)^B = -1$, if $B$ is a function of an odd products of $\gamma^a$'s,
otherwise $(-)^B = 1$) enables that  $2^{\frac{d}{2}-1}$ irreducible representations  
of $S^{ab}= \frac{i}{2}\, \{\gamma^a\,,\, \gamma^b\}_{-}$ (each with the  
$2^{\frac{d}{2}-1}$ members) obtain the family quantum numbers determined by  
$\tilde{S}^{ab}= \frac{i}{2}\, \{\tilde{\gamma}^a\,,\,\tilde{\gamma}^b\}_{-}$.
 
\vspace{2mm}

The decision to use in the {\it spin-charge-family} theory in $d=2(2n +1)$, $n\ge 3$,
the superposition of the odd products of the Clifford algebra elements $\gamma^{a}$'s 
to describe  the internal space of fermions  which interact with the gravity only 
(with the vielbeins, the gauge fields of momenta, and the two kinds of the spin 
connection fields, the gauge fields of  $S^{ab}$ and $\tilde{S}^{ab}$, respectively), Eq.~(\ref{wholeaction}), offers not 
only the explanation for all the assumed properties of fermions and bosons in 
the {\it standard model}, with the appearance of  the families of quarks and leptons 
and antiquarks and antileptons~(\cite{nh2021RPPNP} and the references therein) and 
of the corresponding vector gauge fields  and the Higgs's scalars included~\cite{nd2017},  
but also for the appearance of the dark matter~\cite{gn2009} in the universe, for the explanation of the matter/antimatter asymmetry in the 
universe~\cite{n2014matterantimatter},  and for several other observed phenomena, 
making several predictions~\cite{pikanorma2005,gmdn2007,gmdn2008,gn2013}. 

\vspace{2mm}

The recognition that the superposition of the even products of the Clifford 
algebra elements $\gamma^{a}$'s offers the description of the internal space  
of boson fields manifesting all the properties of the observed boson fields, as 
demonstrated in this article, makes clear that the Clifford algebra offers the 
explanation for the postulates of the second quantized anticommuting fermion 
and commuting boson fields.

The relations in Eq.~(\ref{relationomegaAmf0}) 
\begin{eqnarray}
\label{relationomegaAmf01}
\{\frac{1}{2}  \sum_{ab} S^{ab}\, \omega_{ab \alpha} \} 
\sum_{m } \beta^{m f}\, \hat{\bf b}^{m \dagger}_{f }(\vec{p}) &{\rm relate\,\, to}&
\{ \sum_{m' f '} {}^{I}{\hat{\cal A}}^{m' \dagger}_{f '} \,
{\cal C}^{m' f '}_{\alpha} \}
\sum_{m } \beta^{m f} \, \hat{\bf b}^{m \dagger}_{f }(\vec{p}) \,, \nonumber\\
 &&\forall f \,{\rm and}\,\forall \, \beta^{m f}\,, \nonumber\\
{\bf \cal S}^{cd} \,\sum_{ab} (c^{ab}{}_{mf}\, \omega_{ab \alpha})  &{\rm relate\,\, to}& 
{\bf \cal S}^{cd}\, ({}^{I}{\hat{\cal A}}^{m \dagger}_{f}\, {\cal C}^{m f}_{\alpha})\,, \nonumber\\
&& \forall \,(m,f), \nonumber\\
&&\forall \,\,{\rm Cartan\,\,subalgebra\, \, \, member}  \,{\bf \cal S}^{cd} \,,
\nonumber
\end{eqnarray}
offers the possibility to replace the covariant derivative 
$ p_{0 \alpha }$
  $$p_{0\alpha} = p_{\alpha}  - \frac{1}{2}  S^{ab} \omega_{ab \alpha} - 
                    \frac{1}{2}  \tilde{S}^{ab}   \tilde{\omega}_{ab \alpha} 
                    \quad \quad \quad\quad\;$$
in Eq.~(\ref{wholeaction}) with 

$$ p_{0\alpha}  = p_{\alpha}  - 
\sum_{m f}   {}^{I}{ \hat {\cal A}}^{m \dagger}_{f}
{}^{I}{\cal C}^{m}_{f \alpha}   - 
 \sum_{m f} {}^{I}{\hat{\widetilde{\cal A}}}^{m \dagger}_{f}\,
{}^{I}{\widetilde{\cal C}}^{m}_{f \alpha}\,, $$ 

where the ''basis vectors'' ${}^{I}{\hat{\widetilde{\cal A}}}^{m \dagger}_{f}
{}^{I}{\widetilde{\cal C}}^{m}_{f \alpha}$ are related to 
${}^{II}{\hat{\cal A}}^{m \dagger}_{f}\,
{}^{II}{\cal C}^{m}_{f \alpha}$.

\vspace{2mm}

In this article the properties of the Clifford even ''basis vectors'' in relation to the 
Clifford odd ''basis vectors'', and correspondingly the transfer of the commutativity 
and anticommutativity from the Clifford ''basis vectors'' to the creation and 
annihilation operators for boson and fermion fields,  respectively, are demonstrated, 
offering the explanation for the second quantization postulates for boson and fermion 
fields. 

These relations need further study to find out what new can the proposed new insight 
into the internal space of fermions and bosons bring into understanding the second 
quantized fermion and boson fields. 

The Einstein gravity is known as a gauge theory based on the Abelian group of local translations, for which vielbein is the corresponding gauge field. It is also known that 
that the action of Eq.~(\ref{wholeaction}), in which there appear besides vielbeins
also the spin connection fields (two kinds of the spin connection fields in the 
{\it spin-charge-family} theory, which are the gauge fields of $S^{ab}$'s and 
$\tilde{S}^{ab}$'s, respectively,  manifest in  $d=(3+1)$ as ordinary gravity and all
the known gauge fields~(\cite{nd2017} and references therein)~\footnote{
If there are no fermions present the spin connection fields of both kinds are
expressible with the vielbeins~(\cite{nh2021RPPNP}, Eq.~(103)).}

 The study of properties  of the second quantized boson fields, the internal space of 
  which is described by the Clifford even algebra, has just started. It is needed to 
find out whether Eq.~(\ref{relationomegaAmf0}) is really able to describe gravity and 
correspondingly unify all the gauge fields, with the scalar fields included.

\appendix

%
\section{One family representation in  $d=(13+1)$-dimensional space with 
$2^{\frac{d}{2}-1}$ members representing quarks and leptons and antiquarks and 
antileptons in the {\it spin-charge-family} theory } 
\label{13+1representation}
%


This appendix illustrates  the family members of one family of the Clifford odd
''basis vectors'', written as products of  (odd number of) 
nilpotents and of projectors, which are chosen to be the eigenvectors of the Cartan 
subalgebra members, Eq.~(\ref{eigencliffcartan}, \ref{cartangrasscliff}) of 
the Lorentz algebra in the internal space of fermions. Analysing the group $SO(13,1)$
with respect to the subgroups $SO(3,1)\times SU(2)\times SU(2)\times SU(3) \times
U(1)$ with the same number of commuting operators as has the group $SO(13,1)$,
one can see in Table~\ref{Table so13+1.} that the ''basis vectors'' of one irreducible 
representation, one family, of the Clifford odd basis vectors of left handedness, 
$\Gamma^{(13+1)}=-1$,   includes all the quarks and the leptons as well as the 
antiquarks and antileptons of the {\it standard model}, with the right handed neutrino
and left handed antineutrino included.\\
While the starting ''basis vectors'' can be either left or right handed, the subgroups of the starting group contain left and right handed members, just as required by the {\it standard model}~\footnote{
The breaks of the symmetries, manifested in Eqs.~(\ref{so1+3}, \ref{so42}, \ref{so64}), 
are in the {\it spin-charge-family} theory caused by the condensate and by the non zero 
vacuum expectation values (constant values) of the scalar fields carrying the space index 
$(7,8)$ (Refs.~\cite{normaJMP2015,IARD2016,nh2021RPPNP} and the references therein), 
all originating in the vielbeins and the two kinds of the spin connection fields. The space 
breaks first to $SO(7,1)$ $\times SU(3) \times U(1)_{II}$ and then further to 
$SO(3,1)\times SU(2)_{I} \times U(1)_{I}$ $\times SU(3) \times U(1)_{II}$, what 
explains the connections between the weak and the hyper charges and the handedness 
of spinors.}.\\
The needed definitions of the quantum numbers are presented in 
App.~\ref{grassmannandcliffordfermions}.

\bottomcaption{\label{Table so13+1.}%
\begin{small}
The left handed ($\Gamma^{(13,1)} = -1$~\cite{IARD2016}) irreducible representation 
of one family of spinors --- the product of the  odd number of nilpotents and of projectors, which are eigenvectors of the Cartan subalgebra of the $SO(13,1)$ group~\cite{n2014matterantimatter,normaJMP2015,nh02,nh03,IARD2016}, manifesting 
the subgroup $SO(7,1)$  of the colour charged quarks and antiquarks and the colourless leptons and antileptons --- is presented. 
It contains the left handed  ($\Gamma^{(3,1)}=-1$)  weak ($SU(2)_{I}$) charged  
($\tau^{13}=\pm \frac{1}{2}$, Eq.~(\ref{so42})), and $SU(2)_{II}$ chargeless 
($\tau^{23}=0$, Eq.~(\ref{so42})) quarks and leptons and the right handed  
($\Gamma^{(3,1)}=1$) weak  ($SU(2)_{I}$) chargeless and $SU(2)_{II}$ charged 
($\tau^{23}=\pm \frac{1}{2}$) quarks and leptons, both with the spin $ S^{12}$  up 
and down ($\pm \frac{1}{2}$, respectively). 
Quarks distinguish from leptons only in the $SU(3) \times U(1)$ part: Quarks are triplets
of three colours  ($c^i$ $= (\tau^{33}, \tau^{38})$ $ = [(\frac{1}{2},\frac{1}{2\sqrt{3}}),
(-\frac{1}{2},\frac{1}{2\sqrt{3}}), (0,-\frac{1}{\sqrt{3}}) $, 
carrying  the "fermion charge" ($\tau^{4}=\frac{1}{6}$, Eq.~(\ref{so64})).
The colourless leptons carry the "fermion charge" ($\tau^{4}=-\frac{1}{2}$).
The same multiplet contains also the left handed weak ($SU(2)_{I}$) chargeless and 
$SU(2)_{II}$ charged antiquarks and antileptons and the right handed weak 
($SU(2)_{I}$) charged and $SU(2)_{II}$ chargeless antiquarks and antileptons.
Antiquarks distinguish from antileptons again only in the $SU(3) \times U(1)$ part: 
Antiquarks are antitriplets,  carrying  the "fermion charge" ($\tau^{4}=-\frac{1}{6}$).
The anticolourless antileptons carry the "fermion charge" ($\tau^{4}=\frac{1}{2}$).
 $Y=(\tau^{23} + \tau^{4})$ is the hyper charge, the electromagnetic charge
is $Q=(\tau^{13} + Y$).
%
\end{small}
}

\tablehead{\hline
i&$$&$|^a\psi_i>$&$\Gamma^{(3,1)}$&$ S^{12}$&
$\tau^{13}$&$\tau^{23}$&$\tau^{33}$&$\tau^{38}$&$\tau^{4}$&$Y$&$Q$\\
\hline
&& ${\rm (Anti)octet},\,\Gamma^{(7,1)} = (-1)\,1\,, \,\Gamma^{(6)} = (1)\,-1$&&&&&&&&& \\
&& ${\rm of \;(anti) quarks \;and \;(anti)leptons}$&&&&&&&&&\\
\hline\hline}
\tabletail{\hline \multicolumn{12}{r}{\emph{Continued on next page}}\\}
\tablelasttail{\hline}
\begin{tiny}
\begin{supertabular}{|r|c||c||c|c||c|c||c|c|c||r|r|}
1&$ u_{R}^{c1}$&$ \stackrel{03}{(+i)}\,\stackrel{12}{[+]}|
\stackrel{56}{[+]}\,\stackrel{78}{(+)}
||\stackrel{9 \;10}{(+)}\;\;\stackrel{11\;12}{[-]}\;\;\stackrel{13\;14}{[-]} $ &1&$\frac{1}{2}$&0&
$\frac{1}{2}$&$\frac{1}{2}$&$\frac{1}{2\,\sqrt{3}}$&$\frac{1}{6}$&$\frac{2}{3}$&$\frac{2}{3}$\\
\hline
2&$u_{R}^{c1}$&$\stackrel{03}{[-i]}\,\stackrel{12}{(-)}|\stackrel{56}{[+]}\,\stackrel{78}{(+)}
||\stackrel{9 \;10}{(+)}\;\;\stackrel{11\;12}{[-]}\;\;\stackrel{13\;14}{[-]}$&1&$-\frac{1}{2}$&0&
$\frac{1}{2}$&$\frac{1}{2}$&$\frac{1}{2\,\sqrt{3}}$&$\frac{1}{6}$&$\frac{2}{3}$&$\frac{2}{3}$\\
\hline
3&$d_{R}^{c1}$&$\stackrel{03}{(+i)}\,\stackrel{12}{[+]}|\stackrel{56}{(-)}\,\stackrel{78}{[-]}
||\stackrel{9 \;10}{(+)}\;\;\stackrel{11\;12}{[-]}\;\;\stackrel{13\;14}{[-]}$&1&$\frac{1}{2}$&0&
$-\frac{1}{2}$&$\frac{1}{2}$&$\frac{1}{2\,\sqrt{3}}$&$\frac{1}{6}$&$-\frac{1}{3}$&$-\frac{1}{3}$\\
\hline
4&$ d_{R}^{c1} $&$\stackrel{03}{[-i]}\,\stackrel{12}{(-)}|
\stackrel{56}{(-)}\,\stackrel{78}{[-]}
||\stackrel{9 \;10}{(+)}\;\;\stackrel{11\;12}{[-]}\;\;\stackrel{13\;14}{[-]} $&1&$-\frac{1}{2}$&0&
$-\frac{1}{2}$&$\frac{1}{2}$&$\frac{1}{2\,\sqrt{3}}$&$\frac{1}{6}$&$-\frac{1}{3}$&$-\frac{1}{3}$\\
\hline
5&$d_{L}^{c1}$&$\stackrel{03}{[-i]}\,\stackrel{12}{[+]}|\stackrel{56}{(-)}\,\stackrel{78}{(+)}
||\stackrel{9 \;10}{(+)}\;\;\stackrel{11\;12}{[-]}\;\;\stackrel{13\;14}{[-]}$&-1&$\frac{1}{2}$&
$-\frac{1}{2}$&0&$\frac{1}{2}$&$\frac{1}{2\,\sqrt{3}}$&$\frac{1}{6}$&$\frac{1}{6}$&$-\frac{1}{3}$\\
\hline
6&$d_{L}^{c1} $&$ - \stackrel{03}{(+i)}\,\stackrel{12}{(-)}|\stackrel{56}{(-)}\,\stackrel{78}{(+)}
||\stackrel{9 \;10}{(+)}\;\;\stackrel{11\;12}{[-]}\;\;\stackrel{13\;14}{[-]} $&-1&$-\frac{1}{2}$&
$-\frac{1}{2}$&0&$\frac{1}{2}$&$\frac{1}{2\,\sqrt{3}}$&$\frac{1}{6}$&$\frac{1}{6}$&$-\frac{1}{3}$\\
\hline
7&$ u_{L}^{c1}$&$ - \stackrel{03}{[-i]}\,\stackrel{12}{[+]}|\stackrel{56}{[+]}\,\stackrel{78}{[-]}
||\stackrel{9 \;10}{(+)}\;\;\stackrel{11\;12}{[-]}\;\;\stackrel{13\;14}{[-]}$ &-1&$\frac{1}{2}$&
$\frac{1}{2}$&0 &$\frac{1}{2}$&$\frac{1}{2\,\sqrt{3}}$&$\frac{1}{6}$&$\frac{1}{6}$&$\frac{2}{3}$\\
\hline
8&$u_{L}^{c1}$&$\stackrel{03}{(+i)}\,\stackrel{12}{(-)}|\stackrel{56}{[+]}\,\stackrel{78}{[-]}
||\stackrel{9 \;10}{(+)}\;\;\stackrel{11\;12}{[-]}\;\;\stackrel{13\;14}{[-]}$&-1&$-\frac{1}{2}$&
$\frac{1}{2}$&0&$\frac{1}{2}$&$\frac{1}{2\,\sqrt{3}}$&$\frac{1}{6}$&$\frac{1}{6}$&$\frac{2}{3}$\\
\hline\hline
\shrinkheight{0.25\textheight}
9&$ u_{R}^{c2}$&$ \stackrel{03}{(+i)}\,\stackrel{12}{[+]}|
\stackrel{56}{[+]}\,\stackrel{78}{(+)}
||\stackrel{9 \;10}{[-]}\;\;\stackrel{11\;12}{(+)}\;\;\stackrel{13\;14}{[-]} $ &1&$\frac{1}{2}$&0&
$\frac{1}{2}$&$-\frac{1}{2}$&$\frac{1}{2\,\sqrt{3}}$&$\frac{1}{6}$&$\frac{2}{3}$&$\frac{2}{3}$\\
\hline
10&$u_{R}^{c2}$&$\stackrel{03}{[-i]}\,\stackrel{12}{(-)}|\stackrel{56}{[+]}\,\stackrel{78}{(+)}
||\stackrel{9 \;10}{[-]}\;\;\stackrel{11\;12}{(+)}\;\;\stackrel{13\;14}{[-]}$&1&$-\frac{1}{2}$&0&
$\frac{1}{2}$&$-\frac{1}{2}$&$\frac{1}{2\,\sqrt{3}}$&$\frac{1}{6}$&$\frac{2}{3}$&$\frac{2}{3}$\\
\hline
11&$d_{R}^{c2}$&$\stackrel{03}{(+i)}\,\stackrel{12}{[+]}|\stackrel{56}{(-)}\,\stackrel{78}{[-]}
||\stackrel{9 \;10}{[-]}\;\;\stackrel{11\;12}{(+)}\;\;\stackrel{13\;14}{[-]}$
&1&$\frac{1}{2}$&0&
$-\frac{1}{2}$&$ - \frac{1}{2}$&$\frac{1}{2\,\sqrt{3}}$&$\frac{1}{6}$&$-\frac{1}{3}$&$-\frac{1}{3}$\\
\hline
12&$ d_{R}^{c2} $&$\stackrel{03}{[-i]}\,\stackrel{12}{(-)}|
\stackrel{56}{(-)}\,\stackrel{78}{[-]}
||\stackrel{9 \;10}{[-]}\;\;\stackrel{11\;12}{(+)}\;\;\stackrel{13\;14}{[-]} $
&1&$-\frac{1}{2}$&0&
$-\frac{1}{2}$&$-\frac{1}{2}$&$\frac{1}{2\,\sqrt{3}}$&$\frac{1}{6}$&$-\frac{1}{3}$&$-\frac{1}{3}$\\
\hline
13&$d_{L}^{c2}$&$\stackrel{03}{[-i]}\,\stackrel{12}{[+]}|\stackrel{56}{(-)}\,\stackrel{78}{(+)}
||\stackrel{9 \;10}{[-]}\;\;\stackrel{11\;12}{(+)}\;\;\stackrel{13\;14}{[-]}$
&-1&$\frac{1}{2}$&
$-\frac{1}{2}$&0&$-\frac{1}{2}$&$\frac{1}{2\,\sqrt{3}}$&$\frac{1}{6}$&$\frac{1}{6}$&$-\frac{1}{3}$\\
\hline
14&$d_{L}^{c2} $&$ - \stackrel{03}{(+i)}\,\stackrel{12}{(-)}|\stackrel{56}{(-)}\,\stackrel{78}{(+)}
||\stackrel{9 \;10}{[-]}\;\;\stackrel{11\;12}{(+)}\;\;\stackrel{13\;14}{[-]} $&-1&$-\frac{1}{2}$&
$-\frac{1}{2}$&0&$-\frac{1}{2}$&$\frac{1}{2\,\sqrt{3}}$&$\frac{1}{6}$&$\frac{1}{6}$&$-\frac{1}{3}$\\
\hline
15&$ u_{L}^{c2}$&$ - \stackrel{03}{[-i]}\,\stackrel{12}{[+]}|\stackrel{56}{[+]}\,\stackrel{78}{[-]}
||\stackrel{9 \;10}{[-]}\;\;\stackrel{11\;12}{(+)}\;\;\stackrel{13\;14}{[-]}$ &-1&$\frac{1}{2}$&
$\frac{1}{2}$&0 &$-\frac{1}{2}$&$\frac{1}{2\,\sqrt{3}}$&$\frac{1}{6}$&$\frac{1}{6}$&$\frac{2}{3}$\\
\hline
16&$u_{L}^{c2}$&$\stackrel{03}{(+i)}\,\stackrel{12}{(-)}|\stackrel{56}{[+]}\,\stackrel{78}{[-]}
||\stackrel{9 \;10}{[-]}\;\;\stackrel{11\;12}{(+)}\;\;\stackrel{13\;14}{[-]}$&-1&$-\frac{1}{2}$&
$\frac{1}{2}$&0&$-\frac{1}{2}$&$\frac{1}{2\,\sqrt{3}}$&$\frac{1}{6}$&$\frac{1}{6}$&$\frac{2}{3}$\\
\hline\hline
17&$ u_{R}^{c3}$&$ \stackrel{03}{(+i)}\,\stackrel{12}{[+]}|
\stackrel{56}{[+]}\,\stackrel{78}{(+)}
||\stackrel{9 \;10}{[-]}\;\;\stackrel{11\;12}{[-]}\;\;\stackrel{13\;14}{(+)} $ &1&$\frac{1}{2}$&0&
$\frac{1}{2}$&$0$&$-\frac{1}{\sqrt{3}}$&$\frac{1}{6}$&$\frac{2}{3}$&$\frac{2}{3}$\\
\hline
18&$u_{R}^{c3}$&$\stackrel{03}{[-i]}\,\stackrel{12}{(-)}|\stackrel{56}{[+]}\,\stackrel{78}{(+)}
||\stackrel{9 \;10}{[-]}\;\;\stackrel{11\;12}{[-]}\;\;\stackrel{13\;14}{(+)}$&1&$-\frac{1}{2}$&0&
$\frac{1}{2}$&$0$&$-\frac{1}{\sqrt{3}}$&$\frac{1}{6}$&$\frac{2}{3}$&$\frac{2}{3}$\\
\hline
19&$d_{R}^{c3}$&$\stackrel{03}{(+i)}\,\stackrel{12}{[+]}|\stackrel{56}{(-)}\,\stackrel{78}{[-]}
||\stackrel{9 \;10}{[-]}\;\;\stackrel{11\;12}{[-]}\;\;\stackrel{13\;14}{(+)}$&1&$\frac{1}{2}$&0&
$-\frac{1}{2}$&$0$&$-\frac{1}{\sqrt{3}}$&$\frac{1}{6}$&$-\frac{1}{3}$&$-\frac{1}{3}$\\
\hline
20&$ d_{R}^{c3} $&$\stackrel{03}{[-i]}\,\stackrel{12}{(-)}|
\stackrel{56}{(-)}\,\stackrel{78}{[-]}
||\stackrel{9 \;10}{[-]}\;\;\stackrel{11\;12}{[-]}\;\;\stackrel{13\;14}{(+)} $&1&$-\frac{1}{2}$&0&
$-\frac{1}{2}$&$0$&$-\frac{1}{\sqrt{3}}$&$\frac{1}{6}$&$-\frac{1}{3}$&$-\frac{1}{3}$\\
\hline
21&$d_{L}^{c3}$&$\stackrel{03}{[-i]}\,\stackrel{12}{[+]}|\stackrel{56}{(-)}\,\stackrel{78}{(+)}
||\stackrel{9 \;10}{[-]}\;\;\stackrel{11\;12}{[-]}\;\;\stackrel{13\;14}{(+)}$&-1&$\frac{1}{2}$&
$-\frac{1}{2}$&0&$0$&$-\frac{1}{\sqrt{3}}$&$\frac{1}{6}$&$\frac{1}{6}$&$-\frac{1}{3}$\\
\hline
22&$d_{L}^{c3} $&$ - \stackrel{03}{(+i)}\,\stackrel{12}{(-)}|\stackrel{56}{(-)}\,\stackrel{78}{(+)}
||\stackrel{9 \;10}{[-]}\;\;\stackrel{11\;12}{[-]}\;\;\stackrel{13\;14}{(+)} $&-1&$-\frac{1}{2}$&
$-\frac{1}{2}$&0&$0$&$-\frac{1}{\sqrt{3}}$&$\frac{1}{6}$&$\frac{1}{6}$&$-\frac{1}{3}$\\
\hline
23&$ u_{L}^{c3}$&$ - \stackrel{03}{[-i]}\,\stackrel{12}{[+]}|\stackrel{56}{[+]}\,\stackrel{78}{[-]}
||\stackrel{9 \;10}{[-]}\;\;\stackrel{11\;12}{[-]}\;\;\stackrel{13\;14}{(+)}$ &-1&$\frac{1}{2}$&
$\frac{1}{2}$&0 &$0$&$-\frac{1}{\sqrt{3}}$&$\frac{1}{6}$&$\frac{1}{6}$&$\frac{2}{3}$\\
\hline
24&$u_{L}^{c3}$&$\stackrel{03}{(+i)}\,\stackrel{12}{(-)}|\stackrel{56}{[+]}\,\stackrel{78}{[-]}
||\stackrel{9 \;10}{[-]}\;\;\stackrel{11\;12}{[-]}\;\;\stackrel{13\;14}{(+)}$&-1&$-\frac{1}{2}$&
$\frac{1}{2}$&0&$0$&$-\frac{1}{\sqrt{3}}$&$\frac{1}{6}$&$\frac{1}{6}$&$\frac{2}{3}$\\
\hline\hline
25&$ \nu_{R}$&$ \stackrel{03}{(+i)}\,\stackrel{12}{[+]}|
\stackrel{56}{[+]}\,\stackrel{78}{(+)}
||\stackrel{9 \;10}{(+)}\;\;\stackrel{11\;12}{(+)}\;\;\stackrel{13\;14}{(+)} $ &1&$\frac{1}{2}$&0&
$\frac{1}{2}$&$0$&$0$&$-\frac{1}{2}$&$0$&$0$\\
\hline
26&$\nu_{R}$&$\stackrel{03}{[-i]}\,\stackrel{12}{(-)}|\stackrel{56}{[+]}\,\stackrel{78}{(+)}
||\stackrel{9 \;10}{(+)}\;\;\stackrel{11\;12}{(+)}\;\;\stackrel{13\;14}{(+)}$&1&$-\frac{1}{2}$&0&
$\frac{1}{2}$ &$0$&$0$&$-\frac{1}{2}$&$0$&$0$\\
\hline
27&$e_{R}$&$\stackrel{03}{(+i)}\,\stackrel{12}{[+]}|\stackrel{56}{(-)}\,\stackrel{78}{[-]}
||\stackrel{9 \;10}{(+)}\;\;\stackrel{11\;12}{(+)}\;\;\stackrel{13\;14}{(+)}$&1&$\frac{1}{2}$&0&
$-\frac{1}{2}$&$0$&$0$&$-\frac{1}{2}$&$-1$&$-1$\\
\hline
28&$ e_{R} $&$\stackrel{03}{[-i]}\,\stackrel{12}{(-)}|
\stackrel{56}{(-)}\,\stackrel{78}{[-]}
||\stackrel{9 \;10}{(+)}\;\;\stackrel{11\;12}{(+)}\;\;\stackrel{13\;14}{(+)} $&1&$-\frac{1}{2}$&0&
$-\frac{1}{2}$&$0$&$0$&$-\frac{1}{2}$&$-1$&$-1$\\
\hline
29&$e_{L}$&$\stackrel{03}{[-i]}\,\stackrel{12}{[+]}|\stackrel{56}{(-)}\,\stackrel{78}{(+)}
||\stackrel{9 \;10}{(+)}\;\;\stackrel{11\;12}{(+)}\;\;\stackrel{13\;14}{(+)}$&-1&$\frac{1}{2}$&
$-\frac{1}{2}$&0&$0$&$0$&$-\frac{1}{2}$&$-\frac{1}{2}$&$-1$\\
\hline
30&$e_{L} $&$ - \stackrel{03}{(+i)}\,\stackrel{12}{(-)}|\stackrel{56}{(-)}\,\stackrel{78}{(+)}
||\stackrel{9 \;10}{(+)}\;\;\stackrel{11\;12}{(+)}\;\;\stackrel{13\;14}{(+)} $&-1&$-\frac{1}{2}$&
$-\frac{1}{2}$&0&$0$&$0$&$-\frac{1}{2}$&$-\frac{1}{2}$&$-1$\\
\hline
31&$ \nu_{L}$&$ - \stackrel{03}{[-i]}\,\stackrel{12}{[+]}|\stackrel{56}{[+]}\,\stackrel{78}{[-]}
||\stackrel{9 \;10}{(+)}\;\;\stackrel{11\;12}{(+)}\;\;\stackrel{13\;14}{(+)}$ &-1&$\frac{1}{2}$&
$\frac{1}{2}$&0 &$0$&$0$&$-\frac{1}{2}$&$-\frac{1}{2}$&$0$\\
\hline
32&$\nu_{L}$&$\stackrel{03}{(+i)}\,\stackrel{12}{(-)}|\stackrel{56}{[+]}\,\stackrel{78}{[-]}
||\stackrel{9 \;10}{(+)}\;\;\stackrel{11\;12}{(+)}\;\;\stackrel{13\;14}{(+)}$&-1&$-\frac{1}{2}$&
$\frac{1}{2}$&0&$0$&$0$&$-\frac{1}{2}$&$-\frac{1}{2}$&$0$\\
\hline\hline
33&$ \bar{d}_{L}^{\bar{c1}}$&$ \stackrel{03}{[-i]}\,\stackrel{12}{[+]}|
\stackrel{56}{[+]}\,\stackrel{78}{(+)}
||\stackrel{9 \;10}{[-]}\;\;\stackrel{11\;12}{(+)}\;\;\stackrel{13\;14}{(+)} $ &-1&$\frac{1}{2}$&0&
$\frac{1}{2}$&$-\frac{1}{2}$&$-\frac{1}{2\,\sqrt{3}}$&$-\frac{1}{6}$&$\frac{1}{3}$&$\frac{1}{3}$\\
\hline
34&$\bar{d}_{L}^{\bar{c1}}$&$\stackrel{03}{(+i)}\,\stackrel{12}{(-)}|\stackrel{56}{[+]}\,\stackrel{78}{(+)}
||\stackrel{9 \;10}{[-]}\;\;\stackrel{11\;12}{(+)}\;\;\stackrel{13\;14}{(+)}$&-1&$-\frac{1}{2}$&0&
$\frac{1}{2}$&$-\frac{1}{2}$&$-\frac{1}{2\,\sqrt{3}}$&$-\frac{1}{6}$&$\frac{1}{3}$&$\frac{1}{3}$\\
\hline
35&$\bar{u}_{L}^{\bar{c1}}$&$ - \stackrel{03}{[-i]}\,\stackrel{12}{[+]}|\stackrel{56}{(-)}\,\stackrel{78}{[-]}
||\stackrel{9 \;10}{[-]}\;\;\stackrel{11\;12}{(+)}\;\;\stackrel{13\;14}{(+)}$&-1&$\frac{1}{2}$&0&
$-\frac{1}{2}$&$-\frac{1}{2}$&$-\frac{1}{2\,\sqrt{3}}$&$-\frac{1}{6}$&$-\frac{2}{3}$&$-\frac{2}{3}$\\
\hline
36&$ \bar{u}_{L}^{\bar{c1}} $&$ - \stackrel{03}{(+i)}\,\stackrel{12}{(-)}|
\stackrel{56}{(-)}\,\stackrel{78}{[-]}
||\stackrel{9 \;10}{[-]}\;\;\stackrel{11\;12}{(+)}\;\;\stackrel{13\;14}{(+)} $&-1&$-\frac{1}{2}$&0&
$-\frac{1}{2}$&$-\frac{1}{2}$&$-\frac{1}{2\,\sqrt{3}}$&$-\frac{1}{6}$&$-\frac{2}{3}$&$-\frac{2}{3}$\\
\hline
37&$\bar{d}_{R}^{\bar{c1}}$&$\stackrel{03}{(+i)}\,\stackrel{12}{[+]}|\stackrel{56}{[+]}\,\stackrel{78}{[-]}
||\stackrel{9 \;10}{[-]}\;\;\stackrel{11\;12}{(+)}\;\;\stackrel{13\;14}{(+)}$&1&$\frac{1}{2}$&
$\frac{1}{2}$&0&$-\frac{1}{2}$&$-\frac{1}{2\,\sqrt{3}}$&$-\frac{1}{6}$&$-\frac{1}{6}$&$\frac{1}{3}$\\
\hline
38&$\bar{d}_{R}^{\bar{c1}} $&$ - \stackrel{03}{[-i]}\,\stackrel{12}{(-)}|\stackrel{56}{[+]}\,\stackrel{78}{[-]}
||\stackrel{9 \;10}{[-]}\;\;\stackrel{11\;12}{(+)}\;\;\stackrel{13\;14}{(+)} $&1&$-\frac{1}{2}$&
$\frac{1}{2}$&0&$-\frac{1}{2}$&$-\frac{1}{2\,\sqrt{3}}$&$-\frac{1}{6}$&$-\frac{1}{6}$&$\frac{1}{3}$\\
\hline
39&$ \bar{u}_{R}^{\bar{c1}}$&$\stackrel{03}{(+i)}\,\stackrel{12}{[+]}|\stackrel{56}{(-)}\,\stackrel{78}{(+)}
||\stackrel{9 \;10}{[-]}\;\;\stackrel{11\;12}{(+)}\;\;\stackrel{13\;14}{(+)}$ &1&$\frac{1}{2}$&
$-\frac{1}{2}$&0 &$-\frac{1}{2}$&$-\frac{1}{2\,\sqrt{3}}$&$-\frac{1}{6}$&$-\frac{1}{6}$&$-\frac{2}{3}$\\
\hline
40&$\bar{u}_{R}^{\bar{c1}}$&$\stackrel{03}{[-i]}\,\stackrel{12}{(-)}|\stackrel{56}{(-)}\,\stackrel{78}{(+)}
||\stackrel{9 \;10}{[-]}\;\;\stackrel{11\;12}{(+)}\;\;\stackrel{13\;14}{(+)}$
&1&$-\frac{1}{2}$&
$-\frac{1}{2}$&0&$-\frac{1}{2}$&$-\frac{1}{2\,\sqrt{3}}$&$-\frac{1}{6}$&$-\frac{1}{6}$&$-\frac{2}{3}$\\
\hline\hline
41&$ \bar{d}_{L}^{\bar{c2}}$&$ \stackrel{03}{[-i]}\,\stackrel{12}{[+]}|
\stackrel{56}{[+]}\,\stackrel{78}{(+)}
||\stackrel{9 \;10}{(+)}\;\;\stackrel{11\;12}{[-]}\;\;\stackrel{13\;14}{(+)} $
&-1&$\frac{1}{2}$&0&
$\frac{1}{2}$&$\frac{1}{2}$&$-\frac{1}{2\,\sqrt{3}}$&$-\frac{1}{6}$&$\frac{1}{3}$&$\frac{1}{3}$\\
\hline
42&$\bar{d}_{L}^{\bar{c2}}$&$\stackrel{03}{(+i)}\,\stackrel{12}{(-)}|\stackrel{56}{[+]}\,\stackrel{78}{(+)}
||\stackrel{9 \;10}{(+)}\;\;\stackrel{11\;12}{[-]}\;\;\stackrel{13\;14}{(+)}$
&-1&$-\frac{1}{2}$&0&
$\frac{1}{2}$&$\frac{1}{2}$&$-\frac{1}{2\,\sqrt{3}}$&$-\frac{1}{6}$&$\frac{1}{3}$&$\frac{1}{3}$\\
\hline
43&$\bar{u}_{L}^{\bar{c2}}$&$ - \stackrel{03}{[-i]}\,\stackrel{12}{[+]}|\stackrel{56}{(-)}\,\stackrel{78}{[-]}
||\stackrel{9 \;10}{(+)}\;\;\stackrel{11\;12}{[-]}\;\;\stackrel{13\;14}{(+)}$
&-1&$\frac{1}{2}$&0&
$-\frac{1}{2}$&$\frac{1}{2}$&$-\frac{1}{2\,\sqrt{3}}$&$-\frac{1}{6}$&$-\frac{2}{3}$&$-\frac{2}{3}$\\
\hline
44&$ \bar{u}_{L}^{\bar{c2}} $&$ - \stackrel{03}{(+i)}\,\stackrel{12}{(-)}|
\stackrel{56}{(-)}\,\stackrel{78}{[-]}
||\stackrel{9 \;10}{(+)}\;\;\stackrel{11\;12}{[-]}\;\;\stackrel{13\;14}{(+)} $
&-1&$-\frac{1}{2}$&0&
$-\frac{1}{2}$&$\frac{1}{2}$&$-\frac{1}{2\,\sqrt{3}}$&$-\frac{1}{6}$&$-\frac{2}{3}$&$-\frac{2}{3}$\\
\hline
45&$\bar{d}_{R}^{\bar{c2}}$&$\stackrel{03}{(+i)}\,\stackrel{12}{[+]}|\stackrel{56}{[+]}\,\stackrel{78}{[-]}
||\stackrel{9 \;10}{(+)}\;\;\stackrel{11\;12}{[-]}\;\;\stackrel{13\;14}{(+)}$
&1&$\frac{1}{2}$&
$\frac{1}{2}$&0&$\frac{1}{2}$&$-\frac{1}{2\,\sqrt{3}}$&$-\frac{1}{6}$&$-\frac{1}{6}$&$\frac{1}{3}$\\
\hline
46&$\bar{d}_{R}^{\bar{c2}} $&$ - \stackrel{03}{[-i]}\,\stackrel{12}{(-)}|\stackrel{56}{[+]}\,\stackrel{78}{[-]}
||\stackrel{9 \;10}{(+)}\;\;\stackrel{11\;12}{[-]}\;\;\stackrel{13\;14}{(+)} $
&1&$-\frac{1}{2}$&
$\frac{1}{2}$&0&$\frac{1}{2}$&$-\frac{1}{2\,\sqrt{3}}$&$-\frac{1}{6}$&$-\frac{1}{6}$&$\frac{1}{3}$\\
\hline
47&$ \bar{u}_{R}^{\bar{c2}}$&$\stackrel{03}{(+i)}\,\stackrel{12}{[+]}|\stackrel{56}{(-)}\,\stackrel{78}{(+)}
||\stackrel{9 \;10}{(+)}\;\;\stackrel{11\;12}{[-]}\;\;\stackrel{13\;14}{(+)}$
 &1&$\frac{1}{2}$&
$-\frac{1}{2}$&0 &$\frac{1}{2}$&$-\frac{1}{2\,\sqrt{3}}$&$-\frac{1}{6}$&$-\frac{1}{6}$&$-\frac{2}{3}$\\
\hline
48&$\bar{u}_{R}^{\bar{c2}}$&$\stackrel{03}{[-i]}\,\stackrel{12}{(-)}|\stackrel{56}{(-)}\,\stackrel{78}{(+)}
||\stackrel{9 \;10}{(+)}\;\;\stackrel{11\;12}{[-]}\;\;\stackrel{13\;14}{(+)}$
&1&$-\frac{1}{2}$&
$-\frac{1}{2}$&0&$\frac{1}{2}$&$-\frac{1}{2\,\sqrt{3}}$&$-\frac{1}{6}$&$-\frac{1}{6}$&$-\frac{2}{3}$\\
\hline\hline
49&$ \bar{d}_{L}^{\bar{c3}}$&$ \stackrel{03}{[-i]}\,\stackrel{12}{[+]}|
\stackrel{56}{[+]}\,\stackrel{78}{(+)}
||\stackrel{9 \;10}{(+)}\;\;\stackrel{11\;12}{(+)}\;\;\stackrel{13\;14}{[-]} $ &-1&$\frac{1}{2}$&0&
$\frac{1}{2}$&$0$&$\frac{1}{\sqrt{3}}$&$-\frac{1}{6}$&$\frac{1}{3}$&$\frac{1}{3}$\\
\hline
50&$\bar{d}_{L}^{\bar{c3}}$&$\stackrel{03}{(+i)}\,\stackrel{12}{(-)}|\stackrel{56}{[+]}\,\stackrel{78}{(+)}
||\stackrel{9 \;10}{(+)}\;\;\stackrel{11\;12}{(+)}\;\;\stackrel{13\;14}{[-]} $&-1&$-\frac{1}{2}$&0&
$\frac{1}{2}$&$0$&$\frac{1}{\sqrt{3}}$&$-\frac{1}{6}$&$\frac{1}{3}$&$\frac{1}{3}$\\
\hline
51&$\bar{u}_{L}^{\bar{c3}}$&$ - \stackrel{03}{[-i]}\,\stackrel{12}{[+]}|\stackrel{56}{(-)}\,\stackrel{78}{[-]}
||\stackrel{9 \;10}{(+)}\;\;\stackrel{11\;12}{(+)}\;\;\stackrel{13\;14}{[-]} $&-1&$\frac{1}{2}$&0&
$-\frac{1}{2}$&$0$&$\frac{1}{\sqrt{3}}$&$-\frac{1}{6}$&$-\frac{2}{3}$&$-\frac{2}{3}$\\
\hline
52&$ \bar{u}_{L}^{\bar{c3}} $&$ - \stackrel{03}{(+i)}\,\stackrel{12}{(-)}|
\stackrel{56}{(-)}\,\stackrel{78}{[-]}
||\stackrel{9 \;10}{(+)}\;\;\stackrel{11\;12}{(+)}\;\;\stackrel{13\;14}{[-]}  $&-1&$-\frac{1}{2}$&0&
$-\frac{1}{2}$&$0$&$\frac{1}{\sqrt{3}}$&$-\frac{1}{6}$&$-\frac{2}{3}$&$-\frac{2}{3}$\\
\hline
53&$\bar{d}_{R}^{\bar{c3}}$&$\stackrel{03}{(+i)}\,\stackrel{12}{[+]}|\stackrel{56}{[+]}\,\stackrel{78}{[-]}
||\stackrel{9 \;10}{(+)}\;\;\stackrel{11\;12}{(+)}\;\;\stackrel{13\;14}{[-]} $&1&$\frac{1}{2}$&
$\frac{1}{2}$&0&$0$&$\frac{1}{\sqrt{3}}$&$-\frac{1}{6}$&$-\frac{1}{6}$&$\frac{1}{3}$\\
\hline
54&$\bar{d}_{R}^{\bar{c3}} $&$ - \stackrel{03}{[-i]}\,\stackrel{12}{(-)}|\stackrel{56}{[+]}\,\stackrel{78}{[-]}
||\stackrel{9 \;10}{(+)}\;\;\stackrel{11\;12}{(+)}\;\;\stackrel{13\;14}{[-]} $&1&$-\frac{1}{2}$&
$\frac{1}{2}$&0&$0$&$\frac{1}{\sqrt{3}}$&$-\frac{1}{6}$&$-\frac{1}{6}$&$\frac{1}{3}$\\
\hline
55&$ \bar{u}_{R}^{\bar{c3}}$&$\stackrel{03}{(+i)}\,\stackrel{12}{[+]}|\stackrel{56}{(-)}\,\stackrel{78}{(+)}
||\stackrel{9 \;10}{(+)}\;\;\stackrel{11\;12}{(+)}\;\;\stackrel{13\;14}{[-]} $ &1&$\frac{1}{2}$&
$-\frac{1}{2}$&0 &$0$&$\frac{1}{\sqrt{3}}$&$-\frac{1}{6}$&$-\frac{1}{6}$&$-\frac{2}{3}$\\
\hline
56&$\bar{u}_{R}^{\bar{c3}}$&$\stackrel{03}{[-i]}\,\stackrel{12}{(-)}|\stackrel{56}{(-)}\,\stackrel{78}{(+)}
||\stackrel{9 \;10}{(+)}\;\;\stackrel{11\;12}{(+)}\;\;\stackrel{13\;14}{[-]} $&1&$-\frac{1}{2}$&
$-\frac{1}{2}$&0&$0$&$\frac{1}{\sqrt{3}}$&$-\frac{1}{6}$&$-\frac{1}{6}$&$-\frac{2}{3}$\\
\hline\hline
57&$ \bar{e}_{L}$&$ \stackrel{03}{[-i]}\,\stackrel{12}{[+]}|
\stackrel{56}{[+]}\,\stackrel{78}{(+)}
||\stackrel{9 \;10}{[-]}\;\;\stackrel{11\;12}{[-]}\;\;\stackrel{13\;14}{[-]} $ &-1&$\frac{1}{2}$&0&
$\frac{1}{2}$&$0$&$0$&$\frac{1}{2}$&$1$&$1$\\
\hline
58&$\bar{e}_{L}$&$\stackrel{03}{(+i)}\,\stackrel{12}{(-)}|\stackrel{56}{[+]}\,\stackrel{78}{(+)}
||\stackrel{9 \;10}{[-]}\;\;\stackrel{11\;12}{[-]}\;\;\stackrel{13\;14}{[-]}$&-1&$-\frac{1}{2}$&0&
$\frac{1}{2}$ &$0$&$0$&$\frac{1}{2}$&$1$&$1$\\
\hline
59&$\bar{\nu}_{L}$&$ - \stackrel{03}{[-i]}\,\stackrel{12}{[+]}|\stackrel{56}{(-)}\,\stackrel{78}{[-]}
||\stackrel{9 \;10}{[-]}\;\;\stackrel{11\;12}{[-]}\;\;\stackrel{13\;14}{[-]}$&-1&$\frac{1}{2}$&0&
$-\frac{1}{2}$&$0$&$0$&$\frac{1}{2}$&$0$&$0$\\
\hline
60&$ \bar{\nu}_{L} $&$ - \stackrel{03}{(+i)}\,\stackrel{12}{(-)}|
\stackrel{56}{(-)}\,\stackrel{78}{[-]}
||\stackrel{9 \;10}{[-]}\;\;\stackrel{11\;12}{[-]}\;\;\stackrel{13\;14}{[-]} $&-1&$-\frac{1}{2}$&0&
$-\frac{1}{2}$&$0$&$0$&$\frac{1}{2}$&$0$&$0$\\
\hline
61&$\bar{\nu}_{R}$&$\stackrel{03}{(+i)}\,\stackrel{12}{[+]}|\stackrel{56}{(-)}\,\stackrel{78}{(+)}
||\stackrel{9 \;10}{[-]}\;\;\stackrel{11\;12}{[-]}\;\;\stackrel{13\;14}{[-]}$&1&$\frac{1}{2}$&
$-\frac{1}{2}$&0&$0$&$0$&$\frac{1}{2}$&$\frac{1}{2}$&$0$\\
\hline
62&$\bar{\nu}_{R} $&$ - \stackrel{03}{[-i]}\,\stackrel{12}{(-)}|\stackrel{56}{(-)}\,\stackrel{78}{(+)}
||\stackrel{9 \;10}{[-]}\;\;\stackrel{11\;12}{[-]}\;\;\stackrel{13\;14}{[-]} $&1&$-\frac{1}{2}$&
$-\frac{1}{2}$&0&$0$&$0$&$\frac{1}{2}$&$\frac{1}{2}$&$0$\\
\hline
63&$ \bar{e}_{R}$&$\stackrel{03}{(+i)}\,\stackrel{12}{[+]}|\stackrel{56}{[+]}\,\stackrel{78}{[-]}
||\stackrel{9 \;10}{[-]}\;\;\stackrel{11\;12}{[-]}\;\;\stackrel{13\;14}{[-]}$ &1&$\frac{1}{2}$&
$\frac{1}{2}$&0 &$0$&$0$&$\frac{1}{2}$&$\frac{1}{2}$&$1$\\
\hline
64&$\bar{e}_{R}$&$\stackrel{03}{[-i]}\,\stackrel{12}{(-)}|\stackrel{56}{[+]}\,\stackrel{78}{[-]}
||\stackrel{9 \;10}{[-]}\;\;\stackrel{11\;12}{[-]}\;\;\stackrel{13\;14}{[-]}$&1&$-\frac{1}{2}$&
$\frac{1}{2}$&0&$0$&$0$&$\frac{1}{2}$&$\frac{1}{2}$&$1$\\
\hline
\end{supertabular}
\end{tiny}


%

%
\section{Some useful relations in Grassmann and Clifford algebras, needed also in App.~\ref{13+1representation} }
\label{grassmannandcliffordfermions}

The generator of the Lorentz transformation in Grassmann space is defined as follows~\cite{norma93}
\begin{eqnarray}
\label{Lorentztheta}
{\cal {\bf S}}^{ab} &=&  (\theta^a p^{\theta b} - \theta^b p^{\theta a})\,\nonumber\\
&=& S^{ab} +\tilde{S}^{ab} \,, \quad  \{S^{ab}, \tilde{S}^{cd}\}_{-} =0\,,
\end{eqnarray}
where $S^{ab}$ and $\tilde{S}^{ab}$ are the two corresponding generators of the 
Lorentz transformations of the two Clifford sub algebras, forming orthogonal 
representations with respect to each other, Eq.~(\ref{gammatildeantiher}).

The infinitesimal generators of the Lorentz transformations in the two 
 subspaces, defined with the two Clifford subalgebras are 
\begin{eqnarray}
\label{Lorentzgammatilde}
S^{ab} &=& \frac{i}{4} (\gamma^a \gamma^b - \gamma^b \gamma^a)\,, \quad
S^{ab \dagger} = \eta^{aa} \eta^{bb} S^{ab}\,,\nonumber\\
\tilde{S}^{ab} &=& \frac{i}{4} (\tilde{\gamma}^a \tilde{\gamma}^b -
\tilde{\gamma}^b\tilde{\gamma}^a) \,,  \quad \tilde{S}^{ab \dagger} =
\eta^{aa} \eta^{bb} \tilde{S}^{ab}\,,
\end{eqnarray}
where $\gamma^a$ and $\tilde{\gamma}^a$ are defined in Eqs.~(\ref{clifftheta1},
\ref{gammatildeantiher}). 
The commutation relations for either ${\cal {\bf S}}^{ab}$ or $S^{ab}$ or 
$\tilde{S}^{ab}$, ${\cal {\bf S}}^{ab} = S^{ab} + \tilde{S}^{ab}$, are 
%
\begin{eqnarray}
\label{LorentzthetaCliffcom}
\{S^{ab}, \tilde{S}^{cd}\}_{-}&=& 0\,, \nonumber\\
\{S^{ab},S^{cd}\}_{-} &=& i (\eta^{ad} S^{bc} + \eta^{bc} S^{ad} -
 \eta^{ac} S^{bd} - \eta^{bd} S^{ac})\,,\nonumber\\
\{\tilde{S}^{ab},\tilde{S}^{cd}\}_{-} &=& i(\eta^{ad} \tilde{S}^{bc} + 
\eta^{bc} \tilde{S}^{ad} 
- \eta^{ac} \tilde{S}^{bd} - \eta^{bd} \tilde{S}^{ac})\,,\nonumber\\
\{{\cal {\bf S}}^{ab}, {\cal {\bf S}}^{cd}\}_{-}&=& i (\eta^{ad} {\cal {\bf S}}^{bc}
+ \eta^{bc} {\cal {\bf S}}^{ad} - \eta^{ac} {\cal {\bf S}}^{bd}-
\eta^{bd} {\cal {\bf S}}^{ac}\,.
\end{eqnarray}
The infinitesimal generators of the two invariant subgroups of the group $SO(3,1)$ can be expressed as follows
\begin{eqnarray}
\label{so1+3}
\vec{N}_{\pm}(= \vec{N}_{(L,R)}): &=& \,\frac{1}{2} (S^{23}\pm i S^{01},S^{31}\pm i S^{02}, 
S^{12}\pm i S^{03} )\,.
\end{eqnarray}
The infinitesimal generators of the two invariant subgroups of the group $SO(4)$ are expressible with
$S^{ab}, (a,b) = (5,6,7,8)$ as follows 
 \begin{eqnarray}
 \label{so42}
 \vec{\tau}^{1}:&=&\frac{1}{2} (S^{58}-  S^{67}, \,S^{57} + S^{68}, \,S^{56}-  S^{78} )\,,
\nonumber\\
 \vec{\tau}^{2}:&=& \frac{1}{2} (S^{58}+  S^{67}, \,S^{57} - S^{68}, \,S^{56}+  S^{78} )\,,
 \end{eqnarray}
while the generators of the $SU(3)$ and  $U(1)$ subgroups of the group $SO(6)$ can be expressed by
$S^{ab}, (a,b) = (9,10,11,12,13,14)$
 \begin{eqnarray}
 \label{so64}
 \vec{\tau}^{3}: = &&\frac{1}{2} \,\{  S^{9\;12} - S^{10\;11} \,,
  S^{9\;11} + S^{10\;12} ,\, S^{9\;10} - S^{11\;12} ,\nonumber\\
 && S^{9\;14} -  S^{10\;13} ,\,  S^{9\;13} + S^{10\;14} \,,
  S^{11\;14} -  S^{12\;13}\,,\nonumber\\
 && S^{11\;13} +  S^{12\;14} ,\, 
 \frac{1}{\sqrt{3}} ( S^{9\;10} + S^{11\;12} - 
 2 S^{13\;14})\}\,,\nonumber\\
 \tau^{4}: = &&-\frac{1}{3}(S^{9\;10} + S^{11\;12} + S^{13\;14})\,.
 \end{eqnarray}
 The group $SO(6)$ has $\frac{d (d-1)}{2}=15$  generators and $\frac{d}{2}=3$ 
 commuting operators.  The subgroups $SU(3)$  $\times U(1)$ have the same number
 of commuting operators, expressed with $\tau^{33}$, $\tau^{38}$ and  $\tau^4$, 
 and $9$ generators, $8$ of $SU(3)$ and one of $U(1)$. The rest of $6$ generators, 
 not included in $SU(3)$  $\times U(1)$, can be expressed as $\frac{1}{2} \,\{  S^{9\;12}   +S^{10\;11},   S^{9\;11} - S^{10\;12}$, $S^{9\;14} + S^{10\;13}, S^{9\;13} - 
 S^{10\;14},   S^{11\;14} +  S^{12\;13},  S^{11\;13} -  S^{12\;14} $.\\

The hyper charge $Y$ can be defined as $Y=\tau^{23} + \tau^{4}$. 

The equivalent expressions for the "family" charges, expressed by $\tilde{S}^{ab},
$ follow if in Eqs.~(\ref{so1+3} - \ref{so64}) $S^{ab}$ are replaced by $\tilde{S}^{ab}$.

\section*{Acknowledgment} 
The author thanks Holger Bech Nielsen for very fruitful discussions, Department of Physics, FMF, University of Ljubljana, Society of Mathematicians, Physicists and Astronomers of Slovenia,  for supporting the research on the {\it spin-charge-family} theory by offering the room and computer facilities and Matja\v z Breskvar of Beyond Semiconductor for donations, in particular for the annual workshops entitled "What comes beyond the standard models". 



\begin{thebibliography}{}




\bibitem{norma92}  N. Manko\v c Bor\v stnik, "Spin connection as a 
              superpartner of a vielbein", {\it Phys. Lett.} {\bf B 292} (1992)  25-29.
\bibitem{norma93} N. Manko\v c Bor\v stnik, "Spinor and vector representations in four dimensional Grassmann
              space", {\it J. of Math. Phys.} {\bf 34} (1993) 3731-3745.
\bibitem{norma95} N. Manko\v c Bor\v stnik, ''Unification of spin and charges 
            in Grassmann space?'', hep-th 9408002, IJS.TP.94/22, 
            Mod. Phys. Lett.{\bf A (10)} No.7 (1995) 587-595.
%
 \bibitem{nh2021RPPNP}   N. S. Manko\v c Bor\v stnik, H. B. Nielsen,
 "How does Clifford algebra 
            show the way to the second quantized fermions with unified spins, 
            charges and families, and with vector and scalar gauge fields beyond 
            the {\it standard model}", Progress in Particle and Nuclear Physics,
           http://doi.org/10.1016.j.ppnp.2021.103890 .  
 %
\bibitem{2020PartIPartII}   N.S. Manko\v c Bor\v stnik, H.B.F. Nielsen, 
"Understanding the second  quantization of fermions in Clifford and in Grassmann 
space",
             {\it New way of second quantization of fermions ---  Part I and Part II}, in this 
             proceedings [arXiv:2007.03517, arXiv:2007.03516]. 
             %
\bibitem{n2021SQ}    N. S. Manko\v c Bor\v stnik, ''How do Clifford algebras show 
             the way to the second quantized fermions with unified spins, charges and 
             families, and to the corresponding second quantized vector and scalar 
             gauge field '', Proceedings  to  the $24^{rd}$ Workshop 
             "What comes    beyond the standard models", 5 - 11 of July, 2021, 
             Ed. N.S. Manko\v c Bor\v stnik, H.B. Nielsen, D. Lukman, A. Kleppe, DMFA  
             Zalo\v zni\v stvo,  Ljubljana, December 2021, [arXiv:2112.04378] .               
\bibitem{n2022SQ}    N. S. Manko\v c Bor\v stnik,  ''How Clifford algebra can help 
             understand second quantization of fermion and boson fields'', 
              [arXiv:2112.04378].
 %
\bibitem{n2019PIPII} N.S. Manko\v c Bor\v stnik, H.B.F. Nielsen, "Understanding the second 
             quantization of fermions in Clifford and in Grassmann space"
             {\it New way of second quantization of fermions ---  Part I and Part II},
              Proceedings  to  the $22^{nd}$ Workshop "What comes 
             beyond the standard models", 6 - 14 of July, 2019, Ed. N.S. Manko\v c Bor\v stnik, 
            H.B. Nielsen, D. Lukman, DMFA  Zalo\v zni\v stvo, Ljubljana, December 2019,
   [arXiv:1802.05554v4,  arXiv:1902.10628].                   
 %
\bibitem{Dirac} P.A.M. Dirac {\it Proc. Roy. Soc. (London)}, {\bf A 117} (1928) 610.
\bibitem{BetheJackiw} H.A. Bethe, R.W. Jackiw, "Intermediate quantum mechanics",
New York : W.A. Benjamin, 1968.
\bibitem{Weinberg} S. Weinberg, "The quantum theory of fields", Cambridge, 
Cambridge University Press, 2015.
%
 \bibitem{prd2018} N.S. Manko\v c Bor\v stnik, H.B.F. Nielsen, "New way of second 
             quantized theory of  fermions with either Clifford or Grassmann coordinates
             and  {\it spin-charge-family} theory " 
             [arXiv:1802.05554v4,arXiv:1902.10628]. 
 \bibitem{ND2018Grass}      D. Lukman, N. S. Manko\v c Bor\v stnik, 
          "Properties of fermions with integer spin described with
            Grassmann algebra", Proceedings  to  the $21^{st}$ Workshop "What comes 
             beyond the standard models", 23 of June - 1 of July, 2018, Ed. N.S. Manko\v c Bor\v stnik, 
            H.B. Nielsen, D. Lukman, DMFA  Zalo\v zni\v stvo, Ljubljana, December 2018         
           [arxiv:1805.06318, arXiv:1902.10628].
\bibitem{nh02}  N.S. Manko\v c Bor\v stnik, H.B.F. Nielsen, {\it J. of Math. Phys.} {\bf 43}, 
5782 (2002) [arXiv:hep-th/0111257].
\bibitem{nh03} N.S. Manko\v c Bor\v stnik, H.B.F. Nielsen,
``How to generate families of spinors'',
{\it J. of Math. Phys.} {\bf 44} 4817 (2003) [arXiv:hep-th/0303224].
%
\bibitem{IARD2016} N.S. Manko\v c Bor\v stnik, "Spin-charge-family theory is offering next step
             in understanding elementary particles and fields and correspondingly universe", 
             Proceedings to the Conference on Cosmology, Gravitational Waves and Particles, 
             IARD conferences, Ljubljana, 6-9 June 2016, The $10^{th}$ Biennial Conference 
             on Classical and Quantum Relativistic Dynamics of Particles and Fields,  
                          J. Phys.: Conf. Ser. 845 012017 
    [arXiv:1409.4981, arXiv:1607.01618v2].
%
\bibitem{IARD2020} N.S. Manko\v c Bor\v stnik, "The attributes of the 
             Spin-Charge-Family theory giving hope that the theory offers the next step 
             beyond the Standard Model", Proceedings to the $12^{th}$ Bienal 
             Conference on Classical and Quantum Relativistic Dynamics of Particles and 
             Fields IARD 2020, Prague, $1 - 4$ June 2020  Journal of Physics, Conference 
             Series volume 1956  012020, 2021, 
              iopscience.iop.org/issue/1742-6596/1956/1.
%
\bibitem{n2014matterantimatter}  N.S. Manko\v c Bor\v stnik, 
"Matter-antimatter asymmetry in the {\it spin-charge-family} theory", 
{\it Phys. Rev.} {\bf D 91} (2015) 065004  [arXiv:1409.7791]. 
\bibitem{normaBled2020} N. S. Manko\v c Bor\v stnik, ''How far has so far the Spin-Charge-Family theory succeeded to explain the Standard Model assumptions, the matter-antimatter asymmetry,
the appearance of the Dark Matter, the second quantized fermion fields...., making several predictions'',  Proceedings  to  the $23^{rd}$ Workshop "What comes 
             beyond the standard models", 4 - 12 of July, 2020 
             Ed. N.S. Manko\v c Bor\v stnik, H.B. Nielsen, D. Lukman, DMFA  Zalo\v zni\v stvo, Ljubljana, December 2020, [arXiv:2012.09640]
\bibitem{nd2017} N.S. Manko\v c Bor\v stnik, D. Lukman, "Vector and scalar gauge fields with
              respect to $d=(3+1)$ in Kaluza-Klein theories and in the {\it spin-charge-family theory}",
              {\it Eur. Phys. J. C} {\bf 77} (2017) 231.
\bibitem{n2012scalars} N.S. Manko\v c Bor\v stnik, "The {\it spin-charge-family} theory explains  
              why  the scalar Higgs carries the weak charge $\pm \frac{1}{2}$ and the hyper charge 
              $ \mp \frac{1}{2}$",
              Proceedings to 
              the $17^{th}$ Workshop "What comes beyond the standard models", Bled, 20-28 of July, 2014, 
              Ed. N.S. Manko\v c Bor\v stnik, H.B. Nielsen, D. Lukman, DMFA  Zalo\v zni\v stvo, 
              Ljubljana December 2014, p.163-82 [ arXiv:1502.06786v1] [arXiv:1409.4981].  
\bibitem{JMP2013} N.S. Manko\v c Bor\v stnik N S, "The spin-charge-family theory is explaining the
 origin of families, of the Higgs and the Yukawa couplings", {\it J. of Modern Phys.} {\bf 4} (2013) 823
[arXiv:1312.1542].
\bibitem{nh2017} N.S. Manko\v c Bor\v stnik, H.B.F. Nielsen, "The spin-charge-family theory 
             offers understanding of the triangle anomalies cancellation in the standard model",
             {\it Fortschritte der Physik, Progress of Physics} (2017) 1700046.
\bibitem{normaJMP2015} N.S. Manko\v c Bor\v stnik, "The explanation for the origin of the 
Higgs  scalar and for the Yukawa couplings by the {\it spin-charge-family} theory", 
{\it J.of Mod. Physics} {\bf 6} (2015) 2244-2274, http://dx.org./10.4236/jmp.2015.615230
              [arXiv:1409.4981].
\bibitem{nh2018}  N.S. Manko\v c Bor\v stnik and H.B.  Nielsen, "Why nature made a choice of 
Clifford and not  Grassmann coordinates",   Proceedings  to  the $20^{th}$ Workshop "What comes 
beyond the standard models", Bled, 9-17 of July, 2017, Ed. N.S. Manko\v c Bor\v stnik, H.B. Nielsen,
D. Lukman, DMFA  Zalo\v zni\v stvo, Ljubljana, December 2017, p. 89-120 
[arXiv:1802.05554v1v2].
%
\bibitem{nhds} N.S. Manko\v c Bor\v stnik and H.B.F. Nielsen,
"Discrete symmetries in the Kaluza-Klein theories",
 {\em JHEP} 04:165,  2014 [arXiv:1212.2362].
%
%
\bibitem{n2021MDPIsymmetry} N. S. Manko\v c Bor\v stnik, Second quantized ''anticommuting integer spin fields'',  
              sent to  arXiv.         
 %
 \bibitem{pikanorma} A. Bor\v stnik,  N.S.  Manko\v c Bor\v stnik, ''Left and
              right handedness of fermions and bosons'', 
              J. of Phys. G:  Nucl. Part. 
              Phys.{\bf 24}(1998) 963-977, hep-th/9707218.  
                      %
 \bibitem{pikanorma2005} A. Bor\v stnik Bra\v ci\v c, N. S. Manko\v c Bor\v stnik, 
               ''On the origin 
               of  families of fermions and their mass matrices'', hep-ph/0512062,  
               Phys Rev. {\bf D 74} 073013-28  (2006).
 %
 \bibitem{mdn2006} M. Breskvar, D. Lukman, N. S. Manko\v c Bor\v stnik, 
                     ''On the Origin of Families of Fermions and Their Mass Matrices\,---\,%
                   Approximate Analyses of Properties of Four Families Within 
                   Approach Unifying Spins and Charges", 
                     Proceedings to the $9^{\rm th}$ Workshop ''What Comes Beyond the Standard 
                     Models'', Bled, Sept. 16 - 26, 2006,  Ed. by Norma Manko\v c Bor\v stnik, 
		     Holger Bech Nielsen, Colin Froggatt, Dragan Lukman, DMFA Zalo\v zni\v stvo, 
                     Ljubljana December 2006, p.25-50, hep-ph/0612250.
%
\bibitem{gmdn2007} G. Bregar, M. Breskvar, D. Lukman, N.S. Manko\v c Bor\v stnik,
                    "Families of Quarks and Leptons and Their Mass Matrices", 
                    Proceedings to the $10^{th}$ international workshop ''What Comes Beyond 
		    the Standard Model'', 17 -27 of July, 2007, Ed. Norma Manko\v c  
		    Bor\v stnik, Holger Bech Nielsen, Colin Froggatt, Dragan Lukman,
		    DMFA  Zalo\v zni\v stvo, Ljubljana December 2007, 
		    p.53-70, hep-ph/0711.4681.
%
\bibitem{gmdn2008} G. Bregar, M. Breskvar, D. Lukman, N.S. Manko\v c Bor\v stnik, 
                     "Predictions for four families by the Approach unifying spins and charges"
                     {\it New J. of Phys.} {\bf 10} (2008) 093002,
                     hep-ph/0606159, hep/ph-07082846.
%
              %
\bibitem{gn2009} G. Bregar, N.S. Manko\v c Bor\v stnik, "Does dark matter consist of baryons 
	        of new stable family quarks?", {\it Phys. Rev. D } {\bf 80}, 083534 (2009), 1-16.
%
\bibitem{gn2013}  G. Bregar, N.S. Manko\v c Bor\v stnik, "Can we predict the fourth family masses 
              for quarks and leptons?", Proceedings (arxiv:1403.4441) to the 16 th Workshop "What comes beyond the 
              standard models", Bled, 14-21 of July, 2013, Ed. N.S. Manko\v c Bor\v stnik, 
              H.B. Nielsen, D. Lukman, DMFA  Zalo\v zni\v stvo, Ljubljana December 2013, p. 31-51, 
              http://arxiv.org/abs/1212.4055.

%
\bibitem{gn2014} G. Bregar, N.S. Manko\v c Bor\v stnik, "The new experimental data 
             for the quarks     mixing matrix are in better agreement with the 
             {\it spin-charge-family} theory predictions",                      
              Proceedings to the $17^th$ Workshop "What comes beyond the 
              standard models", Bled, 20-28 of July, 2014, 
              Ed. N.S. Manko\v c Bor\v stnik, H.B. Nielsen, D. Lukman, DMFA  Zalo\v zni\v stvo, 
              Ljubljana December 2014, p.20-45 [ arXiv:1502.06786v1] [arxiv:1412.5866].  	        
%
\bibitem{nm2015} N.S. Manko\v c Bor\v stnik, M. Rosina, "Are superheavy stable quark 
              clusters viable   candidates  for the dark matter?",
        International Journal of Modern Physics D (IJMPD) {\bf 24} (No. 13) (2015) 1545003.
%
%
	             

   \end{thebibliography}
   \end{document}